\documentclass[fleqn,usenatbib]{mnras}

\usepackage{newtxtext,newtxmath}

\usepackage[T1]{fontenc}
\usepackage{ae,aecompl}
\usepackage[normalem]{ulem}

\usepackage{graphicx}	
\usepackage{amsmath}	

\usepackage{xcolor,colortbl}

\newcommand{\simgt}{\lower.5ex\hbox{$\; \buildrel > \over \sim \;$}}
\newcommand{\simlt}{\lower.5ex\hbox{$\; \buildrel < \over \sim \;$}}

\title[Proper motion measurements for stars in Stripe~82]{
Proper motion measurements for stars up to $100$~kpc with Subaru HSC and SDSS Stripe~82}

\author[Qiu, Wang, Takada et al.]{Tian~Qiu$^{1,2}$\thanks{Contact e-mail: \href{tian.qiu@ipmu.jp}{tian.qiu@ipmu.jp}},
Wenting~Wang$^{3}$\thanks{Contact e-mail: \href{wenting.wang@sjtu.edu.cn}{wenting.wang@sjtu.edu.cn}},
Masahiro~Takada$^{1}$\thanks{Contact e-mail: \href{masahiro.takada@ipmu.jp}{masahiro.takada@ipmu.jp}},
Naoki~Yasuda$^{1}$,
{\v Z}eljko Ivezi{\'c}$^{4}$,
\newauthor
Robert~H.~Lupton$^{5}$,
Masashi~Chiba$^{6}$,
Miho~Ishigaki$^{6}$,
Yutaka~Komiyama$^{7}$
\\
$^{1}$ Kavli Institute for the Physics and Mathematics of the Universe (WPI), The University of Tokyo Institutes for Advanced Study (UTIAS), 
\\The University of Tokyo, 5-1-5 Kashiwanoha, Kashiwa-shi, Chiba, 277-8583, Japan
\\
$^{2}$ Department of Physics, The University of Tokyo, 7-3-1 Hongo, Bunkyo-ku, Tokyo 113-0033 Japan\\
$^{3}$ Department of Astronomy, Shanghai Jiao Tong University, Shanghai 200240, China\\
$^{4}$ Department of Astronomy, University of Washington, Box 351580, Seattle, WA 98195, USA \\
$^{5}$ Department of Astrophysical Sciences, Princeton University, Peyton Hall, Princeton NJ 08544 USA\\
$^{6}$ Astronomical Institute, Tohoku University, Aoba-ku, Sendai 980-8578, Japan\\
$^{7}$ National Astronomical Observatory of Japan, Mitaka, Tokyo 181-8588, Japan
}

\date{Accepted 2020 December 21. Received 2020 December 21; in original form 2020 April 27}

\pubyear{2020}

\begin{document}
\label{firstpage}
\pagerange{\pageref{firstpage}--\pageref{lastpage}}
\maketitle

\begin{abstract}
We present proper motion measurements for more than 0.55 million main-sequence stars, by comparing astrometric positions of matched 
stars between the multi-band imaging datasets from the Hyper Suprime-Cam (HSC) Survey and the SDSS Stripe~82. In doing this we use 3 
million galaxies to recalibrate the astrometry and set up a common reference frame between the two catalogues. The exquisite depth and 
the nearly 12 years of time baseline between HSC and SDSS enable high-precision measurements of statistical proper motions for stars 
down to $i\simeq 24$. A validation of our method is demonstrated by the agreement with the {\it Gaia} proper motions, to the precision 
better than $0.1$~mas~yr$^{-1}$. To retain the precision, we make a correction of the subtle effects due to the differential chromatic 
refraction in the SDSS images based on the comparison with the {\it Gaia} proper motions against colour of stars, 
which is validated using the SDSS spectroscopic quasars. 
Combining with the photometric distance estimates for individual stars based on the precise HSC 
photometry, we show a significant detection of the net proper motions for stars in each bin of distance out to 100~kpc. The two-component 
tangential velocities after subtracting the apparent motions due to our own motion display rich phase-space structures including a 
clear signature of the Sagittarius stream in the halo region of distance range $[10,35]$~kpc. We also measure the tangential 
velocity dispersion in the distance range 5--20~kpc and find that the data are consistent with a constant isotropic dispersion of
$80\pm 10~{\rm km/s}$. More distant stars appear to have random motions with respect to the Galactic centre on average.
\end{abstract}

\begin{keywords}
proper motions -- methods: statistical -- Galaxy: kinematics and dynamics
\end{keywords}



\section{Introduction}
\label{sec:intro}

Our Milky Way (MW) Galaxy provides a wealth of information on the physics of galaxy formation and the underlying cosmology 
\citep[e.g.][]{1999ApJ...524L..19M,1999ApJ...522...82K,2011MNRAS.415L..40B,2015MNRAS.452.3838C,2018Natur.563...85H,
2018MNRAS.481..286L,2019NatAs...3..932G,2019ApJ...880...38B}. The information can be revealed from the positions, 
velocities and chemistry of stars in the Galaxy, and from its satellites, globular clusters and tidal streams, which give us
very powerful diagnostics to probe kinematical structures of the MW halo as well as the assembly history of the MW in phase 
space. The readers can see \cite{2020arXiv200204340H} for a review.

Among the 6-dimensional phase space information, proper motions of stars in the MW is most expensive to be 
measured precisely, especially for faint stars out to large distances. It usually requires long time baselines, 
accurate astrometry and careful corrections of systematic errors. However, full phase-space information is very 
informative for proper scientific inferences. For example, if the velocity anisotropy of stars cannot be directly
inferred from data due to unavailable proper motions, the MW mass is often estimated with large systematic 
uncertainties, which is the so-called ``mass-anisotropy'' degeneracy \citep[e.g][]{2005MNRAS.364..433B,2019arXiv191202599W}. 

Proper motions can be measured by comparing the astrometric positions of stars observed over different years. Early proper motion 
surveys rely on multiple exposures taken at different times of the same survey, which involve the nearly all-sky but shallow 
surveys \citep[e.g.][]{1980PMMin..55....1L,2000A&A...355L..27H,2000AJ....120.2131Z,2003AJ....125..984M}, and the deeper surveys 
covering smaller areas \citep[e.g.][]{2001MNRAS.326.1279H,2001MNRAS.326.1315H,2001ApJS..133..119R,2003AJ....126..921L}. 
Later on, with new wide and deeper surveys such as the Sloan Digital Sky Survey \citep[hereafter SDSS;][]{2000AJ....120.1579Y}, 
proper motions of stars can be further measured by comparing their astrometry in two or multiple surveys spanning even longer time 
baselines \citep[e.g.][]{2004ApJS..152..103G,munn2004improved,2013ApJ...766...79K,2014AJ....148..132M,tian2017gaia,2018MNRAS.473..647D}.

Since proper motions are measured by comparing the astrometric positions, any systematic errors in astrometry can lead to systematics 
in measured proper motions. The most significant error is due to the different reference frames for observations conducted at different 
times, which is caused mainly due to the motion of astrometric reference stars. This is often calibrated against distant quasars or galaxies, 
which have zero proper motions, given a limited accuracy due to statistical errors. Thus apparent ``proper motions'' of galaxies or quasars, 
if observed, reflect the underlying systematics. For example, \cite{2004ApJS..152..103G} used quasars to correct for proper motions in 
SDSS/USNO-A and USNO-B sources. \cite{munn2004improved} further provided an improved proper motion catalogue to \cite{2004ApJS..152..103G}. 
The USNO-B positions were recalibrated using SDSS galaxies. The catalogue is 90\% complete down to $g\sim19.7$, with statistical errors of 
$\sim$3~mas~$\mathrm{yr^{-1}}$. Combined with photometric parallax and metallicity \citep{2008ApJ...673..864J,2008ApJ...684..287I}, this 
catalogue was used to study the kinematics of stars in a distance range from 100~pc to $\sim$13~kpc \citep{bond2010milky}. 

More recently, \cite{2013ApJ...766...79K} used positional offsets between stars and quasars at multiple epochs over 7~years from 
the SDSS Stripe~82 data \citep{2014ApJ...794..120A}, and then measure statistical proper motions for stars at $r\lesssim 21$ in the
Sagittarius stream, with the precision of $\sim0.1~$mas~yr$^{-1}$. \cite{2014AJ....148..132M} provided deeper proper motion measurements 
by matching objects in SDSS DR7 to two independent observations, each covering more than 1000~deg$^2$ down to $r=22$ and $r=20.9$ 
respectively. Systematic errors were estimated to be less than 1~mas~$\mathrm{yr^{-1}}$ and 2--4~mas~$\mathrm{yr^{-1}}$ for the two 
observations, with typical statistical errors of 5 to 15~mas~$\mathrm{yr^{-1}}$ from the bright to faint end. 

Moreover, by combining 2MASS, SDSS, Pan-STARRS1 and {\it Gaia} DR1 astrometry, the {\it Gaia}-PS1-SDSS (hereafter GPS1) proper motion 
catalogue \citep{tian2017gaia} contains 350 million sources down to $r\sim20$, which covers nearly 3/4 of the sky. \cite{tian2017gaia} 
used galaxies to set up a reference frame and bring different observations to this frame. The typical systematic and statistical errors 
are $<0.3$~mas~$\mathrm{yr^{-1}}$ and 1.5--2.0~mas~$\mathrm{yr^{-1}}$, respectively. \cite{2018MNRAS.473..647D} studied the proper motions
across the Galactic anticentre region using  recalibrated SDSS astrometry in combination with positions from {\it Gaia} DR1. 

{\it Gaia} is producing proper motion measurements with unprecedented high precisions. The {\it Gaia} second data release (DR2)
provides position, parallax and proper motion solutions for about 1.3 billion objects \citep{2018A&A...616A...1G}. The proper motion 
uncertainties of {\it Gaia} DR2 are typically 0.06~mas~$\mathrm{yr^{-1}}$ at $G<15$, 0.2~mas~$\mathrm{yr^{-1}}$ at $G=17$ and 
1.2~mas~$\mathrm{yr^{-1}}$ at $G=20$ \citep{2016A&A...595A...1G,2018A&A...616A...1G}. The precision is expected to further 
improve with the increased duration by future {\it Gaia} observations. 

The proper motion measurements/catalogues introduced above, however, are still relatively shallow. For example, the {\it Gaia} proper 
motions are mainly useful for science of nearby main-sequence stars (distances up to a few kpc) and rare populations of
 distant, luminous stars such as RR-Lyrae stars and K-giants (distances up to at most a few tenth of kpc). The upcoming Rubin Observatory's 
 LSST \citep[][Legacy Survey of Space and Time]{2017ASPC..512..279J,2019ApJ...873..111I} survey is able to measure 
proper motions for a sample of $\sim$200 million F/G main-sequence stars out to a distance of $\sim$100~kpc. LSST is designed to be 
able to image $\sim$10,000 square degrees of the southern sky per night, repeatedly over 10-year operation
\citep{2019ApJ...873..111I,2017arXiv170804058L}.

In this paper, we use the ongoing,  deep Hyper Suprime-Cam Subaru Strategic Program Survey \citep[][hereafter HSC-SSP or HSC]
{2018PASJ...70S...4A} to achieve precise measurements of statistical proper motions for main-sequence stars down to much fainter 
magnitudes $i\simeq 24$: we compare the HSC astrometry with the deep SDSS Stripe~82 data, for the overlapping region of about 100~deg$^2$. 
In many aspects, HSC is a pre-implemented version for the future LSST survey, with similar depth and data quality. In our analysis, 
instead of using sources detected on single exposure images, we compare the astrometry of sources detected on coadd images in both HSC 
and SDSS, which enables us to push down to $r\sim24$, i.e., 2 to 4 magnitudes fainter than previous proper motion catalogues\footnote{The 
light-motion catalogue in the SDSS Stripe~82 region is available \citep{bramich2008light}, which is complete down to $r\sim21.5$. Proper 
motions are derived by recalibrating the astrometry against some mid-run data. In this paper, we are not directly using these proper 
motions in our study.}. However, though HSC and SDSS Stripe~82 data are on average separated by 12 years, the SDSS Stripe~82 spans 
7-8 years of observing duration. Thus, we take into account the mean epoch for individual stars taken from each of the HSC and 
Stripe~82 coadd images, by averaging all input exposure images. The difference between HSC and Stripe~82 mean epochs is then adopted 
as the time baseline for the proper motion measurements of individual stars. 

In our analysis, we also use galaxies to do astrometry recalibration and bring the reference frames between HSC and SDSS Stripe~82 
to be the same. Photometric distances/parallaxes are estimated for individual stars based on the precise HSC multi-band photometry
\citep{2008ApJ...673..864J,2008ApJ...684..287I}, which enable us to measure the tangential velocity as a function of the photometric 
distance. We validate our proper motion measurements by comparing with the {\it Gaia} DR2 proper motions for the matched stars. 
We construct judiciously selected sub-samples of stars to harness the power of statistics and present the first significant detection 
of proper motions for main-sequence stars at distances  up to 100~kpc. 

This paper is structured as follows. We introduce the surveys and corresponding source catalogues in Section~\ref{sec:data}, including 
stars and galaxies of HSC and SDSS Stripe 82, {\it Gaia} DR2 stars and SDSS DR14 quasars. In Section~\ref{sec:method} we describe
our methods of matching sources between different surveys, calibration of systematic errors and calculation of photometric distances.
We present detailed results based on the measured proper motions and the comparisons/validations through {\it Gaia} proper motions 
in Section~\ref{sec:result}. In Section~\ref{sec:disc} we discuss an implication of our proper motion measurements for the Sagittarius 
stream, and then discuss possible systematic and statistical uncertainties inherent in our measurements. We conclude in the end 
(Section~\ref{sec:concl}). In Appendix, we also give details of our analysis.

\section{Data}
\label{sec:data}

\begin{table}
 \begin{center}
 \begin{tabular}{lr}
  \hline
  \hline
  Type & Number of objects\\[2pt]
  \hline
  HSC total ($i_\mathrm{HSC}<24.5$)& 8 273 848\\
  S82 total ($i_\mathrm{S82}<24.2$)& 4 841 985\\[2pt]\hline
  Matched total & 4 302 686 \\[2pt]
  Galaxies for recalibration & 3 153 981 \\[2pt]
  Quasars & 8 757 \\[2pt]
  Stars for proper motion measurements & 555 529 \\[2pt] 
  Blue stars & 97 087\\[2pt]
  Stars matched with {\it Gaia} & 117 703 \\[2pt]
  \hline
 \end{tabular}
 \end{center}
 \caption{Number of objects used in this paper. ``HSC total'' denotes the total number of detected objects with $i_{\rm HSC}<24.5$
 in the HSC data and in the overlapping footprints of HSC and S82 (about 100~deg$^2$). ``S82 total'' denotes the total number of detected 
 objects  with $i_{\rm S82}<24.2$ in the S82 data in  the overlapping footprints. Note, however, that the overlapping footprints of HSC 
 and S82 are not exactly the same due to the difference in masked regions (e.g. around bright stars). These are our master samples of 
 objects. Other rows denote the number of different objects: the matched objects (``Matched total''), the matched galaxies (``Galaxies
 for recalibration''), the matched quasars with the SDSS DR14 spectroscopically-confirmed quasars in the S82 region
 (``Quasars''), the 
 stars used for proper motion measurements, the ``blue'' stars with $g_\mathrm{PSF}-r_\mathrm{PSF}<0.6$ among the matched stars, 
 and the number of matched stars with the {\it Gaia} catalogue among the matched stars of HSC and S82.}
 \label{tab:num}
\end{table}
In this section we describe details of data/catalogues, the Hyper Suprime-Cam and the SDSS that we use in this paper.
We summarise the number of objects used in this paper in Table~\ref{tab:num}.

\subsection{SDSS Stripe 82 (S82)}
\label{subsec:S82}
The Sloan Digital Sky Survey (SDSS) used a dedicated wide-field 2.5m telescope \citep{2000AJ....120.1579Y,2006AJ....131.2332G} 
to image the sky in drift-scanning mode with five optical filters, $ugriz$ \citep{1996AJ....111.1748F,1998AJ....116.3040G}. 
The effective single exposure time per filter is 54.1 seconds. For stellar sources, the 50\% completeness limits are estimated
as $u,g,r,i,z\simeq 22.5, 23.2, 22.6, 21.9, 20.8,$ respectively \citep{2003AJ....126.2081A}.

The SDSS imaged a 275~deg$^2$ region, multiple times, in the Celestial Equatorial region, so-called Stripe~82 (hereafter 
we call the data in the SDSS Stripe~82 region for short as S82). The coadd images in $ugriz$ are $\sim 1.5$ magnitudes deeper than 
those in the SDSS single-pass data \citep{2014ApJ...794..120A,2014ApJS..213...12J}.
The region is $2.5^{\circ}$ wide along Dec. and traces the Celestial Equator in the South Galactic Cap, more precisely 
the contiguous region in the ranges of $-50^{\circ}\le {\rm R.A.}\le 60^{\circ}$ and $-1.25^{\circ}\le {\rm Dec.}\le 1.25^{\circ}$. 
The publicly-available data and catalogues of S82 are constructed from the coadds of $\sim 20$ multiple images, taken from 1998 September 
to 2005 November. The median seeing is 1.1$\arcsec$ and the 50\% completeness limits are $r=23.5$ for galaxies and $r=24.3$ for stars, 
respectively.

In this study we download the data from CasJobs, with the detailed sample selection query provided in Appendix~\ref{app:query}.
About 100~deg$^2$ of the S82 footprint overlaps with HSC as we describe below, which gives about 5~million objects with 
$r<24.2$ in total. This is the master S82 catalogue. However, as we will show later, the purity of star catalogues in S82 is low 
for stars at $i \simgt 21$, due to relatively poor seeing conditions compared to HSC.

For the positions of stars/galaxies/quasars, we use their centroid positions in $r$-band as our fiducial choice. We will also 
use the centroid positions of the other filters to test possible effects of systematic errors on our results.

\subsection{Subaru HSC}
\label{subsec:hsc}

The Hyper Suprime-Cam (HSC) camera at the prime focus of the 8.2m Subaru Telescope is a wide-field imaging camera with 1.77~deg$^2$ 
field-of-view \citep{2018PASJ...70S...1M,2018PASJ...70S...2K}. The HSC survey is using about 330 nights of Subaru time, since 2014, to 
conduct a multi-band wide-field imaging survey \citep{2018PASJ...70S...4A}. The HSC data we use in this paper is taken from the ongoing 
HSC-Wide survey, which is a part of the three layers with different combinations of depth and area, and aims at covering about 1,400~deg$^2$ 
of the sky with five filters, $grizy$, with $5\sigma$ point-source depths of $(g,r,i,z,y)\simeq (26.5, 26.1, 25.9, 25.1, 24.4)$ 
measured in $2\arcsec$ aperture \citep{2018PASJ...70...66K,2018PASJ...70S...3F}. 
The total exposure times range from 10~min in the $g$- and $r$-bands to 20~min in the $i$-, $z$-, and $y$-bands, divided into 
individual exposures of $\sim 3$~min each. 

Since $i$-band images are used for galaxy shape measurements in weak lensing analysis, $i$-band images are preferentially taken when the 
seeing is better \citep{2018PASJ...70S..25M}. The HSC $i$-band images have a median PSF FWHM of $0.6\arcsec$. The superb image quality 
brings benefits for our study because the HSC images allow for a robust selection of stars or equivalently a less contamination of galaxies 
to the star catalogue as we will show below. The details of the software used to reduce the data are given in \citet{2018PASJ...70S...5B} 
\citep[also see][]{2018PASJ...70S...3F}. The fundamental calibration of photometry and astrometry was done against the public Pan-STARRS 
catalogue \citep{2012ApJ...756..158S,2012ApJ...750...99T,2013ApJS..205...20M,2016arXiv161205560C,2016arXiv161205242M}.

In this paper we use the primary photometric sources with $i_{\rm HSC}<24.5$ from the S18a internal data release, which is almost identical 
to the second public data release \citep{2019PASJ..tmp..106A}. We have about 8.3~million objects as the master HSC catalogue. We adopt the 
data in the HSC-Wide survey regions that have the 5-filter data and the approximately full depths, in the S82 region. We provide the 
sample selection query in Appendix~\ref{app:query}.

For the positions of stars/galaxies/quasars, we use their centroid positions in $i$-band as our fiducial choice for the 
HSC data.

\subsection{Other datasets}
\label{subsec:other_datasets}

We also use {\it Gaia}\footnote{\url{https://www.cosmos.esa.int/web/gaia/data-release-2}} DR2 \citep{2018A&A...616A...1G} and 
SDSS DR14 quasars \citep{2018A&A...613A..51P} for validation and calibration of our proper motion measurements, as we will describe 
later. The numbers of DR14 quasars and {\it Gaia} stars, used in this paper, are given in Table~\ref{tab:num}.

\section{Methodology}
\label{sec:method}

In the following, we describe how to classify photometric sources as stars or galaxies, and how we can further improve the classification 
to reduce the contamination. After star/galaxy separation, we identify the same stars/galaxies in the HSC and S82 catalogues, by matching the 
positions of stars/galaxies between the two catalogues. We will also introduce our approach of using galaxies to correct for systematics in 
astrometric solutions of the two catalogues. 

\subsection{Star/Galaxy separation}
\label{sec:stargalsep}

\begin{figure}
\begin{center}
 \includegraphics[width=\columnwidth]{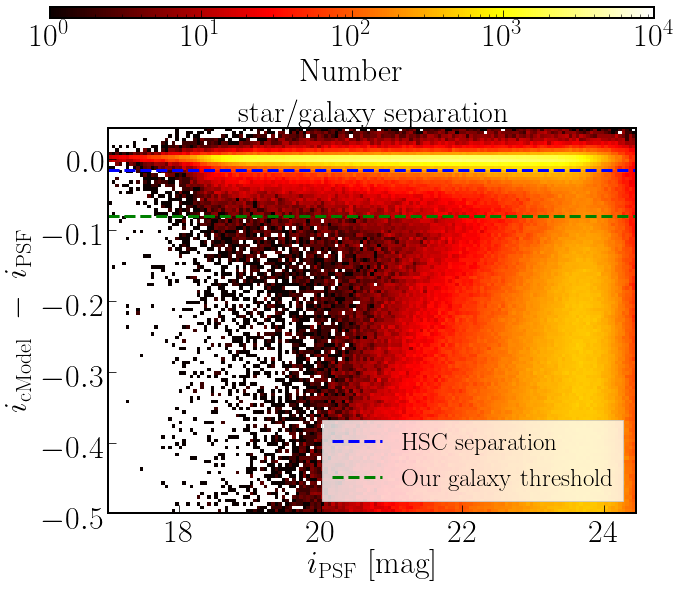}
 \caption{The number distribution of all objects in terms of HSC $i$-band PSF magnitudes ($x$-axis) versus HSC $i$-band 
 cModel minus PSF magnitudes ($y$-axis). Colour denotes the number of objects in each grid of the two dimensional space, 
 as denoted by the upper colour bar. We select stars with the cut of $i_\mathrm{HSC,cModel}-i_\mathrm{HSC,PSF}>-0.015$
 (dark blue dashed line). To ensure a pure sample of galaxies, we adopt a cut of $i_\mathrm{cModel}-i_\mathrm{HSC,PSF}<-0.08$ 
 (green dashed line). 
   }
 \label{fig:sep}
\end{center}
\end{figure}
\begin{figure}
\begin{center}
 \includegraphics[width=\columnwidth]{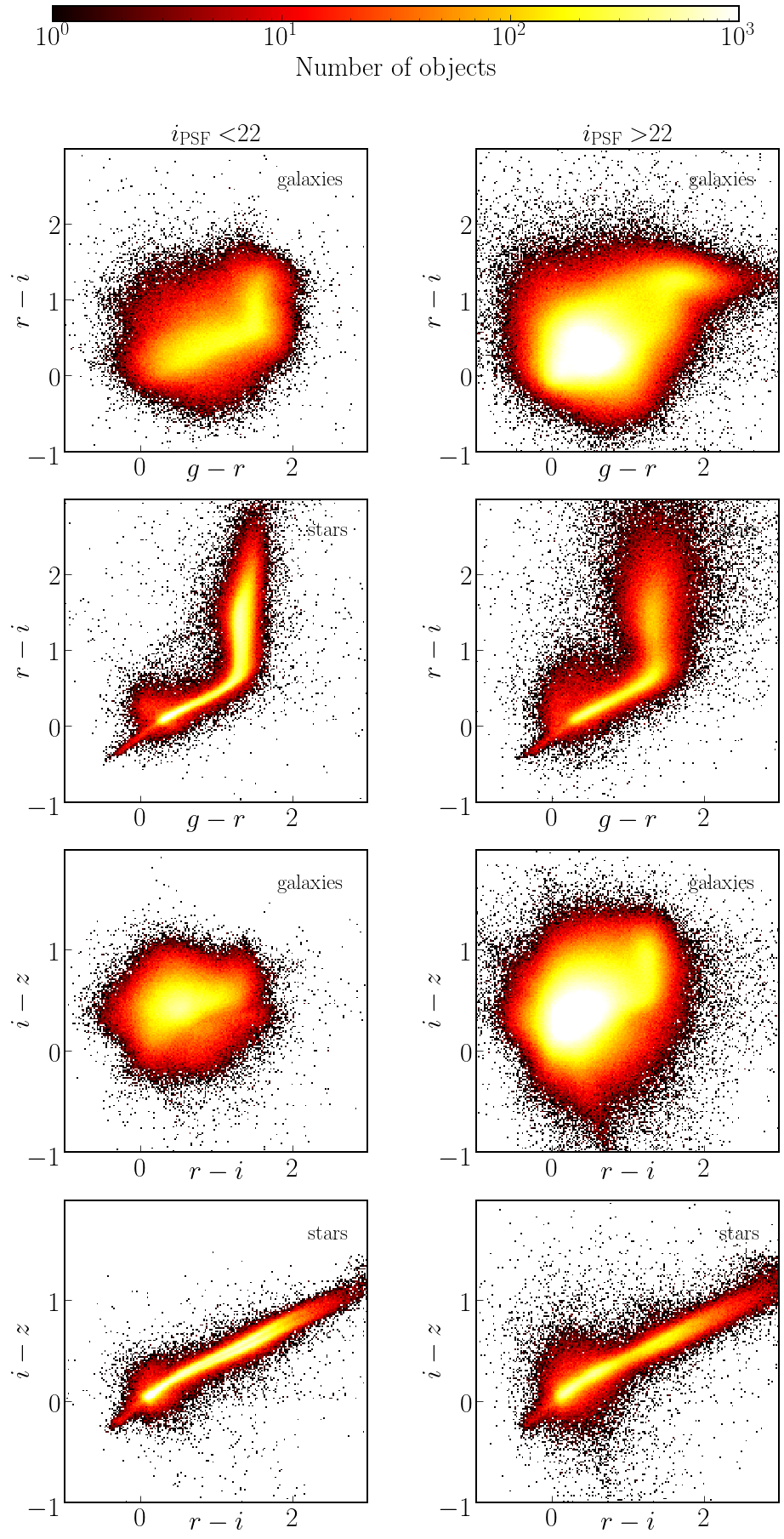}
 \caption{The colour-colour diagrams of stars and galaxies as denoted by the label in each panel. Here colour in each grid of panels 
 represents the number of objects in the grid, as denoted by the upper colour bar. To calculate colour for each object, we use the
 HSC cModel and PSF magnitudes for galaxies and stars, respectively.  The left-column panels show the results for bright objects 
 with $i_\mathrm{HSC,PSF}<22$, while the right panels show the results for faint objects ($i_\mathrm{HSC,PSF}>22$). There is no 
 obvious signs of galaxy contamination around stellar locus.}
 \label{fig:colour}
 \end{center}
\end{figure}
Both HSC and S82 have their own criteria for star/galaxy separation. For both data, a point source is identified on the 
basis of the ratio between PSF and cModel fluxes \citep{2004AJ....128..502A}. For HSC, $i$-band measurements have the 
best seeing so that we adopt the $i$-band ``extendedness'' parameter for the classification. We adopt the criterion 
$i_{\rm HSC, cModel}-i_{\rm HSC, PSF}>-0.015$ to define point sources.

For S82, $r$-band has the best resolution. Point sources are defined from objects satisfying 
$|r_\mathrm{S82,model}$-$r_\mathrm{S82,PSF}|\le 0.03$, which is much more stringent than the threshold 
(0.145) adopted for the nominal SDSS depth thanks to the deeper depth of SDSS S82.

As we will show later, we will also use a catalogue of galaxies to correct for 
systematic errors in astrometric solutions of stars between the HSC and S82 catalogues. 
To define a secure catalogue of galaxies, we use the HSC catalogue and define, as
galaxies, objects satisfying the condition of $i_\mathrm{HSC,cModel}-i_\mathrm{HSC,PSF}<-0.08$. 
The threshold value of $-0.08$ corresponds to the position of a dip in Figure~\ref{fig:sep}, 
which is determined empirically. We believe that the criterion can give a more stringent 
selection of galaxies because the threshold is sufficiently far from the threshold of star separation,
$i_\mathrm{HSC,cModel}-i_\mathrm{HSC,PSF}>-0.015$, and photometric errors are unlikely to cause a scatter 
of stars into this galaxy region thanks to the deep HSC photometry. 

In Figure~\ref{fig:colour} we show the distributions of stars and galaxies in the colour-colour plane, $(g-r)$ vs. 
$(r-i)$ and $(r-i)$ vs. $(i-z)$, for the bright and faint samples that are divided based on the HSC $i$-band apparent 
magnitudes, $i_\mathrm{HSC,PSF}<22$ or $>22$, respectively. Stars are distributed along a narrower locus in each 
colour space, while galaxies display a broader distribution reflecting the diversity of galaxy colours, partly due 
to their shifted colours with redshifts. We can use the colour information to define a more secure sample of galaxies, 
e.g. by using extended objects that are sufficiently away from the stellar locus. Then we will use such a sample to check whether 
our recalibration of astrometry using galaxies are robust against different galaxy samples (see
Section~\ref{sec:astrometry_calibration} for details).

The stellar locus in Figure~\ref{fig:colour} displays a broadened feature at some colour. These are 
either non-main-sequence stars, or caused by failures in photometry \citep[e.g.][]{2008ApJ...673..864J}. As we will introduce 
in Section~\ref{sec:photodis}, we calculate photometric distances for main-sequence stars based on their photometric colour. 
To ensure accurate colour measurements and also guarantee a minimal amount of contamination by non-main-sequence stars in our
measurements, we choose to exclude those stars (or star-like objects) that are
away from the stellar locus for our analysis throughout this paper.
Explicitly, we divide our sample of stars into different bins according to their $g-i$ colour, and for each bin, we make 
3-$\sigma$ clipping based on their $g-r$ colours. This clipping discards about 47,000 objects (discards about 12,000 objects
from the catalogue of blue stars which is the sample for our results beyond 30~kpc). 
Here we define the ``blue''
stars as those stars with colour $g-r<0.6$, which are probing distant halo stars rather than younger disk stars, following \cite{2008ApJ...684..287I}.
In fact, they are closer in colour to yellow stars (like our Sun). 
However, 
we still call them 
``blue''
stars throughout this paper since they are at the blue end of colour range probed by our data, which are also used to differentiate from 
``red''
stars whose colour $g-r$ larger than 0.6.
Besides, we discard those extremely blue stars whose $g-r$ or $g-i$ colour less than 0.2 to avoid the contamination of white dwarfs, 
as we will discuss in Section~\ref{subsec:contamination} in detail.
This cut is also desired because 
the photometric distance estimation also requires a $g-i$ colour limitation from 0.2 to 4.0 as given by Eq.~(A7) in \cite{2008ApJ...673..864J}. 
This is a minor fraction compared to the number of remaining stars used in the following analysis as can be found from Table~\ref{tab:num}.

\subsection{Methodology of proper motion measurements}
\begin{figure}
\begin{center}
 \includegraphics[width=\columnwidth]{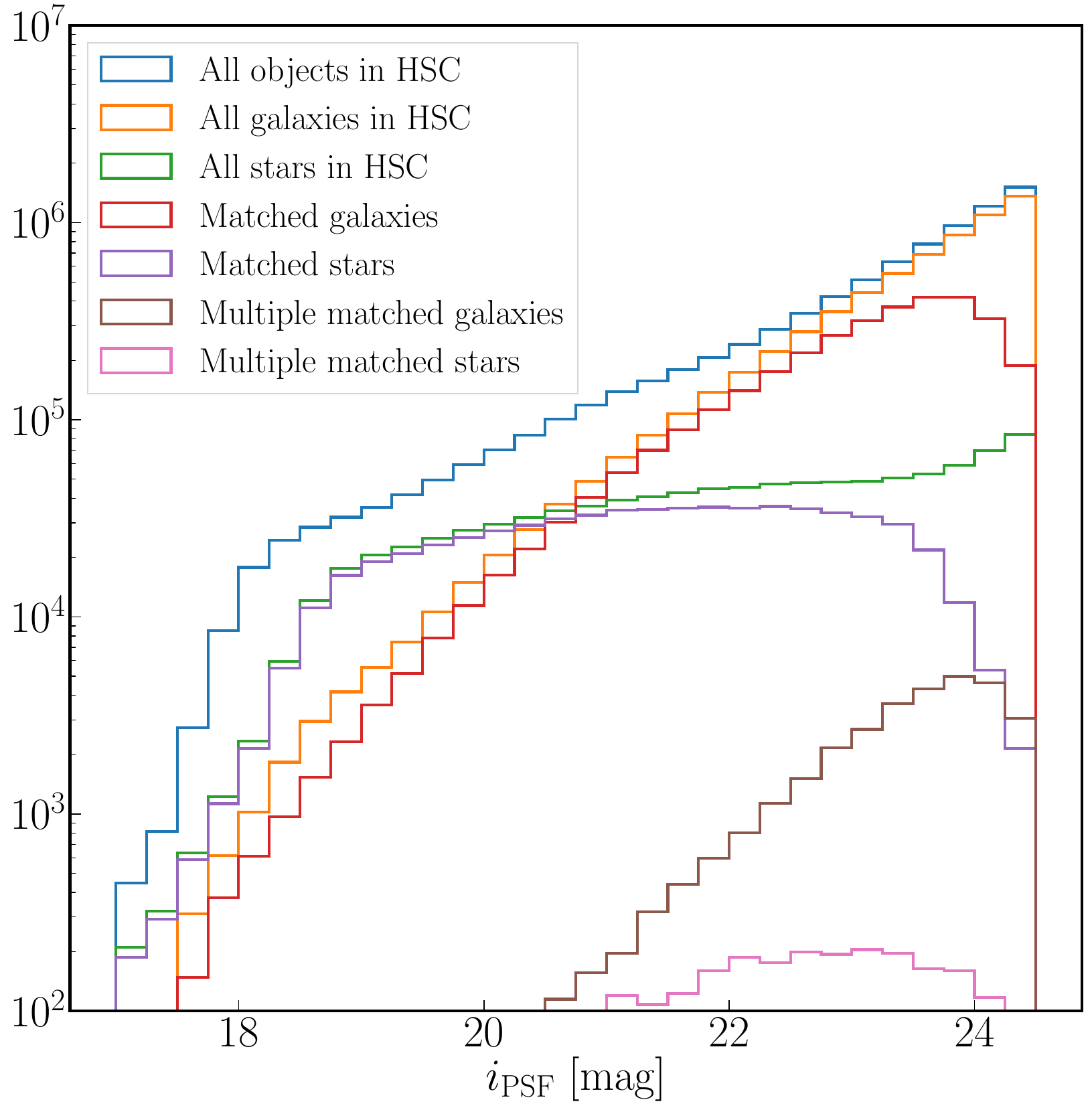}
 \caption{The number of objects against the HSC $i$-band PSF magnitudes. The blue histogram shows the distribution 
 for all detected HSC objects in the overlapping footprints between HSC and S82. The orange and green histograms show the 
 distributions for the HSC-detected galaxies and stars, respectively. In comparison, the red and purple histograms show 
 the distributions for the matched galaxies and stars of HSC and S82 catalogues (see text for details). The brown and magenta 
 histograms denote the galaxies and stars, respectively, where those in HSC catalogue are matched with multiple objects in S82.
 }
 \label{fig:hist}
 \end{center}
\end{figure}
\subsubsection{Matching objects in the HSC and S82 catalogues}
\label{subsec:match}

To measure proper motions of stars from the HSC and S82 data, it is important to construct a matched catalogue of stars 
between the two datasets. Here we describe our method to define the matched catalogues of stars, galaxies and quasars.

The catalogues of objects we use in this paper are constructed based on the coadd images of HSC and S82. The coadd images are contributed 
by multiple single exposures. For HSC, all single exposures overlapped with a particular ``tract'' are resampled to an output plane centred 
on the tract. They are then weighted-averaged to produce the coadd image \citep{2018PASJ...70S...5B,2018PASJ...70S...8A,2019PASJ..tmp..106A}. 
For S82, images from each single run were resampled and coadded following a similar approach \citep{2014ApJ...794..120A}. For both HSC and
S82, objects are detected, deblended and measured from the coadd images.

For HSC, detections are at first made for each band, with the detected footprints and peaks of sources in different bands afterwards 
merged to ensure consistency across different bands\footnote{Also see \url{https://www.astro.princeton.edu/~rhl/photo-lite.pdf}.}. 
These peaks are deblended and a full suite of source measurement algorithm is run on all objects, yielding independent measurements of 
positions and source parameters in each band. A reference band of detection is then defined for each object based on both the signal-to-noise 
and the purpose of maximising the number of objects in the reference band. Finally, the measurement of sources is run again with the position 
and shape parameters fixed to the values from  the measurement in the reference band, which is usually the HSC $i$-band given its 
best image quality \citep{2018PASJ...70S...5B,2018PASJ...70S..25M}. 

For S82, a detailed description about its astrometric calibration can 
be found in \cite{2003AJ....125.1559P}. SDSS $r$-band photometric CCDs are served as the astrometric reference CCDs, which means SDSS 
object positions are based on the $r$-band centroid measurement. 

We construct a catalogue of matched objects by searching for the same objects in the HSC and S82 catalogues with their positions.
To do this we use the master catalogues of objects selected with magnitude cuts, $i_{\rm HSC}<24.5$ and $i_{\rm S82}<24.2$, for HSC 
and S82 respectively. The reason that we stop at $i_{\rm S82}=24.2$ for S82 is because the photometric errors increase very quickly 
beyond $i_{\rm S82}\simeq 24.2$. For the matching, we start with each object in the HSC catalogue and then search for a counterpart 
object(s) in the S82 catalogue within an angular separation of $1\arcsec$ in radius for the HSC object. The search region of 
$1^{\prime\prime}$ radius is chosen because stars with transverse velocity of 220~km/s at 1~kpc leads to an angular offset of 
about $0.5^{\prime\prime}$ or 500~mas over a 10-year baseline, corresponding to a proper motion of $50~$mas~${\rm yr}^{-1}$, so that
most stars moving with typical moderate velocities can be matched properly with the search region. Since we are mainly interested in 
distant stars up to $\sim100$~kpc, the choice of $1^{\prime\prime}$ radius is safe, even if those stars have larger velocities. 
The depth and superb image quality of HSC help construct a secure catalogue of matched objects, especially stars, down to 
$i_{\rm HSC}\simeq 24.5$.

Figure~\ref{fig:hist} shows the matching results. We have about 4.3~million matched objects in total (see the blue histogram). Among 
these, there are about 0.55~million objects which are classified as stars in HSC and 3.2 million objects which are classified as 
galaxies in  both catalogues, as given in Table~\ref{tab:num}. Note that we have excluded stellar or star-like objects
away from the stellar locus (see Section~\ref{sec:stargalsep}), and the fraction of contamination by non-main-sequence stars 
such as giants in our sample would be at most 5\% \citep{2008ApJ...673..864J}.

The sum of ``Galaxies for recalibration'' and ``Stars for proper motion measurements'' in Table~\ref{tab:num} 
is smaller than the total number of matched objects. This is because some of the HSC galaxies are matched to objects
classified as stars in S82, and we have discarded these objects from our sample of galaxies. In addition, there is a 
small fraction of objects which are not classified as either galaxies or stars. 

In this paper, the catalogue of stars is the most fundamental ingredient for our study of proper motion measurements, so we first 
focus on the results for stars. Comparing the green and purple histograms in Figure~\ref{fig:hist}, most of HSC stars 
with $i_\mathrm{HSC,PSF}<22$ have a matched counterpart in S82; more precisely, about 92\% of HSC stars have the matched S82 
object. However, about 8\% of HSC stars are unmatched. We checked that this is mainly because HSC and S82 adopt different masks 
(mainly due to bright stars) in the footprints\footnote{There might be a population of stars that move more than the search radius 
of $1\arcsec$ between the time difference of the two datasets, which would be good candidates of hyper-velocity stars. This is beyond 
the scope of this paper, and will be studied in a separate paper.}. For fainter stars at $i>22$, the matching performance starts 
to degrade. About 80, 70, and 20\% of HSC stars at $i_{\rm HSC}\simeq 22, 23$ and $24$ have matched objects in S82,
respectively, and the matching drops below 10\% at $i_{\rm HSC}\simeq 24.2$, the limit of object detection for S82. 

We should note that some HSC stars are matched only to ``galaxies'' in the S82 catalogue, which occurs due to the relatively poor seeing 
condition of SDSS (typically $1.1^{\prime\prime}$ compared to $0.6^{\prime\prime}$ in the HSC images). However we checked that these 
objects are very likely stars according to their colours and images. More precisely, about 20, 40, 60 and 80\% of the matched stars at
$i\simeq 21, 22, 23$ and 24, respectively, are ``galaxies'' classified in the S82 catalogue. In other words, these results mean that 
the S82 catalogue of stars or galaxies at $21\lesssim i\lesssim 24$ suffer from significant misidentification or non-detection. Also 
note that there are some cases that multiple HSC stars are matched to a single object in S82, as indicated by the magenta histogram. 
This is a minor fraction, and we do not use these stars in the following analysis. In summary, the use of HSC data is crucial for a 
construction of secure star catalogue and therefore for our study.

We also use the matched galaxies, as indicated by the red histogram in Figure~\ref{fig:hist}, for a recalibration of our 
proper motion measurements. Galaxies in the HSC catalogue are identified based on the method described in Section~\ref{sec:stargalsep}. 
We start with each galaxy in HSC and then identify its counterpart in S82. If the matched object is not classified as a 
galaxy in S82, we choose not to include it in order to ensure a secure galaxy sample. The HSC catalogue gives a high number 
density of galaxies, about 15--30~${\rm arcmin}^{-2}$ varying with sky regions due to variations in the depths and observation 
conditions. Thus galaxies are much more abundant than stars on the sky. We have about 3~million matched galaxies. 
In some cases, we have multiple matches: one object in S82 is matched to multiple galaxies in HSC, as shown by the brown
histogram in Figure~\ref{fig:hist}. We do not use the multiply-matched galaxies for the recalibration.

Similarly, we perform a matching of spectroscopically-confirmed quasars in the SDSS DR14 catalogue and stars in the {\it Gaia} DR2 catalogue 
to objects in the HSC and S82 catalogues. We have about 9,000 matched quasars and 0.12~million matched stars with {\it Gaia}, respectively 
(see Table~\ref{tab:num}). We will use the quasars and the {\it Gaia} stars for validation and further tests of our measurements.

\subsubsection{An initial estimate of proper motion for each object}
\label{sec:mjd}

\begin{figure}
\begin{center}
 \includegraphics[width=\columnwidth]{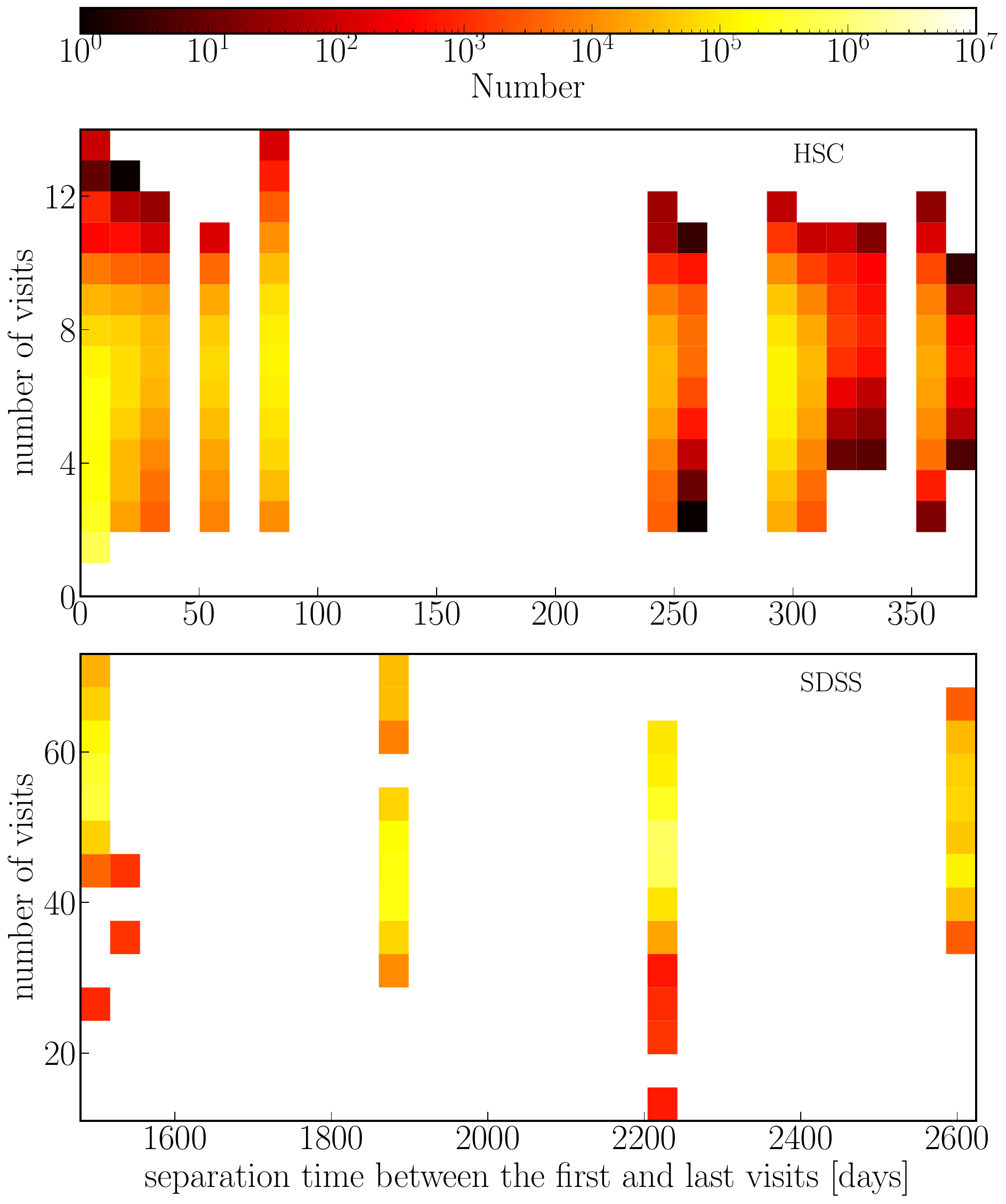}
 \caption{The HSC $i$-band and S82 $r$-band coadd images are constructed from the data taken by
 multiple visits of each field. The coaddition is done based on each HSC ``tract'' region or each SDSS CCD ``frame'', respectively.  
 {\bf Upper}: Colour in each grid denotes the number of HSC tracts lying in a given value of  the 2D space, parametrised by 
 the number of visits of each tract ($y$-axis) and the separation time between the first and last visits of the tract ($x$-axis). 
 {\bf Lower}: The similar result, but for the S82 CCD frames. Based on these information we compute the mean epoch of the coadd 
 image for each matched galaxy/star, and then compute the time baseline for the proper motion measurements. }
 \label{fig:mjd}
 \end{center}
\end{figure}

To measure proper motions of matched stars, we first estimate the angular offset of each star between the HSC and S82
catalogues, and then divide the offset by the time difference between the two datasets as an estimate of its proper motion, by 
making use of the long time baseline, typically 12~years. However, the angular offsets are estimated from the coadd catalogues, 
which are based on multiple exposures, and we need to be careful about how to estimate the representative epoch of each coadd 
image. A reasonable choice would be the mean epoch of input single-exposure images in each field. Since we use the $i$-band 
positions for sources in HSC, we use the multiple $i$-band exposures to calculate the mean epoch. For S82, on the other hand, 
we use $r$-band individual exposures, though given the drift-scanning mode of observation, the epoch difference calculated from 
other SDSS bands are almost the same. 

In Figure~\ref{fig:mjd}, we show the distributions for the number of visits and the time differences between the first and last 
visits for the HSC and S82 coadd catalogues. The HSC $i$-band coadd images are composed of 2 to 14 visit exposures. The spread of
$x$-axis denotes that most of the data are taken within 100~days, i.e. 3~months, and the other data are taken within 360~days, 
i.e. one year \citep[see][for the survey strategy]{2018PASJ...70S...4A}. The S82 coadd images are built from 11 to 73 visit 
exposures (mostly 30 to 60 visits), and the separation time has a wide spread. Many data are taken within 4, 5, 6 and 7~years 
between 1998 and 2005.

For each coadd source, we calculate its mean modified Julian date (MJD) based on epochs of all individual exposures from which its coadd 
image is made, for both HSC and S82. Objects in the same tract of HSC or the same frame of SDSS have the very similar (or same) mean MJD 
dates. We then adopt the separation time between the mean HSC and S82 modified Julian dates for each object as the time baseline 
used for the proper motion measurements. The time baseline ranges from 10.1 to 14.6~years for objects in the HSC-S82 region.
If we instead fix 12~years time baseline  for all objects in the proper motion analysis, it leads to a bias by up to about  
1~mas~yr$^{-1}$,
as we will discuss later and in 
Appendix~\ref{app:calibrole}.

For each matched star (see Section~\ref{sec:stargalsep}), we first estimate its angular offset between the HSC and S82
catalogues, and then divide the angular offset by the time baseline for the star to obtain an ``initial'' estimate of its proper 
motion. Note that neither HSC nor SDSS have corrected for the proper motions of stars before image coaddition, and 
the reference frames for HSC and S82 observations can be different. Although S82 has its own light-motion catalogue 
\citep{bramich2008light}, for which recalibration of the reference system has been done using galaxies and by taking a mid-epoch run 
as reference, the star images are in fact not corrected for their proper motions before coaddition, source detection and solving for 
astrometry. Hence there are likely systematic errors in astrometric solutions in both datasets and also the initial proper 
motion estimates if we do not correct for the difference in reference frames. In order to correct for the systematic errors, we will
use the catalogue of matched galaxies, constructed in Section~\ref{sec:stargalsep}, by making use of the fact that galaxies should not 
move on the sky and can serve to define the reference coordinates. We will describe our recalibration method in the next section.

\begin{figure*}
\begin{center}
 \includegraphics[width=2\columnwidth]{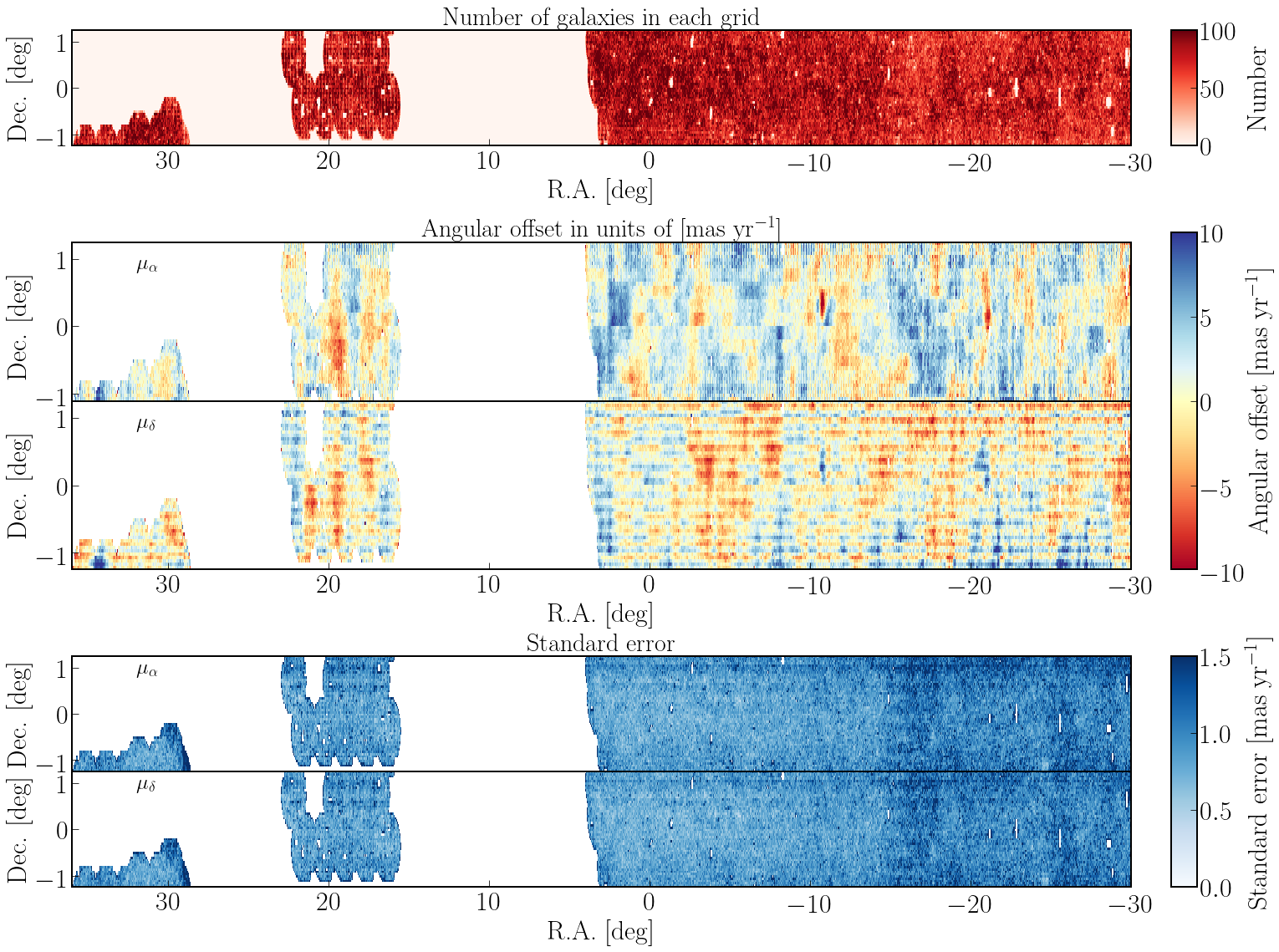} 
 \caption{Recalibration map of astrometric solutions between HSC and S82, obtained using the matched galaxy sample (see text for details).
The map is given by 66,000 grids in total, each of which has a size of $0.05\times0.05$~deg$^2$. \textbf{Upper panel}: The number of 
galaxies in each grid, which contains about 80 galaxies on average. \textbf{Middle}: The mean angular offset of galaxies in each grid 
between HSC and S82, computed by averaging the astrometric R.A. and Dec. positions of galaxies in the grid. Then we convert the angular 
offsets into the unit of proper motion, $[{\rm mas}~{\rm yr}^{-1}]$, using the time baseline estimated for each object in Figure~\ref{fig:mjd}. 
\textbf{Lower}: The 1-$\sigma$ error on the mean angular offset per year in each grid. The error is dominated by the measurement errors of 
galaxy centroid positions. }
 \label{fig:cali}
 \end{center}
\end{figure*}

\subsection{Recalibration}
\label{sec:recalibration}

We use apparent angular offsets of galaxies between the HSC and S82 catalogues to correct for systematic errors in the astrometric 
solutions of the two datasets. Here we describe details of the recalibration method.

\subsubsection{Astrometry recalibration in the proper motion measurements}
\label{sec:astrometry_calibration}

Quasars and galaxies are distant objects that should not have any proper motions. Hence, any apparent angular offsets of galaxy/quasar positions 
between HSC and S82, if found, reflect systematic errors in the astrometric solutions of the two datasets. Conversely, we use the observed 
angular offsets of galaxies/quasars to recalibrate the systematic errors \citep[e.g. see][for a similar method]{2013ApJ...766...79K}.
In order to correctly capture how the systematic errors vary with different regions of the sky, we require a sample of galaxies or quasars 
to have high enough surface number density to measure the apparent angular offsets at each position on the sky at a high statistical significance. 
There are only $\sim$9,000 matched quasars within our footprints (see Table~\ref{tab:num}), which is too sparse, while there are about 3.2 million 
matched galaxies, with the mean surface number density of 32,000~deg$^{-2}$ (Table~\ref{tab:num}). For this reason we use galaxies to perform the
astrometry recalibration.

We build a ``recalibration'' map to correct for the systematics of the angular offset for each matched star at an arbitrary 
position within the overlapping Stripe~82 region of 100~deg$^2$. To do this, we need to well capture possible spatial variations 
in the systematic effects.  The smallest spatial scale would be a size of the CCD chip or frame. The SDSS frame size is about 
$0.225\times0.150$~deg$^2$ (2048$\times$1361~pixels with mean pixel size of about $0.396\arcsec$) \citep{1998AJ....116.3040G}, while 
the HSC CCD chip size is about $0.196\times0.096$~deg$^2$ (4176$\times$2048~pixels with mean pixel size of about $0.168\arcsec$) 
\citep{2018PASJ...70S...1M}. Hence we split the sky footprint into small grids with size of $0.05\times0.05$~deg$^2$ each, which 
is chosen to be smaller than the frame or CCD size, and thus is able to capture spatial variations in the systematic errors within 
one frame or CCD chip, if exists. At the meanwhile, the chosen grid size contains an enough number of galaxies for the recalibration.

We adopt 66,000 grids in total that cover the overlapping region of HSC and S82, including the footprint boundary regions without 
any matched galaxy inside. We have about 80 galaxies per grid on average, with the number density shown in the upper panel of 
Figure~\ref{fig:cali}. For each grid, we compute the mean angular offset of matched galaxies between HSC and S82 after 
3$\sigma$ clipping based on their angular offset distributions in the grid. Then we divide the angular offset by the 
mean time separation, estimated based on the method in Section~\ref{sec:mjd}, to convert the angular offset into units of 
proper motions in R.A. and Dec. directions, i.e. mas~yr$^{-1}$, in each grid\footnote{The division of the galaxy angular 
offsets by the time baseline is just for our convenience. Another way of the recalibration is i) first subtract the galaxy 
angular offset from each star, contained in the corresponding grid of galaxies, and ii) measure the proper motions of stars. 
This is equivalent to our method, and just the order of computation processes is different, because the time 
baselines for galaxies and stars in the same frame/CCD chips are the same.}. We checked that, even if we exclude galaxies 
that are near or within the stellar locus in colour-colour space in Figure~\ref{fig:colour}, to minimise a possible contamination 
of stars to the galaxy sample, the following results are almost unchanged. We also checked that for different grid sizes chosen 
to be smaller than $0.1\times0.1$~deg$^2$, the following results are almost unchanged.

The middle panel of Figure~\ref{fig:cali} shows the map of mean angular offsets. We can see spatial structures for the amplitudes of 
angular offsets. Along the R.A. direction, the structures are smoother and finer, which is the drift-scan direction of SDSS. Along the Dec.
direction, there are about 12 distinct, discontinuous structures. This structure reflects the 2 scan directions by the 6 CCD chips of SDSS 
camera in the S82 region \citep[see Figure~1 in][]{2014ApJS..213...12J}. Thus the astrometry accuracy in our measurements is limited by 
the quality of SDSS data, not by the HSC data. The degree-scale structures seem to reflect depth variations of the data, and the finer 
structures would reflect structures inside the CCD chips. The standard deviation of the offsets among all grids is 2.98~mas~$\mathrm{yr}^{-1}$, 
which quantifies the level of spatial fluctuations in the footprint. In fact, we found that, to measure the tangential velocity dispersions 
of stars at large distances, it is important to take into account the spatial variations of the recalibration map, as we will discuss later 
(see Appendix~\ref{app:measureerr}).

We also estimate the 1-$\sigma$ errors of the mean angular offset in each grid, estimated from the variance of angular offsets of galaxies 
in the grid, as shown in the lower panel. We typically have a $1\sigma$ accuracy of 0.96~mas~$\mathrm{yr^{-1}}$. We checked that the statistical 
error is mainly attributed to measurement errors of the centroid positions for galaxies in the catalogues. However, it turns out that 
the statistical errors quoted in the catalogues underestimate the true errors, and we will later discuss a more careful estimation of the 
statistical errors in the proper motion measurements in Appendix~\ref{app:measureerr}.

To make a recalibration of the proper motion measurements, we first find the closest grid of galaxy offset map (Figure~\ref{fig:cali}) 
to each matched star, and then subtract the mean angular offset per year of galaxies from the measured proper motion of the star.
This is our fiducial dataset of the proper motion measurements.

\subsubsection{Photometric recalibration}
\label{sec:photorecalib}

The HSC $griz$ filter system (transmission curve, wavelength coverage, etc.) is very similar to, but not exactly the same as the SDSS filter system.
As will be described in detail in Section~\ref{sec:photodis}, we adopt the method of \cite{2008ApJ...673..864J} and \cite{2008ApJ...684..287I} to 
infer photometric distance for each star based on its colour, in which the relation linking the colour to intrinsic luminosities is 
calibrated based on the photometry system of SDSS. Hence we need to make colour transformation of the HSC photometry of stars to infer 
the SDSS photometry, i.e., recalibrate the HSC $g$, $r$ and $i$ photometry against the SDSS photometry.

We first pick up stars in the 68\% densest grids over the colour-colour diagram in Figure~\ref{fig:colour}, 
which helps to significantly reduce the number of stars with large photometric errors and hence scattered away from the stellar locus of main-sequence 
stars. We also only focus on stars brighter than $i_\mathrm{HSC,PSF}=23$. For these stars in HSC, we compute their $g$, $r$ and $i$ PSF magnitude differences 
from SDSS PSF magnitudes ($g_\mathrm{HSC,PSF}-g_\mathrm{S82,PSF}$, $r_\mathrm{HSC,PSF}-r_\mathrm{S82,PSF}$ and $i_\mathrm{HSC,PSF}-i_\mathrm{S82,PSF}$) 
against HSC $(g_\mathrm{PSF}-r_\mathrm{PSF})$, $(r_\mathrm{PSF}-i_\mathrm{PSF})$ and $(i_\mathrm{PSF}-z_\mathrm{PSF})$ colours, respectively. We adopt 
the second order polynomial models to fit and describe how the HSC and SDSS PSF magnitude difference in each band depends on colour on average, and use
the best-fit models to recalibrate the HSC photometry of stars against SDSS.

The recalibration not only affects our determination of photometric distance measurements, but also affects the proper motion measurements after 
correcting for our own motion (see Section~\ref{sec:nosolarmotion} for details), because our own motion affects the observed proper motions for stars 
at different distances by different amount. Moreover, it slightly affects how we divide stars into red and blue populations. However, as we have checked, 
the average amount of recalibration in HSC $g$, $r$ and $i$-bands is in fact quite small. For blue stars with $g_\mathrm{PSF}-r_\mathrm{PSF}<0.6$ (after 
photometric recalibration), the average amounts of recalibrations in $g$, $r$ and $i$ PSF magnitudes are 0.0254, 0.0039 and $-0.0126,$ respectively. For red 
stars with $g_\mathrm{PSF}-r_\mathrm{PSF}>0.6$ (after photometric recalibration), the average recalibrations are slightly larger, which are 0.1369, 0.0112 
and 0.0859 in $g$, $r$ and $i$ PSF magnitudes, respectively. We have explicitly checked that, after including the photometric recalibration, the measured 
proper motions change by at most $\sim$0.1~mas~yr$^{-1}$.

\subsection{Photometric distance}
\label{sec:photodis}
\begin{figure}
\begin{center}
 \includegraphics[width=\columnwidth]{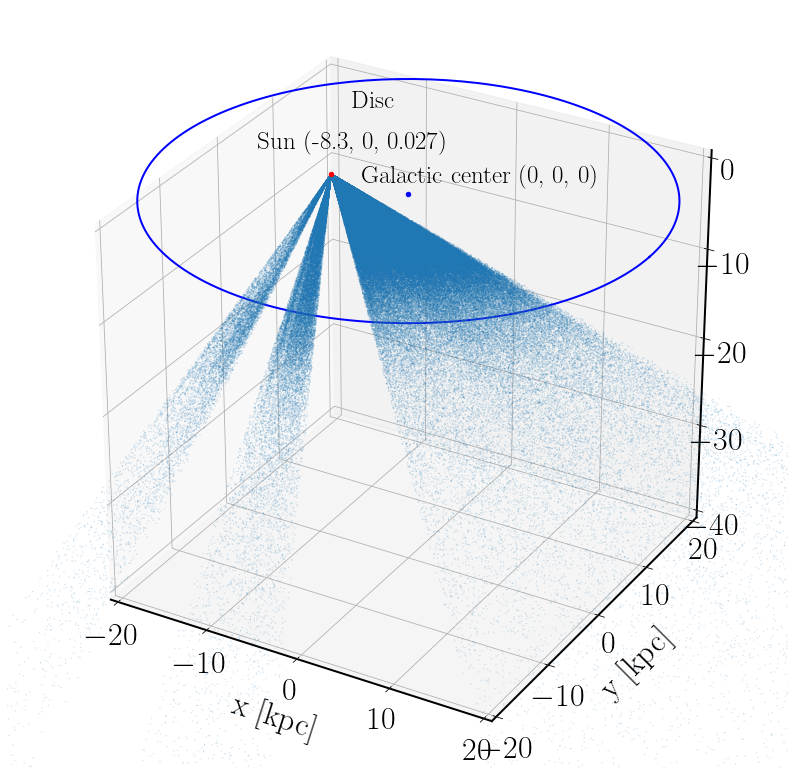}
 \caption{The 3-dimensional distribution of HSC-S82 matched stars (light-blue dots). We derive the 3D positions from the angular position in  
 the Equatorial coordinate (R.A. and Dec.) and photometric distance for each star. Red and dark-blue dots denote the position of 
 the Sun and the Galactic centre. The Galactic disc plane is represented by the blue circle. The positive $x$ and $z$ directions 
 are defined to point towards the Galactic centre and the north Galactic pole.}
 \label{fig:dist}
\end{center}
\end{figure}

We will combine the proper motion measurements with the distance information of each star to exploit the 
three-dimensional structure of the velocity field. We use the method in \cite{2008ApJ...673..864J} and \cite{2008ApJ...684..287I} 
to estimate the photometric distance of each star. The readers can find more details in the two papers, and in this 
subsection we briefly introduce our main approach. 

As mentioned in \cite{2008ApJ...673..864J}, main-sequence stars have a well-defined colour-luminosity relation. One can thus 
estimate an intrinsic luminosity of each main-sequence star and hence the distance from the measured colours and fluxes.
Following the ``photometric parallax'' relation provided by \cite{2008ApJ...673..864J}, we estimate the photometric distances 
of main-sequence stars using their $g_{\rm PSF}-i_{\rm PSF}$ colours and $r_{\rm PSF}$ magnitude. Since we always adopt 
PSF magnitudes in different filters to calculate colours for stars, hereafter we neglect the lower index of ``PSF'' when quoting 
the colours. In addition to the colour, accurate photometric distance estimates also depend on metallicities. \cite{2008ApJ...684..287I} 
developed polynomial models which relate effective temperature and metallicity of main-sequence stars to SDSS $u-g$ and $g-r$ colours. The models are 
based on the effective temperature and metallicity derived from the spectroscopic data of 60,000 F- and G-type main-sequence stars 
with $g-r$ colours in the range of 0.2 and 0.6. 

In principle, one can estimate the metallicity through $u-g$ and $g-r$ colours following the approach of \cite{2008ApJ...684..287I}. 
However, HSC does not have $u$-band data. The SDSS $u$-band filter, unfortunately, is shallow and gives quite large photometric 
errors for faint stars, and thus the errors of photometric distances are dominated by $u$-band photometric errors. The typical S82  
$u$-band PSF magnitude error is as large as 1--1.5~mag at HSC $i$-band magnitudes of $21<i_\mathrm{HSC,PSF}<24.5$, and across this 
magnitude range, SDSS basically measures the sky flux rather than true source detections. The error drops below 1~mag at brighter 
magnitudes, but is still quite large. The typical errors are 0.6, 0.3 and 0.1~mag at $i_\mathrm{HSC,PSF}\sim$20, 19 and 18, respectively. 
For comparison, the median photometry errors of our HSC-S82 matched stars in HSC $g$, $r$ and $i$-band are only about 0.0133, 0.0065 and 
0.0033~mag, respectively; these errors are significantly smaller.

As have been pointed out by \cite{2008ApJ...673..864J}, a 0.1~mag of error in colour can lead to 0.5~dex of error in the metallicity 
estimate, which corresponds to a relative distance error up to 20\%. To ensure accurate photometric distance estimates, the S82 $u$-band data is 
almost useless beyond $i_\mathrm{HSC,PSF}\sim$18. Hence it is difficult to achieve proper metallicity estimates based on the colour for
stars fainter than $i_\mathrm{HSC,PSF}\sim18$. 

\cite{2008ApJ...684..287I} investigated the $[{\rm {\rm Fe/H}}]$ distribution of disc and halo stars out to $\sim$~10~kpc. In their
results, halo stars with $|z|>4$~kpc show a quite homogeneous spatial distribution of $[{\rm Fe/H}]$, and they are dominated by blue stars. 
The value of $[{\rm Fe/H}]$ peaks at $-1.45$ and follows a nearly Gaussian distribution with 1-$\sigma$ scatter of about 0.41~dex 
\citep[see Figure~10 in][]{2008ApJ...684..287I}. On the other hand, nearby disc stars are mostly red stars. It is more difficult 
to properly determine the metallicity for red stars, but the mean $[{\rm Fe/H}]$ of disc stars is about $-0.6$ \citep{2008ApJ...684..287I}.

Thus throughout this paper, we adopt a fixed metallicity of $[{\rm Fe/H}]=-$1.45 to estimate the photometric distances for all stars (both 
red and blue stars)\footnote{We also tried to assign each blue star with a value of $[{\rm Fe/H}]$ from the Gaussian distribution with 
$\langle [{\rm Fe/H}] \rangle=-1.45$ and $\sigma=0.41$~dex, and our results change very little.}, if the initial distance estimate is 
greater than 6.25~kpc; that is, we ignore the metallicity dependence for such halo stars. This mainly picks up stars with $|z|>4$~kpc, i.e., 
stars away from the Galactic disc. For more nearby blue stars, whose photometric colours are more accurately measured, we estimate their 
metallicities based on the $u-g$ and $g-r$ colours, with the $u$-band flux taken from S82 and $g$, 
$r$-band fluxes taken from HSC after photometric recalibrations against SDSS (see Section~\ref{sec:photorecalib}), though the total number 
of nearby blue stars is very small. For nearby red stars at $<6.25~$kpc, we keep their metallicities fixed to $[{\rm Fe/H}]=-0.6$ 
following \citet{2008ApJ...684..287I}.

Figure~\ref{fig:dist} shows the 3-dimensional spatial distribution for our sample of matched stars, based on their Equatorial coordinates 
and photometric distances. The three disjoint HSC-S82 footprints all point towards negative $z$ directions with respect to the Galactic 
disc plane. Thus, except for those very nearby stars, which might include some thick disc stars, the majority of matched stars
are in the halo region. The HSC-S82 regions are distributed in the range of the Galactic latitude, $b\simeq[-64,-40]$~deg.

However, halo stars could have lower metallicities, e.g. [Fe/H]$\sim -2$, than our default value of $-1.45$, as indicated in \citet{2013ApJ...763...65A}. 
In Section~\ref{sec:disc} we will discuss how possible uncertainties in the metallicities affect the photometric distances and in turn 
affect our proper motion measurements for halo stars.

\subsection{Correction for differential chromatic refraction}
\label{subsec:dcr}

After a careful analysis, it turned out that the differential chromatic refraction effect \citep[hereafter DCR;][]{1982PASP...94..715F} by 
the Earth's atmosphere gives a subtle systematic effect in our proper motion measurements. We need to correct for the DCR effect to achieve the 
precision better than 0.1~mas~yr$^{-1}$. In Appendix~\ref{app:dcr}, we describe details of our correction method of the DCR effect, and also give 
a validation of the method using the SDSS DR14 spectroscopic quasar sample.

The HSC data are not affected by DCR, because the HSC wide-field corrector is equipped with the DCR corrector, more exactly 
the Atmospheric Dispersion Compensator (ADC) as described in \citet{2018PASJ...70S...1M}. On the other hand, the SDSS camera does 
not have such a DCR corrector. DCR stretches an observed image of source along the zenith direction due to the wavelength-dependent 
refraction by atmosphere and, as a result, each object appears to be observed at a shifted position in the camera image by a 
different amount in different filters. For SDSS, because of its specific drift-scanning mode of observation, the zenith direction 
is mainly along the Dec. direction for the Stripe~82 region. Since the spectral energy distribution (SED) of galaxies and stars 
are different, our astrometry recalibration method using the galaxy positions (Figure~\ref{fig:cali}) cannot fully correct for 
the DCR effects for stars.

We correct for the DCR effect by using the {\it Gaia} DR2 proper motions as a reference sample. Because the space-based 
{\it Gaia} data are not affected by DCR, the difference between our measured proper motions and {\it Gaia} data reflects the
amount of DCR effect. To do this, we first cross match our sample of stars with the {\it Gaia} DR2 catalogue based on their 
positions. Since {\it Gaia} stars are brighter than $G\sim21$, we can only match bright and nearby stars; the matching gives 
117,703 stars in total in the HSC-S82 region, as given in Table~\ref{tab:num}. As for our default method of correcting for 
the DCR effect, we first measure the  DCR effect as a function of the $g-i$ colours for the matched stars, averaged over 
the survey footprints. When we measure a proper motion for each star, including other stars (non-{\it Gaia} matched 
stars), we use the {\it Gaia} measurement to correct for the DCR effect based on the $g-i$ colour of the star, assuming 
that the DCR effect is characterised by the colour, and does not depend on the distance of the star. The details are 
presented in Appendix~\ref{app:dcr}.

\section{Results}
\label{sec:result}
\begin{figure*}
\begin{center}
\includegraphics[width=12cm]{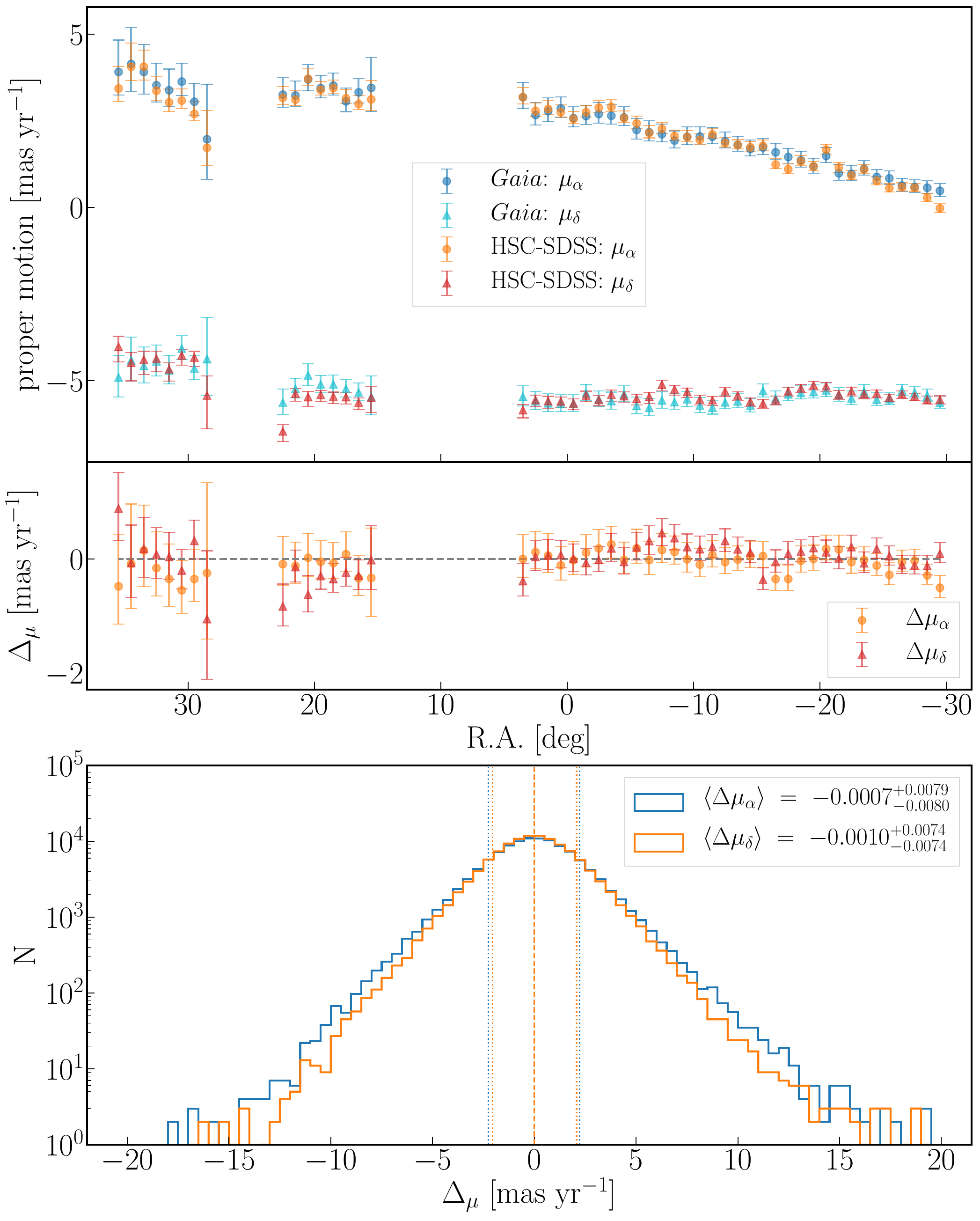}
 \caption{Comparison of our measured proper motions with the {\it Gaia} proper motions for the matched stars, where
 we have about 120,000 matched, bright stars in the overlapping HSC-S82 footprints (see Table~\ref{tab:num}). The 
 \textbf{upper panel} shows the proper motion measurements, ($\mu_\alpha$ and $\mu_\delta$), obtained by taking the 
 median of proper motions for stars in each bin of R.A. region. Our HSC-S82 proper motions are in good agreement with 
 the {\it Gaia} results. The \textbf{middle panel} shows the difference between each component of proper motions in 
 R.A. bins. The asymmetric errorbars show the 1-$\sigma$ errors on the median, estimated using Equation~(\ref{eq:errorbars}).  
 The \textbf{lower panel} shows the distributions of the differences between the proper motion components of HSC-S82 
 and {\it Gaia} for each star in all R.A. regions. The median difference of $\mu_\alpha$ is $-0.0007$~mas~$\mathrm{yr^{-1}}$ 
 while that of $\mu_\delta $ is $-0.001$~mas~$\mathrm{yr^{-1}}$, marked by the vertical dashed lines. The errorbars
 denote the errors of the median computed using Equation~(\ref{eq:errorbars}). The other two vertical dotted lines 
 mark the 68\% width of the whole distribution, which is much wider than the error of the median, reflecting that 
 errors in the proper motion measurements are dominated by the centroid errors of individual stars.}
 \label{fig:gaia}
 \end{center}
\end{figure*}

\begin{figure}
\begin{center}
 \includegraphics[width=\columnwidth]{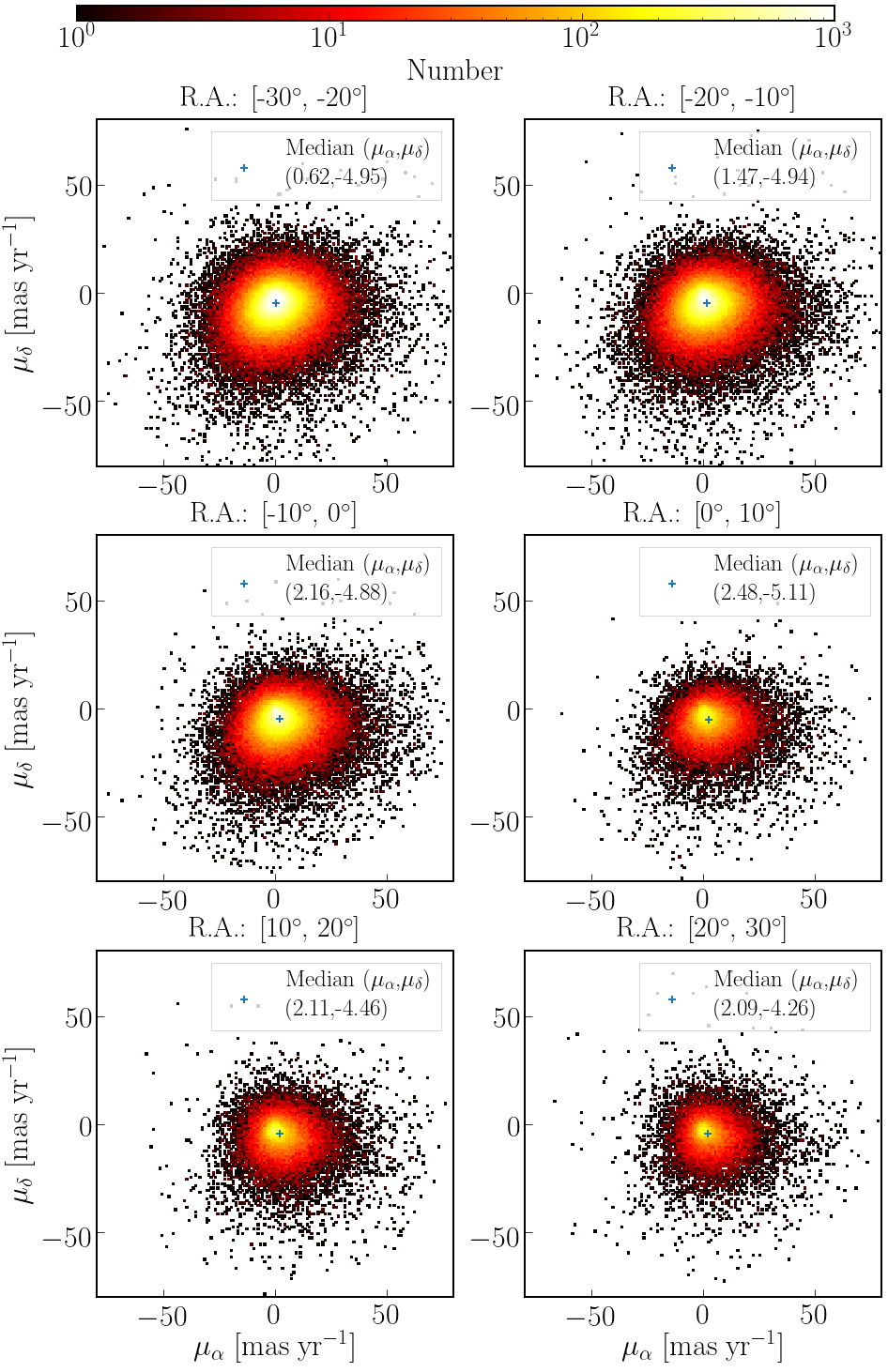}
 \caption{Distribution of stars in the plane of proper motion components, $\mu_\alpha$ versus $\mu_\delta$. Colour in
 each grid denotes the number of stars as indicated by the colour bar. Different panels correspond to the results for 
 the different regions divided by the R.A. range. The numbers in the parenthesis in each panel denote the median values 
 of $\mu_\alpha$ and $\mu_\delta$. 
}
 \label{fig:pmnum}
\end{center}
\end{figure}
\begin{figure}
\begin{center}
\includegraphics[width=\columnwidth]{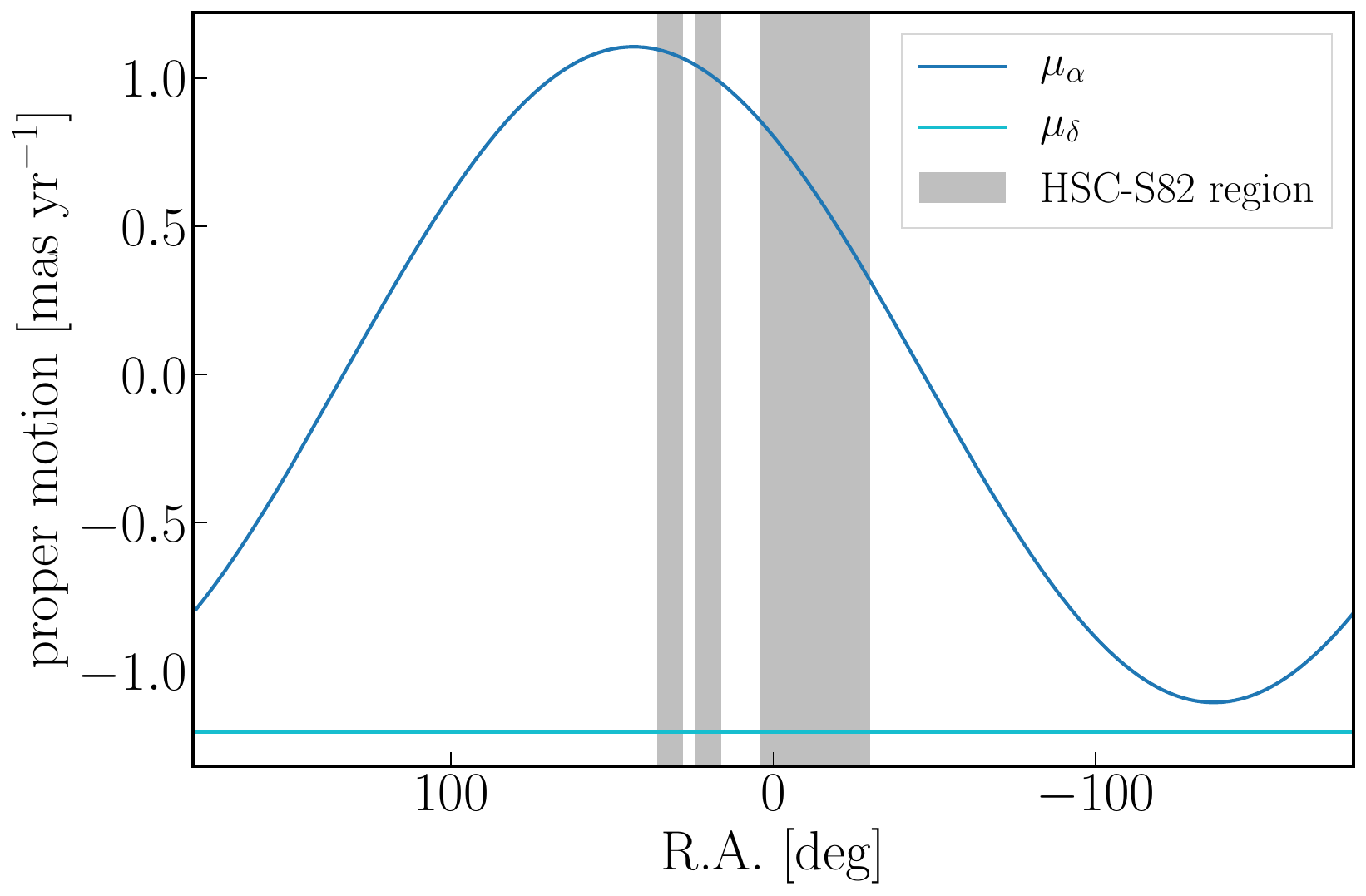}
\caption{The {\it apparent} proper motion of stars due to our own motion if the stars are at rest with respect to 
the Galactic centre. Here we consider stars at $30~$kpc in distance, and we fix Dec. to be zero. We assume 
that our own motion is a sum of the local standard of rest (LSR) motion with respect to the Galactic centre plus the solar 
motion with respect to the LSR; we employ $v_{c}(R_\odot)=220~$km/s for the rotation velocity, and $(U_\odot,V_\odot,
W_\odot)=(11.1,12.24,7.25)~$km/s for the motion of the Sun with respect to LSR. Note the apparent motion scales with 
$1/d$ ($d$ is a distance to each star). The shaded regions denote the overlapping HSC-S82 footprints we study in this 
paper, where the apparent motions in R.A. and Dec. directions are positive and negative on average, respectively. 
$\mu_\delta$ looks like a R.A. independent constant value. This is because the Galactic rotation mainly contributes 
to $\mu_\alpha$ near the Celestial Equatorial region, and $\mu_\delta$ is mainly affected by the solar motion with 
respect to the LSR, which is an order of magnitude smaller.}
\label{fig:solar_motion}
\end{center}
\end{figure}
\begin{figure}
\begin{center}
 \includegraphics[width=\columnwidth]{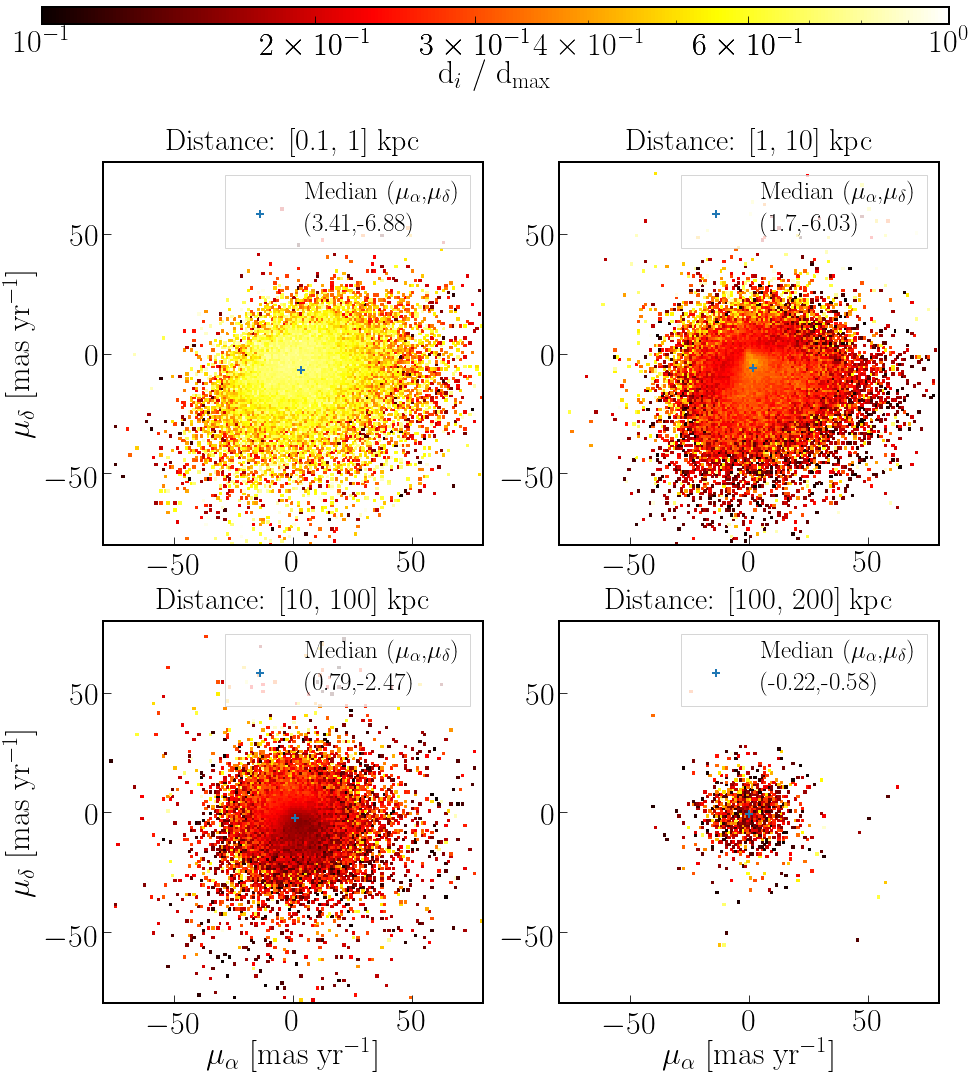}
 \caption{Similar to Figure~\ref{fig:pmnum}, but each panel shows the distribution of stars divided in the different range
 of their photometric distances. Colour in each panel denotes the median of photometric distance of stars in each grid, 
 relative to the maximum distance assumed in each plot, $d_{\rm max}=1, 10, 100$ or 200~kpc, respectively. In each panel, 
 more distant stars tend to display smaller proper motions. Comparing the different panels manifests that the median proper 
 motions and the overall scatters both decrease with the increase in distances.
 }
 \label{fig:pmnum2}
 \end{center}
\end{figure}
\begin{figure}
\begin{center}
\includegraphics[width=\columnwidth]{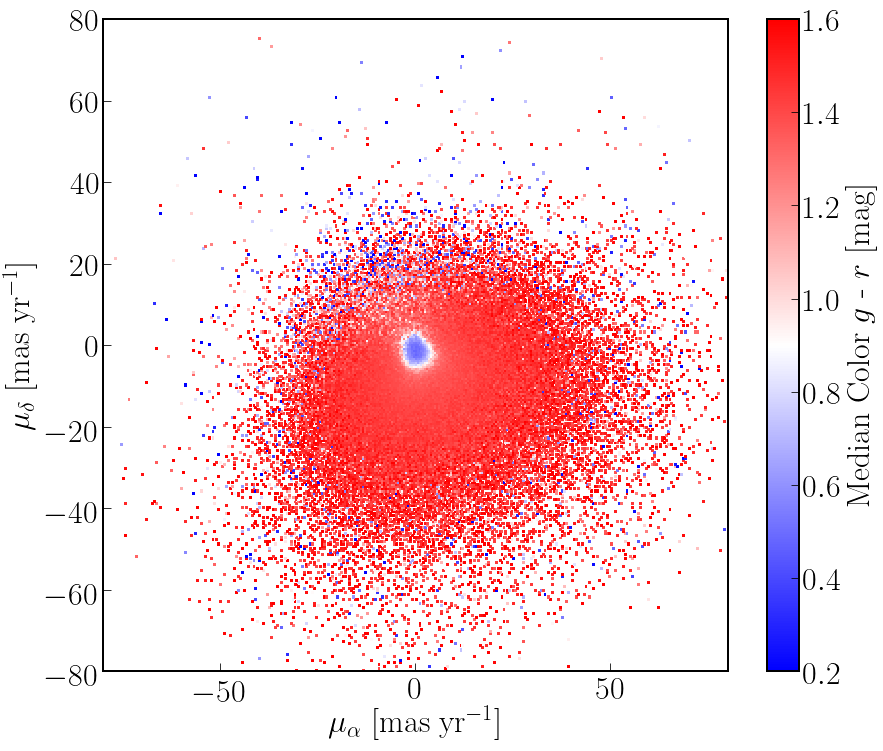}
 \caption{Similar to the previous figure, but the plot shows the distribution for stars divided by their colours, $g-r$. 
 Colour in each grid shows the median of $g-r$ for stars in the grid. Blue stars, which are mainly halo stars, 
 show smaller amplitudes of  proper motions, while red stars are more likely nearby disc stars with larger proper motions.}
 \label{fig:pmco}
 \end{center}
\end{figure}

Following the methods described up to the preceding section, we estimate proper motions for about 0.55 million stars down to  
$i\simeq24.5$. In this section we show the main results of this paper. We first give a validation of our proper motion measurements
by comparing with the {\it Gaia} results in Section~\ref{subsec:gaia}. Hence readers who are interested in the results can directly go 
to Section~\ref{sec:directmotion}.

\subsection{Validation of our method: comparison with {\it Gaia}}
\label{subsec:gaia}

Before going to the main results, we first demonstrate a validation of our method from the comparison with the {\it Gaia} measurements 
of proper motions that have an exquisite precision \citep{2018A&A...616A...1G}. To do this, we first cross-match our sample of stars 
with the {\it Gaia} DR2 catalogue based on their positions. Since {\it Gaia} stars are brighter than $G\sim21$, we can only match bright 
and nearby stars; the matching gives 117,703 stars in total in the HSC-S82 region, as given in Table~\ref{tab:num}. 

In Figure~\ref{fig:gaia}, we show the proper motion measurements for the matched stars, which are estimated by averaging the 
proper motions of stars that reside in each bin of R.A. ($x$-axis). Throughout this paper, for plots showing the averaged proper 
motion measurements, we use the median in a given bin to denote the central value of the measurement, and show the asymmetric 
errorbars to denote statistical errors in the median, computed by
\begin{align}
\sigma_+&\equiv \sqrt{\frac{\pi}{2}}\times \frac{1.428(q_{75}-q_{50})}{\sqrt{N}}, \nonumber\\
\sigma_-&\equiv \sqrt{\frac{\pi}{2}}\times \frac{1.428(q_{50}-q_{25})}{\sqrt{N}}
\label{eq:errorbars}
\end{align}
for the upper- and lower-side errorbars, respectively, where $q_{25}$, $q_{50}$, and $q_{75}$ are the 25, 50, and 75 percentiles,
respectively, and $N$ is the number of stars (or galaxies) in the bin \citep[][]{2019sdmm.book.....I}. 

The top and middle panels of Figure~\ref{fig:gaia} clearly show that our method gives a significant detection of the proper motions 
and our results well reproduce the {\it Gaia} results for both components of $\mu_\alpha$ and $\mu_\delta$. The lower panel quantifies 
the agreement. The averaged difference between the proper motions of HSC-S82 and {\it Gaia} for all the matched stars is remarkably 
close to zero, more precisely only about $\mu_\alpha\simeq -0.0007$~mas~$\mathrm{yr^{-1}}$ and $\mu_\delta\simeq -0.0010$~mas~$\mathrm{yr^{-1}}$. 
We note that the good agreement is a result of the DCR correction, 
which is by design as we used $Gaia$ as a reference (see Section~\ref{subsec:dcr} for details).
A similar plot before correcting for the DCR effect is provided in Appendix~\ref{app:dcr}, which still shows farily good agreement. More 
precisely, 
the averaged difference between the proper motions 
of HSC-S82 and {\it Gaia} is $\sim$0.05~mas~yr$^{-1}$ along Dec., which is about 50 times larger than in Figure~\ref{fig:gaia}.
In addition, the validation test using quasars shown in Section~\ref{app:qso}
is independent and convincing as well.
A closer look at Figure~\ref{fig:gaia} reveals that
the difference from {\it Gaia} is relatively larger
at R.A.$\sim$~20~deg., 
where the region contains more blue stars from the Sagittarius stream.
This is 
because the DCR effect is expected to be stronger for bluer objects in our sample. We also show in Appendix~\ref{app:dcr} that we 
see a more significant difference in the {\it Gaia} comparison for blue stars.

The error of the median in Figure~\ref{fig:gaia} is about 0.008~mas~yr$^{-1}$. Thus we achieve the overall precision 
of $\sim 0.01$~mas~yr$^{-1}$. The figure also shows that the error is much smaller than the width of the difference distribution 
($\sim$2.2~mas~yr$^{-1}$) that mainly reflects the statistical errors of proper motions for individual stars (see below). Thus 
these results demonstrate the power of the statistical method for the accurate proper motion measurements.

Most of the matched stars have $20<G<21$ in the {\it Gaia} band, which is near the flux limit of {\it Gaia} sample. 
The median measurement errors for individual stars in the Gaia data are typically $(\sigma_{\mu_\alpha}, \sigma_{\mu_\delta})=
(1.37, 1.10)$~mas~yr$^{-1}$. Compared to this, the measurement errors for individual stars in the range $20<G<21$ (57,648 
in total) are $(\sigma_{\mu_\alpha},\sigma_{\mu_\delta})=(1.94,1.58)~$mas~yr$^{-1}$. On the other hand, the measurement 
errors for the HSC-S82 proper motions are $(\sigma_{\mu_\alpha},\sigma_{\mu_\delta})=(0.28,0.27)$~mas~yr$^{-1}$, which 
are estimated by propagating the measurement errors of the centroids of stars for both the HSC and S82 data. Note 
that the measurement errors in HSC are one order of magnitude smaller than those in S82, meaning that an accuracy of 
our proper motion measurements is limited by the errors of S82 data. The reader can find more detailed information 
about the measurement errors in Figure~\ref{fig:measureerr}. Thus the statistical error or the width of the proper 
motion difference distribution in Figure~\ref{fig:gaia} is dominated by the measurement errors of {\it Gaia}, not those 
of HSC-S82 data.

We should note that the agreement in Figure~\ref{fig:gaia} cannot be realised unless we implement the proper HSC-S82 time baseline 
(see Section~\ref{sec:mjd}) and the astrometry recalibration using the angular offsets of galaxies in Section~\ref{sec:recalibration}. 
Without the proper time baseline and the astrometry recalibration, our measurement displays a systematic discrepancy from the {\it Gaia} 
results, up to 3~mas~${\rm yr}^{-1}$, as explicitly shown in Figure~\ref{fig:gaia_wo_calibration}. The astrometry recalibration based 
on galaxies gives a more significant correction (about 60\% compared to the signal), but both the two effects are important 
to include. We should also note that the about 12-year time baseline, between HSC and S82, is critical to achieve the 
precise measurement. The angular offsets of stars between HSC and S82, which are our direct observables, are a factor of 12 
greater than the proper motions shown in Figure~\ref{fig:gaia}.

We also comment on a possible effect of parallax on the proper motion measurement. {\it Gaia} DR2 jointly modelled 
parallax and proper motion during the continuous 22 months observation. In our HSC-S82 proper motion measurement we 
do not consider the effect of parallax. The parallax amounts to about 1 or 0.1~mas for a star at distance 1 or 10~kpc, 
respectively, and this corresponds to about 0.1 or 0.01~mas~yr$^{-1}$ in our proper motion measurement for a 10-year time 
baseline. This effect is safely negligible unless we work on very nearby stars, which is not the case for our study.

\subsection{Direct proper motion measurements}
\label{sec:directmotion}

\begin{figure}
\begin{center}
 \includegraphics[width=\columnwidth]{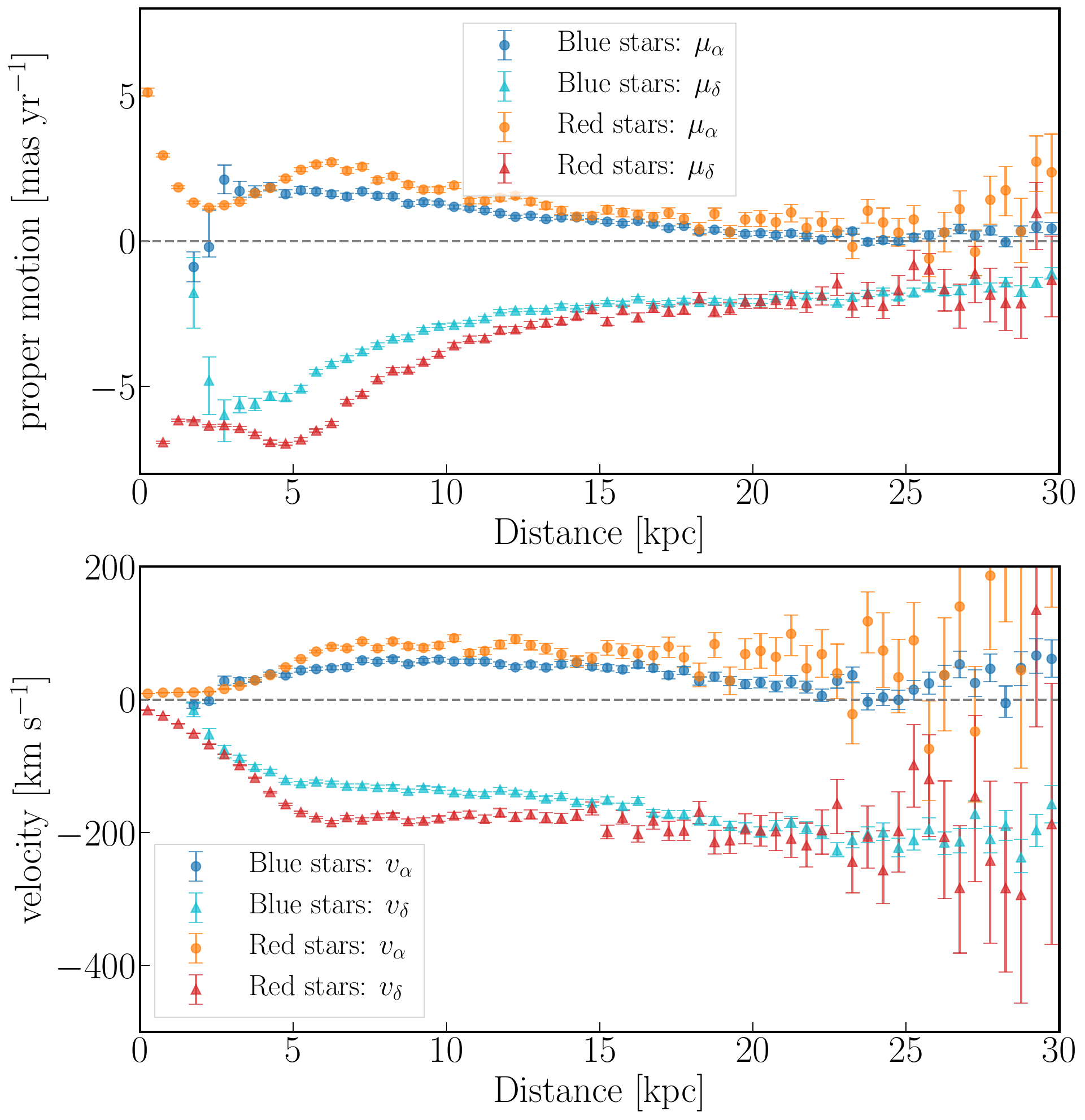}
 \caption{\textbf{Upper panel}: The proper motions of ($\mu_\alpha$,$\mu_\delta$) measured from the red and blue stars divided into different distance 
 bins out to $30~$kpc. Here we take the median of proper motions for stars in each bin, after the 3-$\sigma$ clipping of the distribution in each bin. 
 The red and blue stars are divided based on the colour cut of $g-r=0.6$. The errorbars show the 1-$\sigma$ errors on the median in each bin, computed 
 by Equation~(\ref{eq:errorbars}). The proper motions shown are direct observables of our data, and the apparent motions due to our own motion are not 
 corrected for yet. \textbf{Lower}: The tangential velocities, which are obtained by multiplying the proper motions by the photometric distances for 
 each star. 
 }
 \label{fig:dall}
 \end{center}
\end{figure}
\begin{figure*}
\begin{center}
\includegraphics[width=15cm]{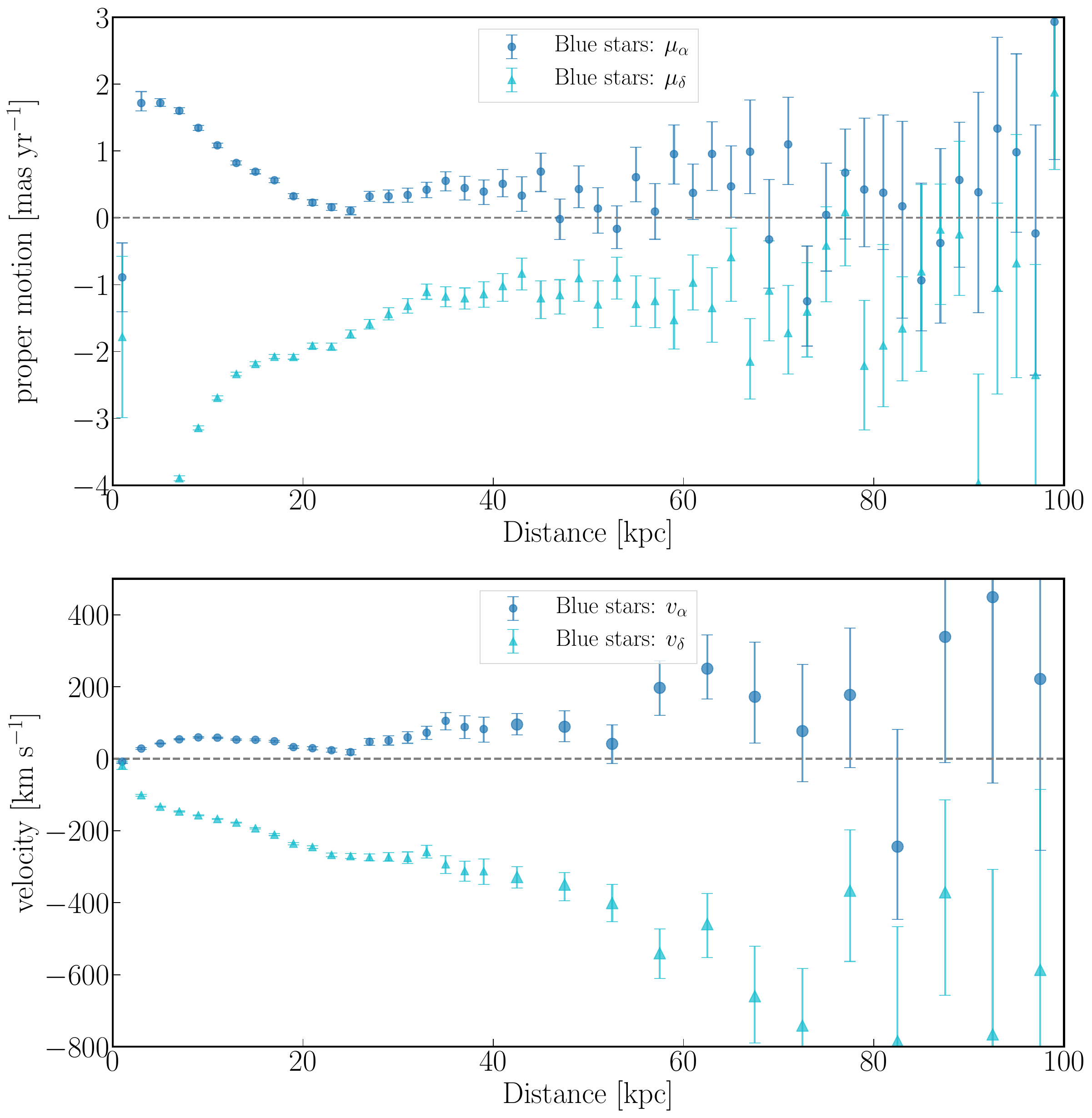}
 \caption{Similar to the previous figure, but the plot shows the {\it net} proper motions for  blue stars out to 100~kpc in Heliocentric distances. This 
 is our direct observable, which shows a significant detection of the proper motions up to 100~kpc. For illustrative purpose, we choose four times wider 
 bin size at $d>40~$kpc in the lower panel than those in the upper panel.
 }
 \label{fig:dblue}
 \end{center}
\end{figure*}
In Figure~\ref{fig:pmnum}, we show the distribution of proper motions\footnote{Generally, $\mu_\alpha$ should be $\mu_\alpha
\cos(\delta)$. Since our region is $-1.25^\circ<\delta<1.25^\circ$, ignoring $\cos(\delta)$ only leads to less than 0.03\% of difference, 
and thus we neglect it.} ($\mu_\alpha$,$\mu_\delta$) for subsamples of stars divided based on different R.A. range.
Note that we use all the matched stars down to $i\simeq 24.5$ for these results. Each panel displays an anisotropic distribution of
stars. While the distribution shows a R.A. dependence, all panels display a coherent proper motion towards the lower right corner; 
the median of the distribution\footnote{The mean shift is very similar to the median after 3-$\sigma$ clipping.} indicates a coherent
motion with positive value of $\mu_\alpha$, and negative value of $\mu_\delta$, respectively. We have explicitly checked that the 
1-$\sigma$ errors on the median values are much smaller than the absolute median values. Thus our detection here is unlikely 
caused by the measurement errors.

The trend is consistent with the effect of our own motion\footnote{We define our own motion to be the motion of our Sun with 
respect to the Galactic centre, which is a combination of the velocity of the Local Standard of Rest (LSR) and the solar motion with
respect to the LSR in the same direction, i.e., $\textbf{v}_\odot= \left(U_\odot,v_c(R_0)+V_\odot,W_\odot \right)$. $v_c(R_0)$ is the
rotation velocity of the LSR at the solar radius, $R_0$. $V_\odot$ is the solar motion with respect to the LSR in the direction of
Galactic rotation. $U_\odot$ is the velocity towards Galactic centre and $W_\odot$ is the velocity component of the solar motion
perpendicular to the Galactic disc. } in terms of both its sign and magnitude. The Sun is rotating around the Galactic centre 
in a counter-rotating manner with the velocity of $\textbf{v}_c(R_0)$. This induces apparent motions of stars with respect to us. 
For nearby disc stars following the Galactic rotation, the observed proper motions are mainly due to the solar motion with 
respect to the Local Standard of Rest (LSR), denoted as ($U_\odot$, $V_\odot$ and $W_\odot$). The magnitude is small ($\sim 10$~km/s), 
but could induce significant proper motions for very nearby stars. In the HSC-S82 footprints, we have explicitly checked that the 
observed proper motions induced by the solar motion correspond to a motion with positive R.A. and negative Dec. directions, though 
there are not many disc stars in our sample. For halo stars whose velocities are close to random, the observed proper 
motions arise from a sum of their true motions with respect to the Galactic centre and the apparent motions due to our own motion. 
As explicitly demonstrated in Figure~\ref{fig:solar_motion}, the effect of our own motion on more distant stars in the HSC-S82 
footprints roughly corresponds to a positive direction in R.A. and a negative direction in Dec., but the exact amount of motions 
projected onto the Equatorial coordinates is R.A. dependent.  

The apparent motion of stars induced by our own motion is expected to show strong dependence on distances of stars, 
with nearby stars having larger apparent proper motions. We investigate this in Figure~\ref{fig:pmnum2}, which shows the 
dependence of measured proper motions on photometric distances. The two upper panels of Figure~\ref{fig:pmnum2} show larger 
median values and more anisotropic distributions, whereas distant stars in the lower panels display a more isotropic distribution, 
with smaller median values. In each panel, more distant stars with lighter colours also show larger proper motions. This is 
consistent with the expectation.

Main-sequence stars with bluer colours tend to be intrinsically brighter, more metal poor and have
 higher effective temperature, 
while stars with redder colours tend to be intrinsically fainter. In addition, stars with different metallicities differ in their 
distances. Those with higher metallicity are younger stars newly formed in the Galactic disc, while the metal-poor stars are old ones, 
most of which are already located in the halo. We represent the metallicity with the $g-r$ colour. Thus we use intrinsically brighter 
blue stars to explore the proper motions at large distances. On the other hand, red stars are relatively fainter, and have a larger 
chance to lie in the Galactic disc at short distances $\lesssim 5$~kpc. Thus, the colour of stars correlates with the distance. 
As shown in Figure~\ref{fig:pmco}, bluer distant stars tend to have smaller proper motions. Interestingly, except for the blue spot in 
the centre of the plot, there are blue dots scattered in outskirts, indicating the possible existence of distant blue stars moving 
with high velocities or the existence of some nearby blue stars. In the following, we describe the proper motion results for blue/red 
stars separately, using the division of $g-r=0.6$.

In Figure~\ref{fig:dall}, we demonstrate how the proper motions and velocities vary with the Heliocentric distances, for red and blue stars. 
Here we show the results only up to 30~kpc because there are few red stars beyond this range. We first measure photometric distances for 
individual stars, and then measure the median proper motions for stars divided into different distance bins. The velocities are estimated by 
multiplying the proper motions by the photometric distances. Thanks to the large sample statistics, we are able to achieve proper motion 
measurements with a high significance at each distance bin. Within 15~kpc, red stars dominate, whereas there are more blue stars beyond 15~kpc. 
Since our own motion has not been corrected here, and those local stars are rotating along with us, this leads to nearly zero velocity at 
very nearby distances (i.e. only small peculiar velocities with respect to us). The velocities increase to $\geq$200 km/s for 
distant stars beyond 5~kpc. As can be seen from Figure~\ref{fig:dist}, those stars are fairly far away from the disc, whose motions are 
closer to random. The $\geq$200 km/s velocity is mainly due to our own rotation motion in the Galactic disc with respect to the Galactic
centre. 

More interestingly, in Figure~\ref{fig:dblue}, we present the results for blue stars out to 100~kpc, where we employ a wider bin width 
of distances for more distant stars to increase the statistics. For more distant stars, the proper motions decrease, but the figure 
displays a coherent motion of about 200 km/s, most of which is caused by our own rotation motion as described below.

\subsection{Measurements after correcting for our own motion}
\label{sec:nosolarmotion}

\begin{figure}
\begin{center}
 \includegraphics[width=\columnwidth]{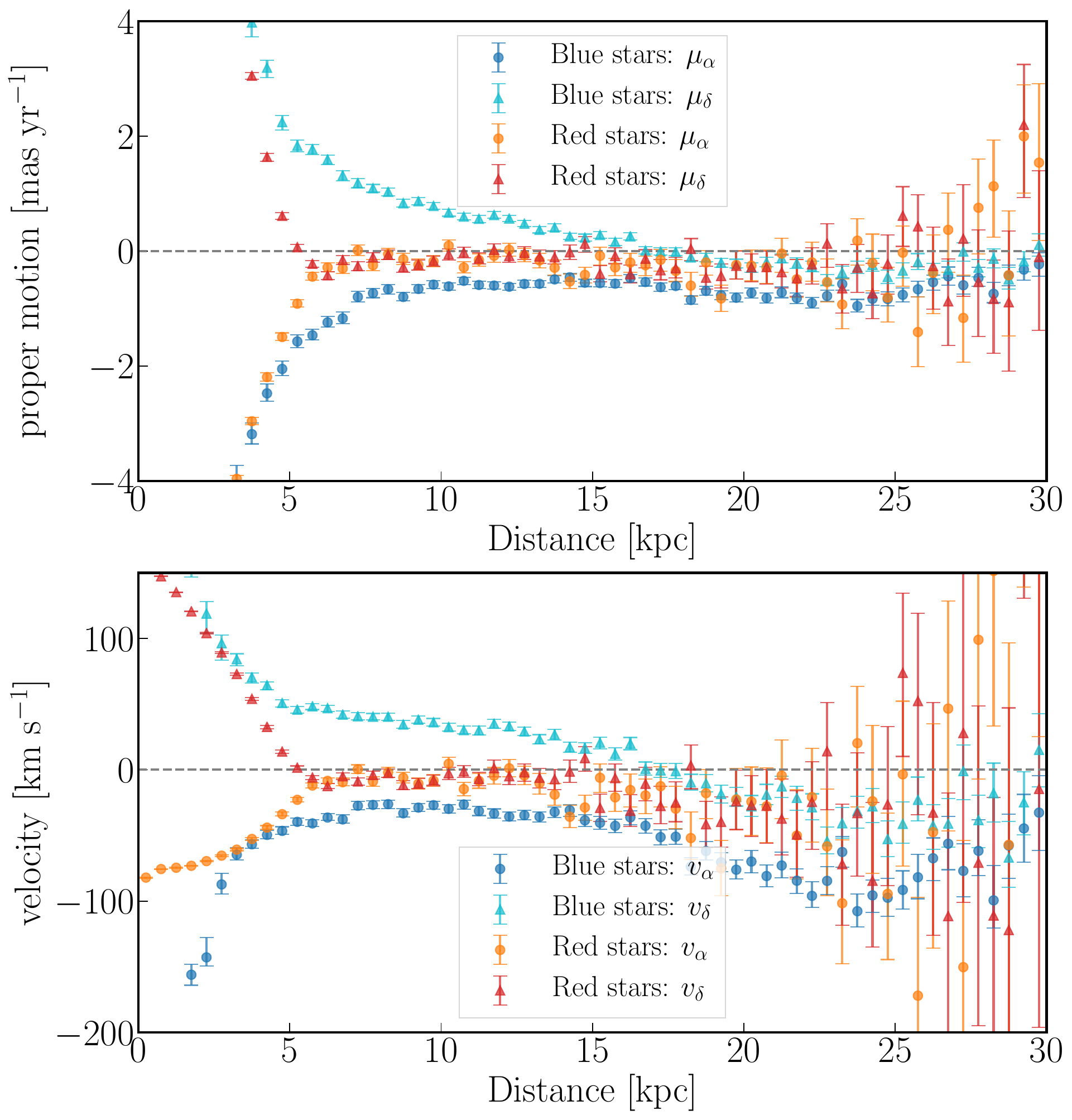}
 \caption{Similar to Figure~\ref{fig:dall}, but the figure shows the 
 the component proper motions and tangential velocities after correcting 
 for the apparent motions due to our own motions for each star.
 }
 \label{fig:dall_sm}
 \end{center}
\end{figure}

\begin{figure*}
\begin{center}
 \includegraphics[width=15cm]{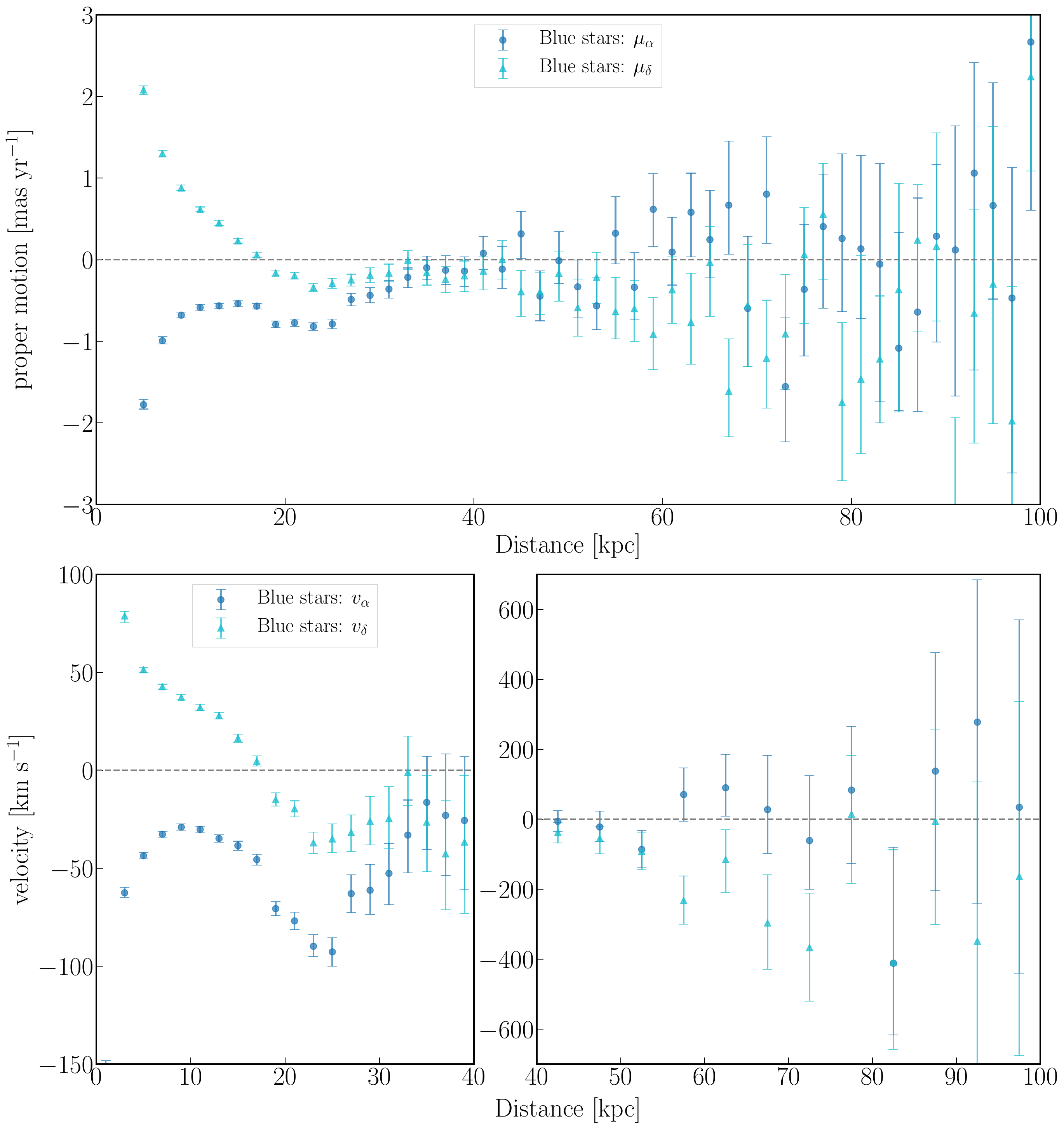}
 \caption{Similar to Figure~\ref{fig:dblue}, but the figure shows the component proper motions and tangential velocities for blue stars 
 out to 100~kpc, after correcting for the effect due to our own motion. For illustrative purpose the lower two panels show the 
 results at distances smaller and greater than 40~kpc separately, where the range of $y$-axis is adjusted to cover the range of data points.
 The averaged proper motions for stars at distances $\ge 40~$kpc are $\langle\mu_\alpha\rangle=-0.091^{+0.090}_{-0.086}$~mas~yr$^{-1}$
 and $\langle \mu_\delta \rangle=-0.398^{+0.079}_{-0.087}$~mas~yr$^{-1}$, or $\langle v_\alpha\rangle=-20.6^{+23.2}_{-22.2}$~km/s and 
 $\langle v_\delta\rangle= -97.1^{+19.9}_{-23.3}$~km/s, respectively.
}
  \label{fig:dblue_sm}
  \end{center}
\end{figure*}

\begin{figure*}
\begin{center}
 \includegraphics[width=16cm]{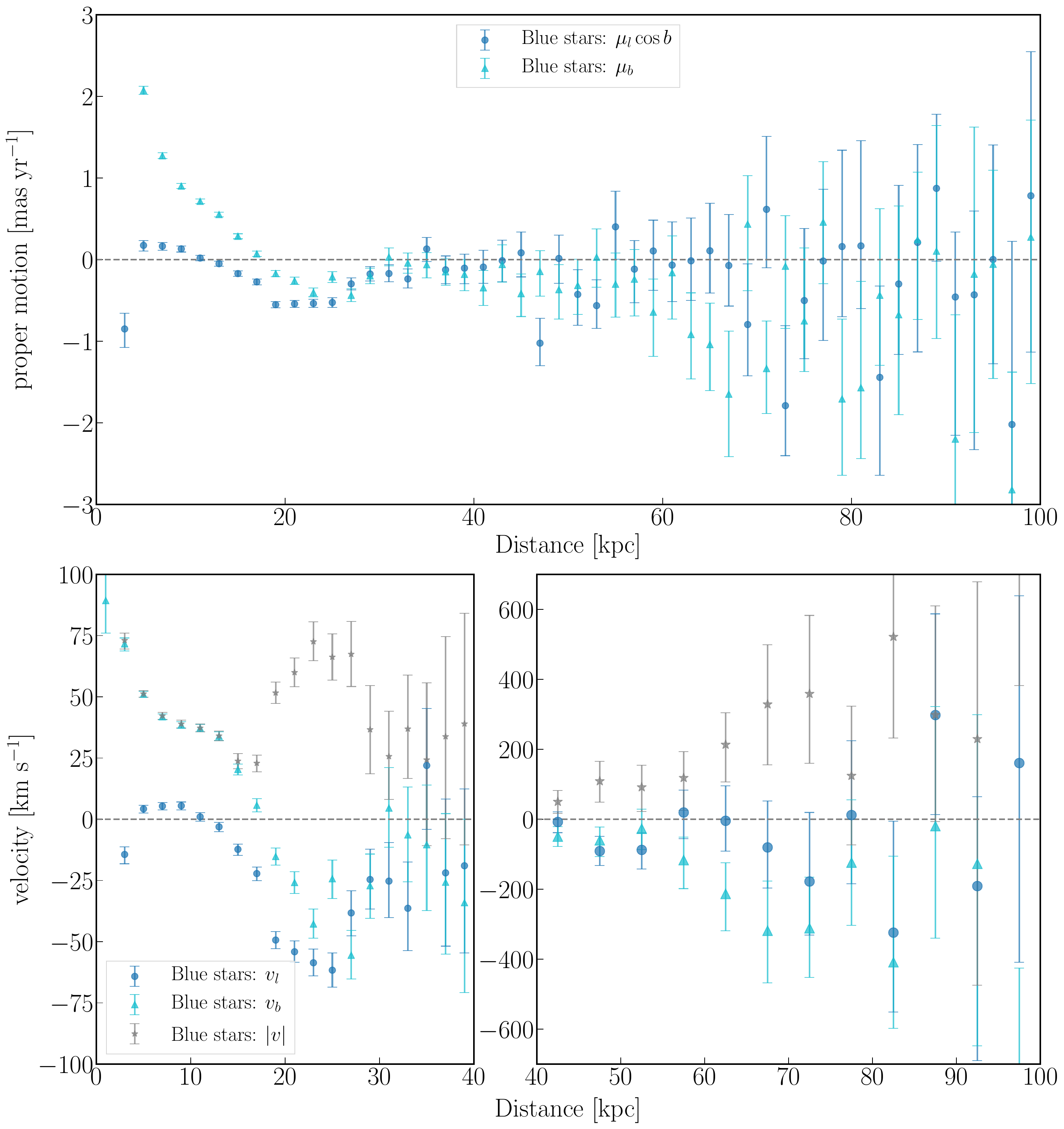}
 \caption{Similar to Figure~\ref{fig:dblue_sm}, but the component proper motions and tangential velocities for blue stars are shown 
 in the Galactic coordinates. The component tangential velocities at distances smaller and larger than 40~kpc are shown in two separate
 panels, in order to more clearly demonstrate the velocity measurements on smaller distance scales. The averaged proper motions for stars
 at $\ge 40$~kpc are: $\langle\mu_\ell\cos b\rangle= -0.166^{+0.083}_{-0.087}$~mas~yr$^{-1}$ and $\langle\mu_b\rangle= 
 -0.355^{+0.081}_{-0.090}$~mas~yr$^{-1}$, while $\langle v_\ell\rangle =-42.2^{+21.5}_{-22.5}$~km/s and $\langle v_b\rangle =
 -87.3^{+20.9}_{-24.4}$~km/s, respectively. We also show the tangential velocity amplitude, $|v|\equiv (v_\ell^2+v_b^2)^{1/2}$, as a 
 function of distances, by the grey star symbols. 
 }
 \label{fig:dall_sm_lb}
 \end{center}
\end{figure*}

To obtain a physical insight on the proper motion measurements presented in the previous subsection,
we need to take into account the effect of our own motion. We use the \texttt{astropy} python package to 
correct for the effect of solar motion via the following steps:

\begin{enumerate}
    \item Input R.A., Dec., proper motion in R.A., proper motion in Dec., line-of-sight velocity
    and photometric distance to generate an \texttt{astropy.SkyCoord} object.
    \item Transfer it to the Galactocentric coordinate system.
    \item Obtain the 3-dimensional positions and velocities with respect to the Galactic centre.
    \item Subtract our own motion (total motion of the LSR motion with respect to the Galactic centre plus the solar motion with respect
    to the LSR) from the measured proper motions of stars
    in 3-dimensional Cartesian coordinates.
    Here we employ $(U_\odot,V_\odot,W_\odot)=(11.1,12.24,7.25)~$km/s for the Galactic (right-handed) velocity components according 
    to \citet{2010MNRAS.403.1829S}, and $v_{c}(R_\odot)=220~$km/s for the disc rotation.
    \item Transfer back the velocities to those in the International Celestial Reference System (ICRS).
\end{enumerate}

By applying the above steps to each star, we obtain its two components of tangential velocity in ICRS. Note without line-of-sight 
velocities, we are not able to transfer the tangential velocities to be with respect to the Galactic centre, and thus tangential velocities 
measured throughout the paper are centred on observers.

In Figures~\ref{fig:dall_sm} and \ref{fig:dblue_sm}, we show the median proper motions and velocities after subtracting our 
own motion up to 30 and 100~kpc, respectively. These results can be compared with the results in Section~\ref{sec:directmotion}. 
We have a clear detection of the coherent proper motions up to $\sim$30--40~kpc, even after the effect of our own motion is corrected 
for.

Within 1~kpc, there are almost no blue stars. The total tangential velocities of these very nearby red stars 
are dominated by the Galactic rotation in the thick disc region. Between 1 and 5~kpc, the motions of red and blue stars show quite 
different behaviours. Blue stars show larger tangential velocities, and with the increase in distance, the tangential velocities of 
both red and blue stars drop, due to the gradually decreased contribution by the Galactic rotation.

Between 5 and 30~kpc of Figure~\ref{fig:dall_sm}, the kinematics of red and blue stars look different. Red stars show
very close to zero tangential velocities, while blue stars show non-zero tangential velocities. Stars beyond 5~kpc are expected 
to be mostly halo stars dominated by random motions, because our HSC-S82 footprints all point towards negative $z$-directions 
(see Figure~\ref{fig:dist}). The none-zero median velocities of blue stars indicate there are some substructures in phase 
space which contribute to the coherent motion. As we discuss below, the non-zero motion is mainly due to the Sagittarius (Sgr)
stream. A part of the Sgr stream is covered within our footprint. The Sgr stream wraps around the Galaxy in an almost polar 
orbit which is partly overlapped with our observed region. Including the stream in our analysis can significantly affect the 
median motion of all stars to a specific direction. The Sgr stream mostly locates at 15--30~kpc in our region with 
$\mu_\alpha\sim-$2~mas~yr$^{-1}$ and $\mu_\delta\sim -2$~mas~yr$^{-1}$ \citep[e.g.][]{2013ApJ...766...79K}. 
We will discuss in Section~\ref{sec:disc} how the proper motion measurements are affected by the Sgr stream.

Figures~\ref{fig:dblue_sm} and \ref{fig:dall_sm_lb} show the proper motion measurements for blue stars at larger distances up to 
100~kpc in the Equatorial and Galactic coordinates, respectively. The measurements at $\gtrsim 40~$kpc are noisy. After the correction 
of our own motion, both the velocity in the Dec. direction ($v_\delta$) and in the $b$ direction ($v_b$) appear to show non-zero
negative values, but the significance in each distance bin is not high. More quantitatively, the averaged proper motions for all 
these blue stars at $\ge 40$~kpc are: 
$\langle\mu_\alpha\rangle=-0.091^{+0.090}_{-0.086}$~mas~yr$^{-1}$ 
 and $\langle \mu_\delta \rangle= -0.398^{+0.079}_{-0.087}$~mas~yr$^{-1}$, or 
 $\langle v_\alpha\rangle=-20.6^{+23.2}_{-22.2}$~km/s and $\langle v_\delta\rangle= -97.1^{+19.9}_{-23.3}$~km/s, respectively. 
 In the Galactic coordinate, 
$\langle\mu_\ell\cos b\rangle= -0.166^{+0.083}_{-0.087}$~mas~yr$^{-1}$ 
and 
$\langle\mu_b\rangle=  -0.355^{+0.081}_{-0.090}$~mas~yr$^{-1}$, while
$\langle v_\ell\rangle =-42.2^{+21.5}_{-22.5}$~km/s
and  
$\langle v_b\rangle = -87.3^{+20.9}_{-24.4}$~km/s, respectively.
Thus the averaged proper motions between 40 and 100~kpc indicate a perhaps $5\sigma$-level detection of the coherent motion 
in the outer MW stellar halo. In Figure~\ref{fig:dall_sm_lb} we also show the absolute tangential velocity, $v\equiv [(v_\ell)^2 +(v_b)^2]^{1/2}$, 
as a function of distances, again indicating a hint of coherent motion up to $\sim 100~$km~s$^{-1}$. If this is genuine, it 
might indicate a rotation of the halo region or the existence of global motions. For example, the Large Magellanic Cloud might 
imprint such a motion in the halo region  \citep[e.g.][]{2016MNRAS.456L..54P,2019MNRAS.487.2685E,2019arXiv190709484E,2019ApJ...884...51G}. 
However, this might be due to unknown systematics such as an imperfect correction for the DCR effect
and/or a contamination of non main-sequence, faint stars such as white dwarfs (see below). 
We need more data to 
derive a more robust conclusion.

\section{Discussion}
\label{sec:disc}

\subsection{The Sagittarius stream}
\label{subsec:sgr}

\begin{figure}
\begin{center}
 \includegraphics[width=\columnwidth]{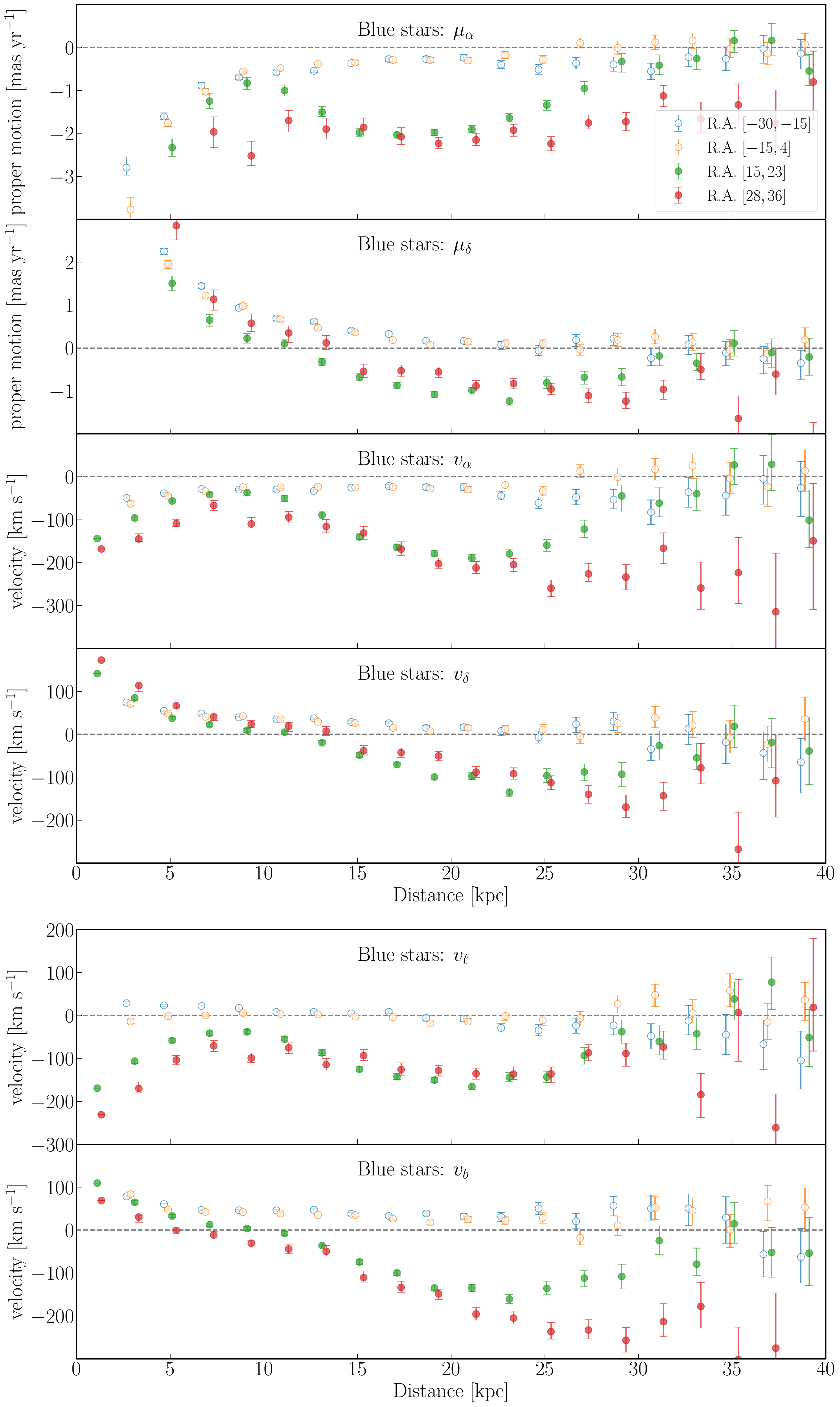}
 \caption{Similar to Figure~\ref{fig:dblue_sm}, but the figure shows the component proper motions for blue stars out to 40~kpc 
 and for four subregions of HSC-S82 footprints, divided by the different range of R.A.$=[-30,-15]$, $[-15,4$], [$15,23$] and $[28,36]$, 
 respectively. Here the two subregions with R.A.$>0$, denoted by the filled symbols, are the regions where member stars in the Sagittarius 
 stream are distributed, while the subregions with R.A$<0$, denoted by the open symbols, are not affected by the Sagittarius stream.
 The two panels at bottom show the velocity components, $v_\ell$ and $v_b$, in the Galactic coordinates.
 }
 \label{fig:sagstream}
 \end{center}
\end{figure}

\begin{figure}
\begin{center}
 \includegraphics[width=\columnwidth]{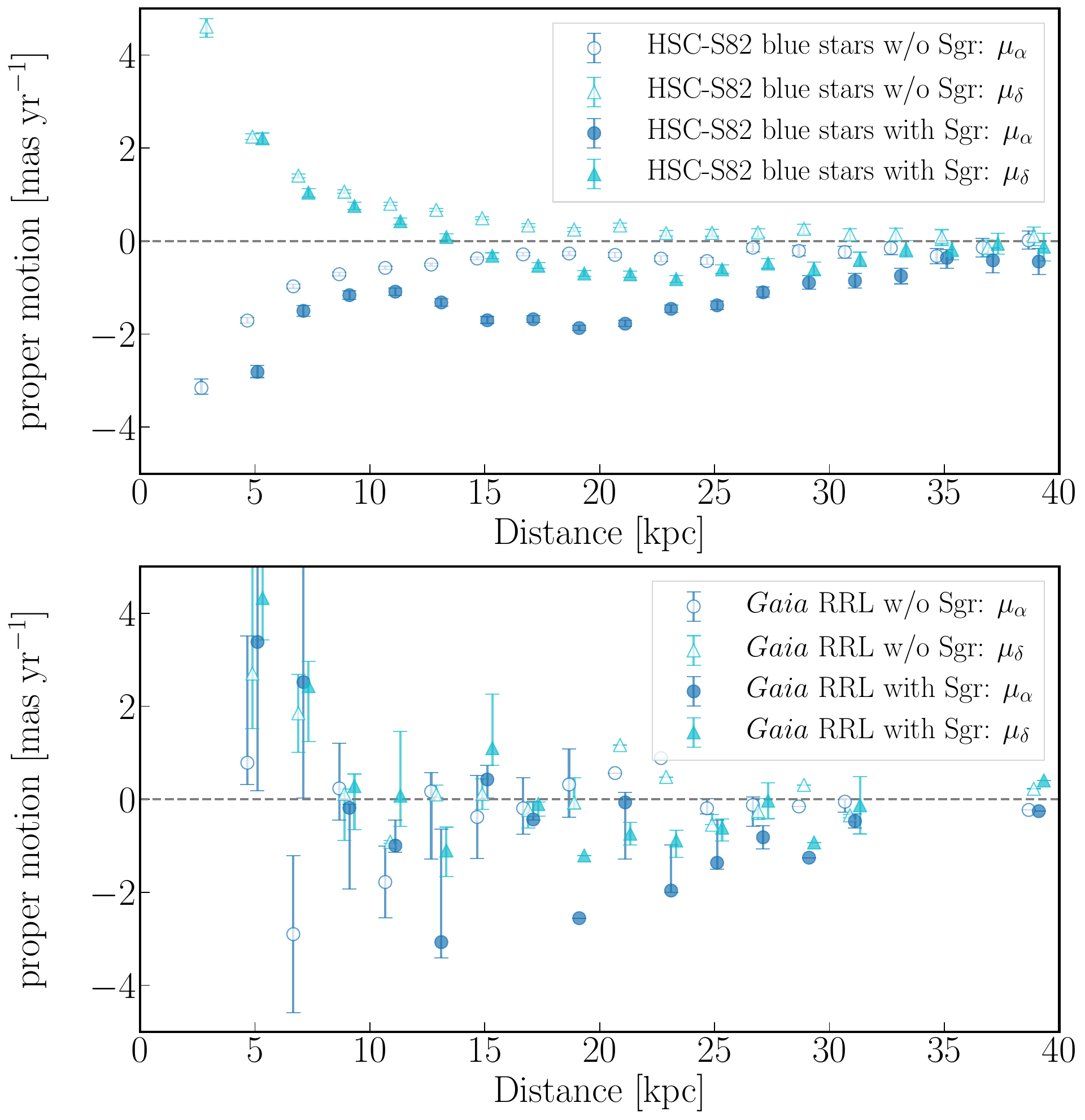}
 \caption{Comparison of the HSC-S82 proper motion measurements with RR-Lyrae proper motions from {\it Gaia}. Here we use 96~RR-Lyrae 
 stars in the overlapping region of  HSC and S82, and assumed $M_G=0.63$ for the absolute magnitude in the {\it Gaia} filter to estimate 
 distances for individual RR-Lyrae stars. To increase statistics of the {\it Gaia} results, here we show the results for the two subfields, 
 divided by R.A.$>0$ (including Sgr) or R.A.$<0$. Although the {\it Gaia} results are noisy due to low statistics, our HSC-SDSS results 
 are qualitatively in good agreement with the {\it Gaia} results. 
 }
 \label{fig:gaia_RRL}
 \end{center}
\end{figure}

The Sgr stream is covered by our HSC-S82 footprints, and in this subsection we investigate how the proper motion measurements 
are affected by the stream. Within the HSC-S82 footprints, the region with R.A.$>0$ contains the Sgr stream, while the other with R.A.$<0$ 
is not affected by the Sgr stream. This can be found from the comparison of our survey footprints (Figure~\ref{fig:cali}) with the left panel 
of Figure~2 in \citet{2013ApJ...766...79K}.

Figure~\ref{fig:sagstream} shows a comparison between the proper motion measurements in four subregions in distances up to 40~kpc. 
The two subregions containing the Sgr stream at R.A.$>0$ show coherent proper motions which are negative in both $\mu_\alpha$ and $\mu_\delta$ 
beyond $\sim$15~kpc. This is in good agreement with the sign of the component proper motions for the Sgr stream in the S82 footprint 
\citep[e.g.][]{2013ApJ...766...79K}. Comparing the two subregions with the Sgr stream show that the coherent motions in the range $[15,35]$~kpc 
look similar to each other. On the other hand, the two subregions without the Sgr stream do not display significant proper motions 
between 10 and 35~kpc, implying that blue stars have random motions on average in these regions. However, the results shown here could have 
underestimated the proper motions of Sgr tracers. If we include non-member stars in the analysis of the Sgr stream subregions and if such 
non-member stars do not have significant proper motions as indicated in the non-Sgr sub-regions, it would underestimate a genuine amplitude 
of the Sgr proper motions. Hence a further study using a cleaner sample of the Sgr tracers would be necessary, and this will be presented separately.

In Figure~\ref{fig:gaia_RRL} we compare our results with the proper motions for the {\it Gaia} RR-Lyrae (RRL) stars in the two subfields with 
the Sgr stream (R.A.$>0$) or without the Sgr stream (R.A.$<0$) in the overlapping region of HSC and S82. RRL stars are relatively bright and have a 
similar intrinsic luminosity (or have a narrow distribution of the absolute magnitudes), so serve as useful tracers to probe the spatial structures 
and proper motions at relatively large distances \citep{2020A&A...635L...3A,2020arXiv200307871B}. We identified 96 RRL stars in the HSC-S82 fields, 
and assume $M_G=0.63$ for the absolute magnitude in the {\it Gaia} filter \citep{2018MNRAS.481.1195M} to estimate the distance for individual RRL. 
Our proper motion measurements are qualitatively in good agreement with the {\it Gaia} proper motions, for the sign and amplitudes, for both the 
two subfields with or without the Sgr stream. Our measurements show much more significant detections of the proper motions in this field thanks 
to the large statistics using many stars. The agreement is quite encouraging because the comparison is a totally independent test of our measurements, 
e.g., {\it Gaia} RRL stars are free from the DCR systematics.

\subsection{Tangential velocity dispersion}

\begin{figure*}
\begin{center}
 \includegraphics[width=1.5\columnwidth]{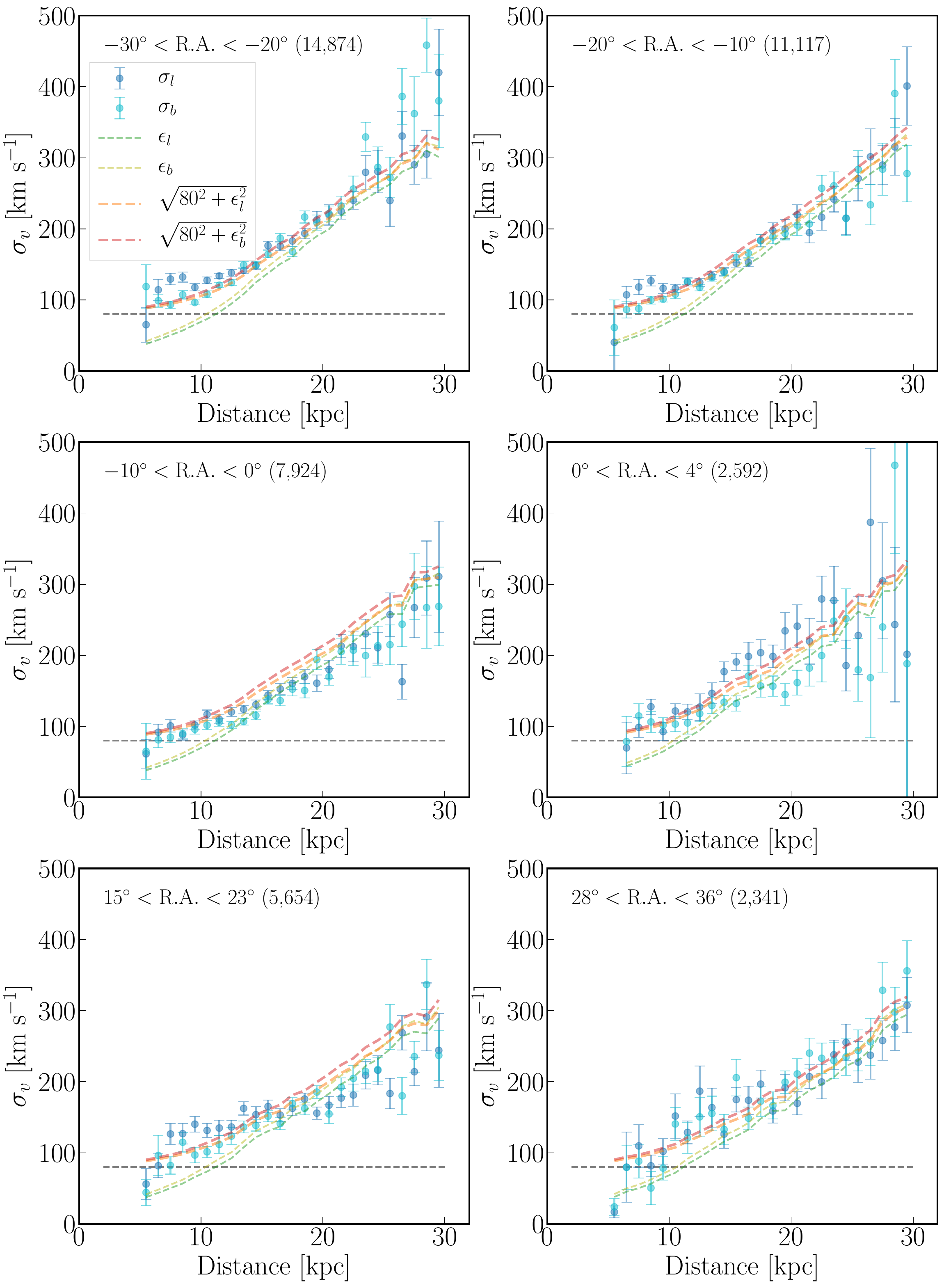}
 \caption{Tangential velocity dispersions of the two components, $v_l$ and $v_b$, in the Galactic coordinates and measured from the blue stars,
 for six subregions of the HSC-S82 footprints, divided by different ranges of R.A.. The tangential velocity dispersions are contributed by both the
 true intrinsic velocity dispersions and the measurement errors. Dark and light green curves in each panel show the amount of measurement errors. 
 Red and orange dashed curves show the predicted measurement for a constant velocity dispersion of 80~km/s plus the measurement errors. The black 
 horizontal dashed line marks the location of 80~km/s. The value in the round parentheses in each panel denotes the number 
 of blue stars in each region. The errorbars in each bin are estimated from the 200 bootstrap realizations in the bin.
 }
 \label{fig:vdisp}
 \end{center}
\end{figure*}

Velocity dispersion measurements are very important for properly understanding the dynamical properties of our MW. Line-of-sight velocity dispersions 
have been measured in many previous studies \citep[e.g.][]{2008ApJ...684.1143X,2013ApJ...766...24D,2014ApJ...794...59K,2016MNRAS.463.2623H}. 
However, much less studies have investigated tangential velocity dispersions out to large distances due to the expenses of measuring proper 
motions and also because tangential velocities usually have larger measurement errors than line-of-sight velocities. In previous sections, we 
have provided statistical studies on the averaged proper motions and tangential velocities. With our proper motion measurements, we are also able 
to take a close look at the two-component tangential velocity dispersions. To measure the tangential velocity dispersions, we focus on a bright 
subsample of 44,502 blue stars brighter than $r_\mathrm{S82,PSF}=22.5$. Red stars show similar results, and thus we avoid repeatedly showing them. 
Stars fainter than $r_\mathrm{S82,PSF}=22.5$ are excluded because they have larger measurement errors, and hence properly estimating their intrinsic 
dispersions is more difficult. 

Figure~\ref{fig:vdisp} shows the tangential velocity dispersions for $v_l$ and $v_b$ in the Galactic coordinates. We divide the HSC-S82 footprints 
into six subregions according to R.A.. Because the measurement errors dominate over the proper motion signal on individual star basis, 
the directly measured velocity dispersions contain both true intrinsic dispersions and the measurement errors. The measurement errors are mainly 
contributed by the centroid position errors from S82. However, as we will show in Appendix~\ref{app:measureerr}, the centroid errors 
underestimate the true measurement errors, and thus alternatively, we estimate the measurement errors from the dispersions in 
apparent proper motions of quasars and small-size galaxies \citep[e.g.][]{bond2010milky}. The readers can find details in Appendix~\ref{app:measureerr}.  

The predicted measurement errors are shown as dark and light green curves in each panel of Figure~\ref{fig:vdisp}. Following \cite{bond2010milky}, 
we choose to not subtract the measurement errors, because for our noisy measurements any small over-estimates in measurement errors can cause 
significant over-subtractions in true signals. To guide the eye, we also plot as red and orange dashed curves the prediction of a constant velocity 
dispersion of 80~km/s, combined with the measurement errors. In the top two panels, where the regions are not contaminated by the Sgr 
stream, our data are in good agreement with a constant velocity dispersion of 80~km/s in the distance range 10--20 kpc. The uncertainty of 
this dispersion estimate increases with distance due to increasing measurement errors and is about~10 km/s at distances below 15 kpc, and 
about 20~km/s at 20~kpc. These uncertainties are dominated by distance errors at smaller distances and proper motion errors at large distances. 
Note that at 95\% confidence level, we cannot rule out a velocity dispersion gradient of about 4.4~km/s/kpc in the distance range 10--20 kpc. 
The data in the other four panels also show mild agreement with a constant velocity dispersion of 80~km/s. Our measurement of the velocity 
dispersion for halo stars at 10~kpc is consistent with the results from \citet{2009MNRAS.399.1223S} and \citet{bond2010milky}, which did not 
reach out to 20 kpc though. Beyond 20~kpc, the measurement errors dominate. 

The tangential velocity dispersions for $v_l$ and $v_b$ are quite similar with each other, indicating isotropic dispersions.
In addition, the measured velocity dispersions are slightly smaller in the three panels with R.A.$>0$, i.e., the footprints containing 
the Sgr stream, which is true even after considering the measurement errors. This is because the orbital motions of stars stripped from 
the Sgr satellite are correlated along the stream track. Such coherent motions decrease the velocity dispersions.

\subsection{Photometric distance errors}

As have been discussed by \cite{2008ApJ...673..864J}, the photometric distance estimation can have relative errors up to at most 20\%. 
In addition, the photometric error affects the photometric distance estimates.
As we mentioned in Section~\ref{sec:photodis}, the median photometric errors
in $g$, $r$, $i$-band are 0.0133, 0.0065 and 0.0033~mag for most of the stars thanks to the deep HSC data.  
Even at the very faint end (for stars with magnitude 23.5-24.5 in $i$-band), the median errors 
are 0.062, 0.033, 0.023 mag for $g$, $r$, $i$-band. 
The uncertainty for $g-i$ colour which our photometric distances mostly rely on is at most 0.07~mag at the very faint end ($g, r, i\simeq24$). 
The photometric errors of $0.013$~mag cause only 3\% in the photometric distances on individual star basis, which is not significant compared to the errors due to uncertainty in the metallicity of stars.  
Fainter stars with errors of 0.07~mag could cause about 26\% errors in the photometric distances, but these stars are not a majority of our sample
(see Figure~\ref{fig:hist}). We should note that the systematic error in distances, after the average of many stars, should be reduced by 
$1/\sqrt{N}$ ($N$ is the number of stars taken for the average in each bin).
In our analysis, errors in photometric distances not only just shift results presented from Figures~\ref{fig:dall} to \ref{fig:vdisp} horizontally, 
but it can also affect the correction for our own motion. If the photometric distances are either systematically under- or over-estimated, we 
might have made under- or over-corrections for our own motion. As an extreme test, even if we input 20\% smaller or larger distances to correct 
for our own motion, it almost unchanges the results beyond 30~kpc. Within 30~kpc, increasing (or decreasing) the input distances by 20\% would 
decrease (or increase) the proper motion measurements after correcting for our own motion by 0.2~mas~yr$^{-1}$ on average, while the general 
trend of median proper motions as a function of distances and the relative difference between red and blue stars keep the same. 

Besides the above uncertainty, we fixed the values of $[{\rm Fe/H}]$ to $-1.45$ for stars at $\ge 6.25$~kpc in distances as our default analysis. 
As an alternative approach, we have tried to assign each blue star with a value of $[{\rm Fe/H}]$ drawn randomly from the Gaussian distribution 
with $\langle [{\rm Fe/H}] \rangle=-1.45$ and $\sigma=0.41$~dex. The Gaussian distribution is based on the real $[{\rm Fe/H}]$ distribution of 
blue stars \citep[see Figure~10 in][]{2008ApJ...684..287I}. We confirmed that the randomly assigned metallicity leads to a small change by less
than 0.1~mas~yr$^{-1}$ in our averaged proper motion measurements  at fixed distances. On the other hand, because some blue stars in the inner 
stellar halo are found to have extreme values of $[{\rm Fe/H}]$ as low as -2 \citep[e.g.]{2007Natur.450.1020C,2010ApJ...714..663D,2013ApJ...763...65A,
2019arXiv191206847A,2019arXiv191111140D}, as an extreme test, we fix ${\rm [Fe/H]}=-0.95$ or $-1.95$, i.e., shifted from the fiducial value by $\pm0.5$, 
for all red and blue stars beyond 6.25~kpc, it causes roughly $20\%$ or $-20\%$ shifts in photometric distances, but both cause mainly a horizontal 
shift in the proper motion profile against distances at $\gtrsim 10~$kpc. For nearby blue stars within 6.25~kpc, we calculated their metallicity based 
on their $(u-g)$ and $(g-r)$ colours for results in the main text. In this way, their mean metallicity is about $[{\rm Fe/H}]\sim -1.2$. After fixing 
their $[{\rm Fe/H}]$ to $-1.95$, the proper motions after correcting for our own motion change by about 25\%, while our main conclusions 
are not affected. Hence our results are qualitatively valid. Nevertheless a further careful study, e.g. using spectroscopic observations for such 
distant red stars, would be needed to understand the impact of metallicities for halo stars.

\subsection{Contamination by non-main-sequence stars}
\label{subsec:contamination}
The focus of this paper is to measure and investigate the proper motions of main-sequence stars. Moreover, our approach of photometric distance 
estimates are based on the luminosity-colour relation of main-sequence stars. Hence, if our star sample is contaminated by other types of 
non-main-sequence stars, our estimated photometric distances and thus the results from Figures~\ref{fig:dall} to \ref{fig:vdisp} might be 
affected, even though main stars are a majority population of stellar objects.

If our star sample is contaminated by giants, these intrinsically bright giants treated as fainter main-sequence stars can cause under-estimates in 
photometric distances and thus over-corrections of our own motion. \cite{2008ApJ...673..864J} have removed stars scattered by 0.3 magnitudes away from 
the stellar locus in colour-colour space for main-sequence stars, which helped to eliminate the contamination by giants. In addition, \cite{2008ApJ...673..864J} 
built a fiducial Galaxy model to investigate the contamination by giants in their sample, and they reported a worst fraction of about 5\%. In our analysis, 
the HSC flux and colour are more precise than those of SDSS, and we have applied 3-$\sigma$ clipping to our sample of stars in colour-colour space. The 
contamination by giants in our analysis is not expected to be higher than that of \cite{2008ApJ...673..864J}. Furthermore, we have repeated our analysis 
by adopting 1-$\sigma$ clipping  to our sample of stars in colour-colour space, to throw away even more stars scattered away from the main-sequence stellar 
locus, and our results presented from Figures~\ref{fig:dall} to \ref{fig:vdisp} are proved to be robust against different levels of $\sigma$ clippings. 
We have also looked at the number density profile for our sample of stars as a function of $z$ distance to the disc plane, and there is no indication of 
flattening that can be considered as a sign of the contamination by red giants. In the end, we emphasise that giant stars are usually more distant and hence 
have smaller observed proper motions, the inclusion of a small fraction of giant stars in our sample are thus not expected to significantly change our 
median/mean proper motion measurements. 

\begin{figure}
\begin{center}
 \includegraphics[width=\columnwidth]{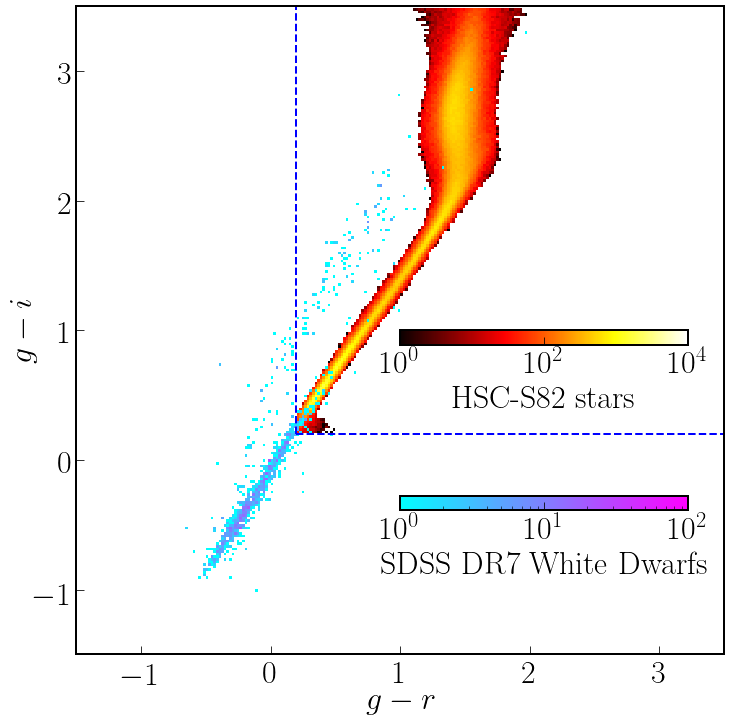}
 \caption{The distribution of main-sequence stars used in our fiducial analysis in the colour-colour diagram of $(g-r)$ vs. $(g-i)$, compared 
 to that of white dwarfs that are taken from the SDSS DR7 spectroscopic catalog of white dwarfs.
 The colours in each pixel denote the number of stars or white dwarfs in the pixel. Note that the main-sequence stars are after the 
 3-$\sigma$ clipping around the locus of main-sequence stars in the colour space, and 
 $g-r$ and $g-i$ are the colours used for photometric distance estimation of each star.
 The blue dashed lines denote the colour cuts that we use to define our sample of main-sequence stars. 
 Most of white dwarfs are outside the colour cuts, and only 3.5\% of white dwarfs lie into the colour region of main-sequence stars.
 }
 \label{fig:wd}
 \end{center}
\end{figure}

On the other hand, if our star sample is contaminated by intrinsically faint and blue non-main-sequence stars, such as white dwarfs, these intrinsically 
faint white dwarfs treated as brighter main-sequence stars can cause over-estimates in photometric distances and hence under-corrections for our own 
motion. In addition, white dwarfs are usually observed in nearby distances, and thus have large observed proper motions. If our sample of stars is 
contaminated by white dwarfs, the median/mean proper motion measurements might be biased to be larger. However, we believe that a contamination of white dwarfs to our sample is small, as described in the following. First of all, the total number of white dwarfs is only about 10\% of the number of main-sequence stars as predicted by the standard MW model \citep[also see][for the microlensing observations]{2017Natur.548..183M,
2019PhRvD..99h3503N}. Figure~\ref{fig:wd} shows the distribution of white dwarfs in the colour-colour diagram of $g-r$ vs. $g-i$, compared to that 
of main-sequence stars of the sample used in our proper motion measurements. For this comparison, we use a secure sample of white dwarfs that is taken from the SDSS DR7 spectroscopic sample of white dwarfs \citep{2013ApJS..204....5K}. The figure shows that only about 3.5\% of white dwarfs lie into the region of colours which we use to define our sample of main-sequence stars (after including 3-$\sigma$ clipping around the locus of main-sequence stars). 
Hence we conclude that white dwarfs, even if exist, give only about 0.35\%($=0.1\times 0.035$) contamination to our sample (only for the blue star sample). 
Although we think that 
our results are unlikely significantly 
affected by the contamination of white dwarfs, the possible contamination could affect the proper motion measurements at large distances, especially 
$d>40~$kpc, where we found an apparent non-zero proper motion. This requires a more careful study, and is our future study by using more datasets. 
In addition, 
searching for white dwarfs themselves from our deep photometry and proper motion measurements, e.g. through hyper velocities, would be interesting, and this
will be our future study. 

\section{Conclusion}
\label{sec:concl}

By taking advantage of the nearly 12~years of time baseline leverage between SDSS and HSC, we measure the proper motions for more than 0.55 million stars. 
Thanks to the superb image quality and the deep coadd photometric sources from SDSS Stripe~82 (S82) and HSC, we are able to make proper motion measurements 
for stars down to $\sim$24 in $i$-band, which is 3 magnitudes deeper than the nominal SDSS depth, and is deeper than almost all of the previous proper 
motion catalogues. The net proper motions of stars, including the contribution of the Galactic rotation, can be measured out to 100~kpc. In particular the 
use of HSC data was crucial for our study; for example, the HSC data allows for a secure identification of stars from the matched catalogues.

We adopt the mean epoch difference for each individually matched source between S82 and HSC as its time baseline, to calculate the proper motion in unit of 
mas~yr$^{-1}$. The matched galaxies are used to recalibrate systematic errors in astrometry and to bring the observations taken at different epochs to be in 
the same reference frame. 

We first compare our measured proper motions with the {\it Gaia} DR2 proper motions, which shows very good overall agreement. The median systematic bias 
of our proper motion measurements only differ from {\it Gaia} by at most 0.052~mas~yr$^{-1}$ (see Figure~\ref{fig:gaia_nocorr}). The scatter is mainly 
contributed by the proper motion errors from {\it Gaia} itself, because the matched stars are at faint end of the Gaia catalogue, $20<G<21$, where the 
{\it Gaia} proper motion errors are relatively large. 

However, after dividing stars into red and blue populations, each of them shows a non-negligible median offset from {\it Gaia} DR2 proper motions 
in $\mu_\delta$, which is in good agreement with the effect of differential chromatic diffraction (DCR) by the Earth's atmosphere for the SDSS images, 
where the {\it Gaia} space observation is free from this kind of systematics and the effect in the HSC images is corrected by the Atmospheric 
Dispersion Compensator. The offsets are $\Delta \mu_\delta\simeq -0.538$~mas~yr$^{-1}$ for blue stars and $\Delta \mu_\delta\simeq$0.115~mas~yr$^{-1}$ 
for red stars, respectively. In addition, an independent test using the matched quasars, which have bluer spectra from galaxies, reveal systematic 
errors of $\simeq-0.566$~mas~yr$^{-1}$ in $\mu_\delta$, which is consistent with the DCR effect in  comparison with the {\it Gaia} results. 
Hence, as our default analysis method, we corrected for the DRC effects for individual stars using the measured offsets in the {\it Gaia} 
comparison as a function of colours of stars ($g-i$). This DCR correction is important to achieve the precision better than $0.1$~mas~yr$^{-1}$,
that is needed to study the proper motions at large distances in the Galactic halo region after subtracting the contribution due to own own motion.

We estimate the photometric distance of individual stars, following the method in \cite{2008ApJ...673..864J} and \cite{2008ApJ...684..287I}. The directly 
measured proper motions show coherent R.A. and distance dependent motions, most of which are due to our own motion (the Galactic rotation plus the Sun's 
motion with respect to the Local Standard of Rest). After correcting for our own motion, we find the proper motions are dominated by the Galactic disc 
rotation within 1~kpc, and gradually decrease out to 5~kpc. Our measurements are affected by the Sgr stream, which have negative proper motions in both 
$\mu_\alpha$ and $\mu_\delta$ beyond 13~kpc. By looking into the footprint without the Sgr stream, we do not find any significant coherent proper motions 
in the distance range between 10 and 40~kpc (Figures~\ref{fig:dall_sm} and \ref{fig:sagstream}). 
The kinematical structures in the halo region are useful to study the assembly history of the MW, and a further careful study including properties of the Sgr 
stream will be presented elsewhere. As a prospect for future observation, the Subaru Prime Focus Spectrograph \citep{2014PASJ...66R...1T}, which is currently 
under construction, would be powerful to study radial motions and chemical compositions for distant stars that provide very complementary information to 
the proper motions and other information from imaging datasets.  

Beyond 40~kpc, the measurements are noisy, but the motions of  blue stars in the more distant stellar halo are close to random. After adopting wider bins 
in distance, we find the velocity in the Dec. direction ($v_\delta$) or in the $b$ direction ($v_b$) both appears to show non-zero negative values, but the 
significance in each distance bin is not high due to the limited statistics. If this is genuine, it might indicate the existence of global motions 
in the distant stellar halo, but we need more data to reach a rigorous conclusion. 

The two-component tangential velocity dispersions are measured for a subsample of blue stars brighter than $r_\mathrm{S82,PSF}=22.5$. Following \cite{bond2010milky},
we estimate the measurement errors through the dispersion in apparent proper motions of small-size galaxies and quasars after astrometry recalibrations to remove the 
spatial variations of astrometric systematics, i.e., the middle panel of Figure~\ref{fig:cali}. We measured the tangential velocity dispersion in the distance range 
5-20 kpc and find that the data are consistent with a constant isotropic dispersion of $80\pm 10~{\rm km/s}$, which is consistent with the earlier works 
\citep{2009MNRAS.399.1223S,bond2010milky} for the shorter distances of $<10~{\rm kpc}$. The footprints containing the Sgr stream tend to show slightly smaller 
velocity dispersions, which is due to the correlation in orbital motions of stripped stars from the Sgr dwarf.

Tangential velocity dispersions are crucial for properly constraining the velocity anisotropy of halo stars. Early studies circumvent the unknown anisotropy 
by adopting arbitrary values, treating it as a free but constant parameter, assuming a certain functional form, constraining it from numerical simulations or 
modelling the velocity ellipsoid \citep[e.g.][]{2005MNRAS.364..433B,2006MNRAS.369.1688D,2008ApJ...684.1143X,2010MNRAS.406..264W,2014ApJ...794...59K}. These 
approaches, however, unavoidably introduce additional uncertainties \citep{2018MNRAS.476.5669W}. The situation has been significantly improved by the advent of 
{\it Gaia} proper motions for bright halo stars, globular clusters and satellite galaxies \citep[e.g.][]{2018ApJ...862...52S,2019ApJ...873..118W,2019ApJ...886...69L,
li2019constraining,2019MNRAS.484.5453C,2019A&A...621A..56P,2019MNRAS.484.2832V,2019ApJ...875..159E,2020arXiv200102651F}. However, {\it Gaia} proper motions for 
individual main-sequence stars are still confined within very nearby distances. The tangential velocity dispersion for our sample of main-sequence stars is thus
very helpful. Although currently we do not have line-of-sight velocities, by assuming the velocity anisotropy of different stellar populations are similar to 
each other, we can combine existing line-of-sight velocity dispersion measurements from other types of stars to constrain the anisotropy of the stellar halo. 
However, such an assumption still awaits further tests, and we leave more detailed discussions to a future study. 

Our results prove the power of statistical method for measuring proper motions of distant stars, especially in the halo regions of the MW. In this paper we mainly 
focus on the {\it averaged} proper motions of stars and their dispersions, and did not focus on rare populations of the proper motion distribution in a given bin 
or the spatial structures of the proper motions. For example, a tail population such as hyper-velocity stars is a useful tracer of the gravitational potential strength 
of the MW, and the spatial structures of the proper motions are powerful to study the substructure remnants and assembly histories of the MW. These are our future work, 
and will be presented elsewhere as well. Our method can be applied to ongoing and upcoming surveys such as the DES, LSST, Euclid and WFIRST. The current study is 
the first step of the statistical proper motion measurements, and we hope our work presented in this paper gives a guidance for these surveys.

\section*{Acknowledgements}

The Hyper Suprime-Cam (HSC) collaboration includes the astronomical communities of Japan and Taiwan, and Princeton University. The HSC instrumentation and software 
were developed by the National Astronomical Observatory of Japan (NAOJ), the Kavli Institute for the Physics and Mathematics of the Universe (Kavli IPMU), the University 
of Tokyo, the High Energy Accelerator Research Organization (KEK), the Academia Sinica Institute for Astronomy and Astrophysics in Taiwan (ASIAA), and Princeton University. 
Funding was contributed by the FIRST program from Japanese Cabinet Office, the Ministry of Education, Culture, Sports, Science and Technology (MEXT), the Japan Society for
the Promotion of Science (JSPS), Japan Science and Technology Agency (JST), the Toray Science Foundation, NAOJ, Kavli IPMU, KEK, ASIAA, and Princeton University.

This paper makes use of software developed for the Large Synoptic Survey Telescope. We thank the LSST Project for making their code available as free software at \url{http://dm.lsst.org}.

The Pan-STARRS1 Surveys (PS1) and the PS1 public science archive have been made possible through contributions by the Institute for Astronomy, the University of Hawaii, the 
Pan-STARRS Project Office, the Max-Planck Society and its participating institutes, the Max Planck Institute for Astronomy, Heidelberg and the Max Planck Institute for 
Extraterrestrial Physics, Garching, The Johns Hopkins University, Durham University, the University of Edinburgh, the Queen’s University Belfast, the Harvard-Smithsonian 
Center for Astrophysics, the Las Cumbres Observatory Global Telescope Network Incorporated, the National Central University of Taiwan, the Space Telescope Science Institute, 
the National Aeronautics and Space Administration under Grant No.~NNX08AR22G issued through the Planetary Science Division of the NASA Science Mission Directorate, the National 
Science Foundation Grant No. AST-1238877, the University of Maryland, Eotvos Lorand University (ELTE), the Los Alamos National Laboratory, and the Gordon and Betty Moore Foundation.

This paper is based on data collected at the Subaru Telescope and retrieved from the HSC data archive system, which is operated by Subaru Telescope and Astronomy Data Center 
at National Astronomical Observatory of Japan. Data analysis was in part carried out with the cooperation of Center for Computational Astrophysics, National Astronomical 
Observatory of Japan.

We would like to thank Jim~Annis, Ana~Bonaca, Linghua~Jiang, Chervin~Laporte, Surhud~More, Xiangchong~Li, Hu~Zou, Xu~Zhou, Zhaoyu~Li and Chao~Liu for useful discussions and helps.
This work is supported in part by World Premier International Research Center Initiative (WPI Initiative), MEXT, Japan, and JSPS KAKENHI Grant Numbers JP15H03654, JP15H05887,
JP15H05893, JP15H05896, JP15K21733, and JP19H00677.
Wenting Wang is supported by NSFC (12022307).

\section*{Data Availability}

The data underlying this article are available in HSC-SSP, at \href{url}{https://doi.org/10.1093/pasj/psz103}, SDSS DR7 Stripe82 at \href{url}{https://doi.org/10.1088/0004-637X/794/2/120}, 
$Gaia$ DR2 at \href{url}{https://doi.org/10.1051/0004-6361/201833051} and SDSS DR14 quasar at \href{url}{https://doi.org/10.1051/0004-6361/201732445}. 
The datasets were derived from sources in the public domain: HSC-SSP  \href{url}{https://hsc-release.mtk.nao.ac.jp/doc/}, SDSS Catalogue Archive Server \href{url}{http://skyserver.sdss.org/CasJobs/}, 
$Gaia$ Archive \href{url}{https://gea.esac.esa.int/archive/} and SDSS Quasar Catalogue \href{url}{https://www.sdss.org/dr14/algorithms/qso\_catalog/}.




\bibliographystyle{mnras}
\bibliography{Ref} 




\appendix
\section{query for samples in HSC and S82}
\label{app:query}

In the following, we provide the queries that we adopted to download our sample of sources from HSC (the HSC-SSP Public Data Release webpage\footnote{\href{url}{https://hsc-release.mtk.nao.ac.jp/doc/}}) 
and S82 (SDSS catalog Archive Server\footnote{\href{url}{http://skyserver.sdss.org/CasJobs/}}).

\subsection{HSC}
\texttt{SELECT \\
        object\_id, ra, dec, i\_extendedness\_value,\\
		i\_sdsscentroid\_ra, i\_sdsscentroid\_dec,\\ i\_sdsscentroid\_rasigma, i\_sdsscentroid\_decsigma,\\
		g\_cmodel\_mag, r\_cmodel\_mag,\\
		i\_cmodel\_mag, z\_cmodel\_mag,\\
		g\_cmodel\_magsigma, r\_cmodel\_magsigma, \\
		i\_cmodel\_magsigma, z\_cmodel\_magsigma \\
		g\_psfflux\_mag, r\_psfflux\_mag,\\
		i\_psfflux\_mag, z\_psfflux\_mag,\\
		g\_psfflux\_magsigma, r\_psfflux\_magsigma,\\ i\_psfflux\_magsigma, z\_psfflux\_magsigma\\
FROM \\
    s18a\_wide.forced \\
LEFT JOIN \\
    s18a\_wide.forced2 USING (object\_id)\\
WHERE \\
    i\_psfflux\_mag>17 and i\_psfflux\_mag<24.5 \\
    and dec>-1.26 and dec<1.26 and (ra<=60 or ra>=300) \\
    and isprimary
    }

\subsection{S82}
\texttt{SELECT\\
  objID, ra, dec, \\
  run, rerun, camcol, field, flags, type,\\
  rowc\_r,colc\_r,rowcErr\_r,colcErr\_r,\\
  u,g,r,i,z,\\
  err\_u, err\_g, err\_r, err\_i, err\_z,\\
  psfmag\_u, psfMag\_g, psfMag\_r, psfMag\_i, psfMag\_z, \\
  psfmagerr\_u,psfmagerr\_g, psfmagerr\_r, psfmagerr\_i, psfmagerr\_z\\
into mydb.all from PhotoPrimary\\
WHERE\\
  ((flags \& 0×10000000) != 0) \\
  AND ((flags \& 0×8100000c00a4) = 0)\\
  AND (((flags \& 0×400000000000) = 0) or \\
  (psfmagerr\_r <= 0.2 and psfmagerr\_i <= 0.2 \\
  and psfmagerr\_g <= 0.2))\\
  AND (((flags \& 0x100000000000) = 0) or (flags \& 0x1000) = 0)\\
  AND (run = 106 or run = 206) \\
  AND ((ra BETWEEN 329 and 360) or (ra BETWEEN 0 and 36))\\
  AND (psfMag\_i BETWEEN 17 and 24.2)
}

\section{The importance of astrometry recalibration and proper estimates of time baselines}
\label{app:calibrole}
\begin{figure}
\begin{center}
\includegraphics[width=0.9\columnwidth]{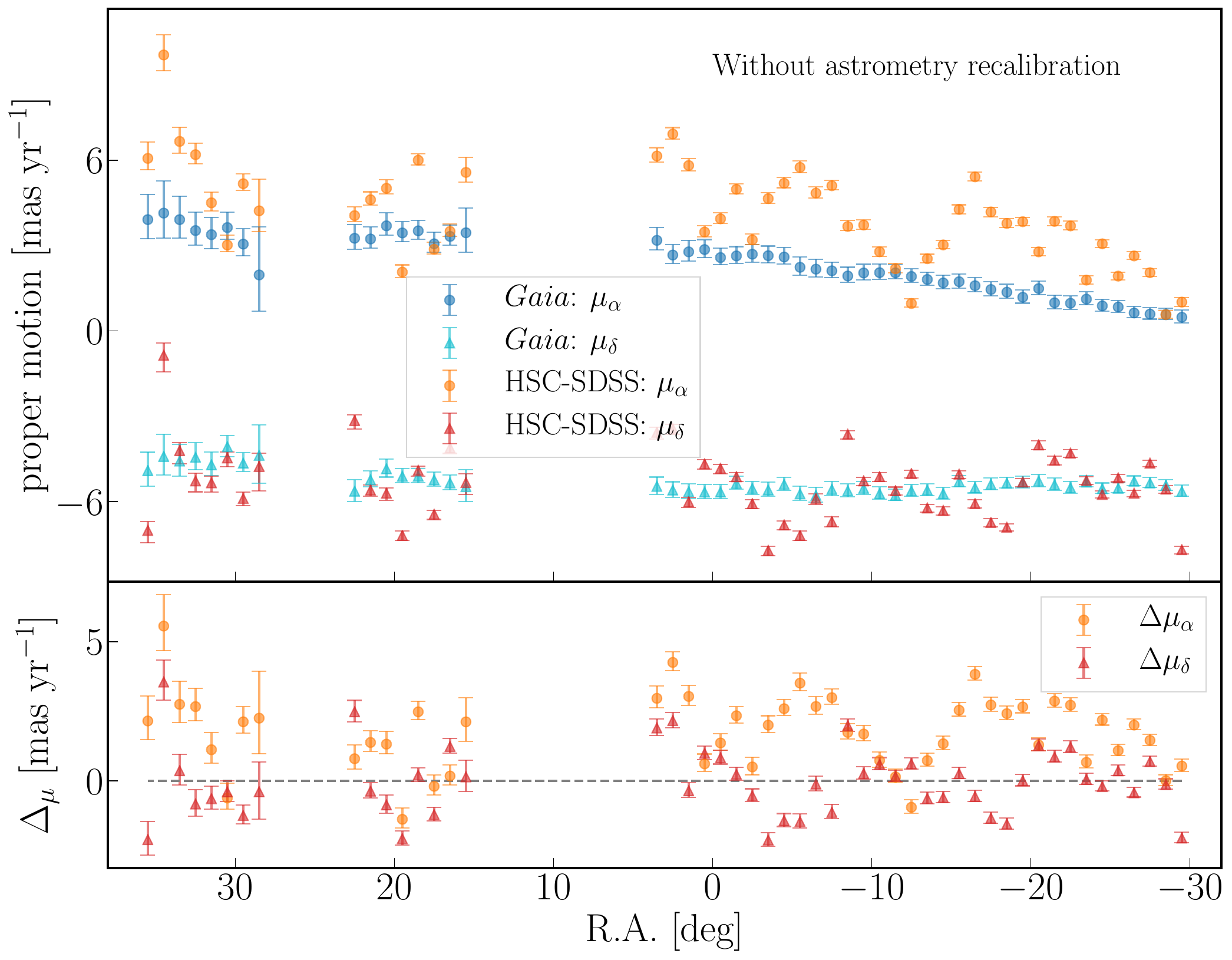}
\includegraphics[width=0.9\columnwidth]{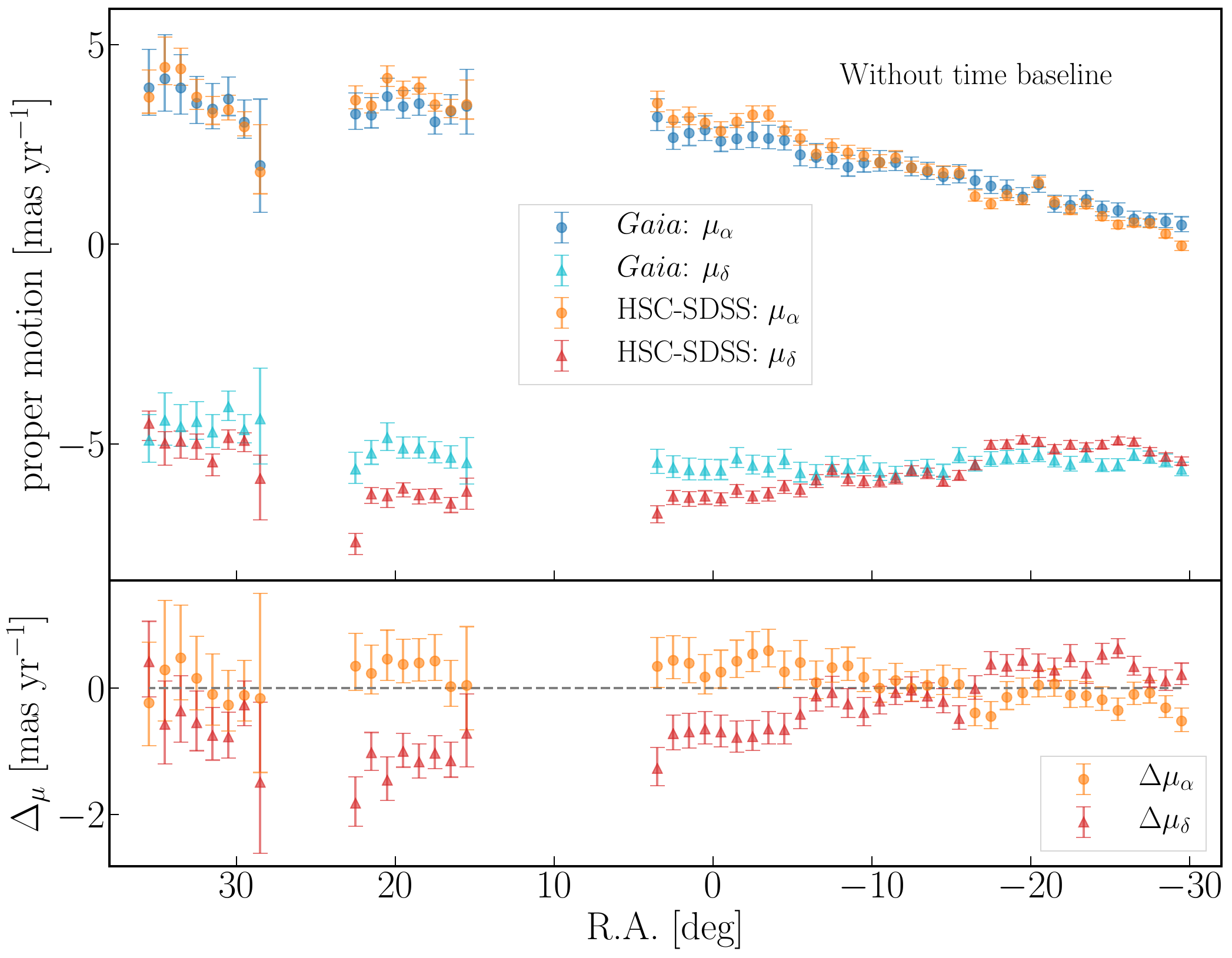}
\caption{Similar to Figure~\ref{fig:gaia}, but here we show the proper motion measurements when either the astrometry recalibration is not 
implemented or the time baseline is not correctly estimated. \textbf{Upper panels}: The results without the astrometry recalibration using 
the angular offsets of galaxies between HSC and S82 (see Section~\ref{sec:astrometry_calibration}), but with the correctly estimated time baseline. 
If we do not use this recalibration, the HSC-S82 proper motions display a significant discrepancy from the {\it Gaia} measurements, up to 5~mas~yr$^{-1}$. 
\textbf{Lower}: The results if we fix 12 years for the HSC-S82 time baseline in the proper motion estimations, rather than using the more exact 
mean epoch difference for each star in Figure~\ref{fig:sep}. The fixed 12-years of time baseline introduces systematic discrepancies with {\it Gaia} 
by up to about 1~mas~yr$^{-1}$.}
\label{fig:gaia_wo_calibration}
\end{center}
\end{figure}

Throughout the main text of the paper, we measure proper motions for main-sequence stars by comparing the R.A. and Dec. coordinates between matched 
photometric sources detected on deep coadd images from the SDSS Stripe 82 region and HSC. We estimate the time baseline from the mean epoch difference 
based on all single exposure images adopted for coadding. On the other hand, galaxies have been used to recalibrate the astrometry of stars, which has 
helped to correct for residual systematics in astrometry and set up the common reference frame. 

Figure~\ref{fig:gaia_wo_calibration} demonstrates that both astrometry recalibration using galaxies and proper estimates of time baselines are very 
important. If we do not include the astrometry recalibration, the measured proper motions can differ from {\it Gaia} DR2 proper motion measurements 
by up to 5~mas~yr$^{-1}$. If we simply fix the time baseline to be 12 years, which is roughly the average time difference between HSC and SDSS DR7, 
our measurements can differ from {\it Gaia} DR2 by up to about 1~mas~yr$^{-1}$.

\section{Measurement errors in the proper motion measurements}
\label{app:measureerr}

\begin{figure}
\begin{center}
 \includegraphics[width=0.9\columnwidth]{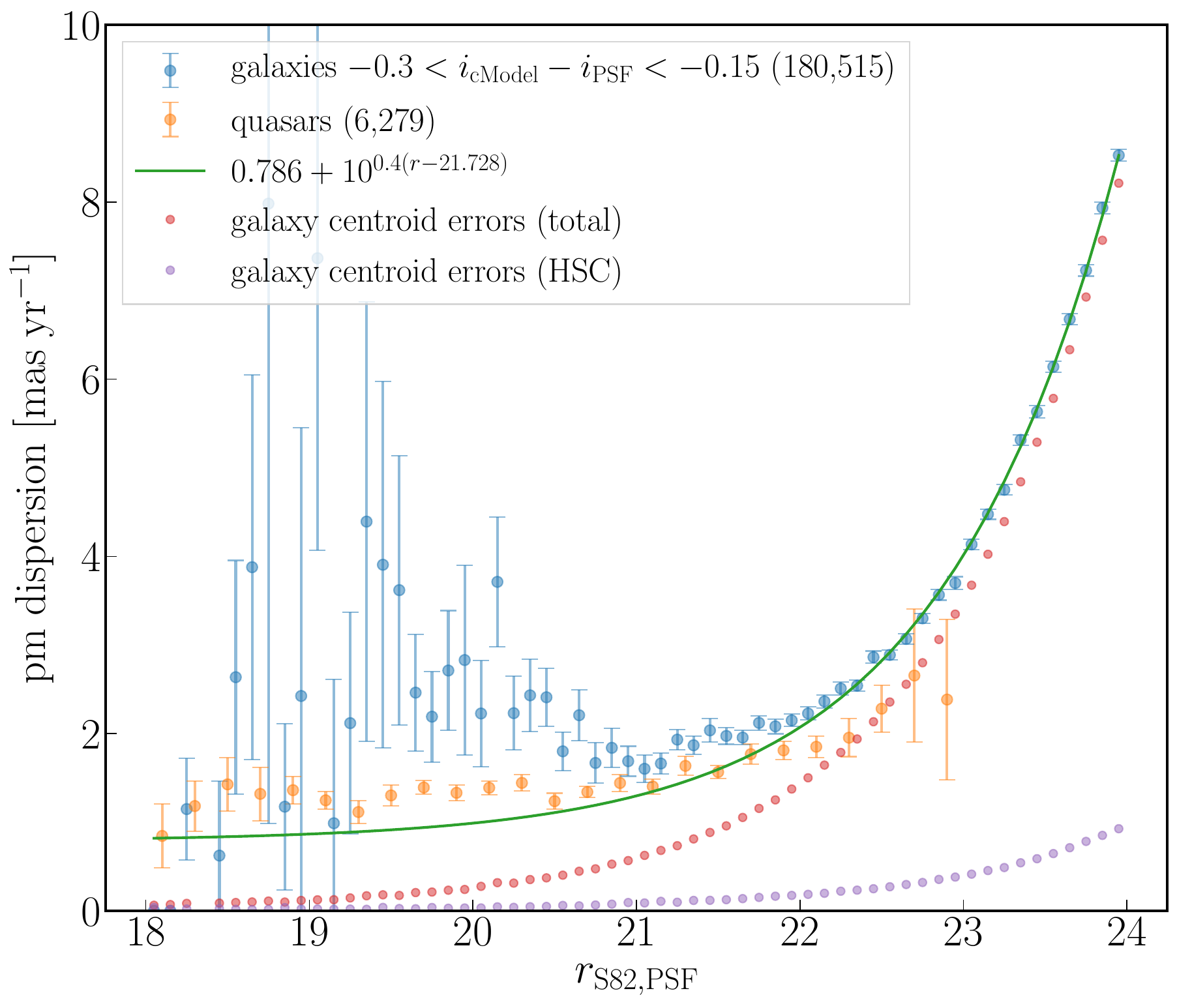}
 \caption{Measurement errors in our proper motion measurements. Blue points are the errors estimated from the dispersion or width of 
 the apparent proper motion distribution of ``small-size'' galaxies in each bin of the SDSS $r$-band magnitudes, 
 where small-size galaxies are defined from the HSC data based on the criterion $-0.3<i_{\rm HSC,cModel} - i_{\rm HSC, PSF}<-0.15$ 
 (see text for details). We have 180,515 small-size galaxies in the HSC-S82 regions, and the errorbars are estimated from the bootstrap method. 
 The orange points are the errors estimated from the quasars that are from the same sample in Figure~\ref{fig:qso_dcr}. The green line 
 shows the expectation if the errors scale with a flux of object as $1/{\rm flux}\propto 10^{0.4(r-21.728)}$, where we find the 
 normalisation and the pivot magnitude ($r=21.728$) so that the line has a good match to the galaxy results at faint magnitudes. 
 The data points of galaxies nicely follow the expectation. The red points are the errors that are estimated from the quadrature 
 sum of the centroid errors of galaxies in the S82 and HSC catalogues, while the purple points are from the centroid errors in the HSC 
 catalogue alone.
 }
 \label{fig:measureerr}
 \end{center}
\end{figure}

In this appendix we quantify the measurement errors in our proper motion measurements. As indicated, e.g. by the bottom panel 
of Figure~\ref{fig:gaia}, the measurement (statistical) errors are dominant over the proper motion signal on individual star basis. 
Naively we expect that the measurement errors are mainly from the uncertainties in centroid determinations of stars, as reported in 
the HSC and S82 catalogues, but it turns out that the centroid errors underestimate the measurement errors as we show below. In addition, 
the measurement errors are from combinations of many effects in our proper motion measurements; photometry, centroid determination, 
astrometry calibrations, spatial variations of the data quality and calibrations, and so on. Hence it is better to estimate the errors 
from data themselves.

Since galaxies and quasars do not have proper motions, we can estimate the measurement errors from the dispersion or width in the distribution 
of apparent proper motions for galaxies/quasars (angular offsets between the HSC and S82 catalogue). However, this is not so easy. For quasars 
we suffer from the low statistics due to a small size of the sample. On the other hand, galaxies are extended objects, and the proper motion 
measurements receive additional scatters from uncertainties in the centroid determination due to the finite size. To overcome this obstacle, 
we choose to use ``small-size'' galaxies. We select such small-size galaxies from galaxies satisfying $-0.3<i_{\rm HSC, cModel} - i_{\rm HSC, PSF}<-0.15$ 
in Figure~\ref{fig:sep}. Here the upper cutoff of $-0.15$ is more conservative than our nominal cut of star/galaxy separation ($-0.08$) to 
have a less contamination of stars. The lower cutoff of $-0.30$ is to remove large-size galaxies. In fact, even if we change these lower and 
upper cutoffs within the range $-0.20<i_{\rm HSC, cModel} - i_{\rm HSC, PSF}<-0.30$, the following results remain almost unchanged. Note that 
we here take advantage of superb image quality of HSC data to identify small-size galaxies. The astrometry of these small-size galaxies 
has been recalibrated in the same way as for stars, using the recalibration map constructed from all galaxies (see
Section~\ref{sec:recalibration}). This removes a contribution of the spatial variations of astrometric systematic effects (the middle panel 
of Figure~\ref{fig:cali}) to the scatters in the apparent proper motions.

Figure~\ref{fig:measureerr} shows the measurement errors estimated from the scatters in the proper motion measurements for the small-size 
galaxies in each bin of the S82 magnitudes, which are estimated from the 25 and 75 percentiles of the proper motion distribution in each bin; 
$\sigma=0.714(q_{\rm 75}-q_{25})$. We divide the galaxies into bins of the SDSS $r$-band magnitudes because the measurement errors are mainly 
from the SDSS data, not the HSC data. The errorbars are estimated through 
bootstrap resampling using 200 realisations in each bin. The orange data points show the results for quasars, which are reasonably similar to 
the galaxy results at magnitude bins down to $r\simeq 22$, with a small offset. The offset indicates the effect of galaxy extendedness; the 
galaxy size larger than the PSF increases scatters in the centroid determination. The error curve for quasars at $r\gtrsim 21.5$ 
are flatter than those of galaxies, but the number of faint quasars in these bins is very small. 
The green solid line shows the theoretical expectation; if the accuracy of the centroid measurements is limited mainly by the photometry accuracy, 
the errors scale with fluxes of objects as $1/{\rm flux}\propto 10^{r-21.729}$, where $r=21.729$ is the best-fit pivot point  to the 
galaxy results at faint magnitudes \citep[also see Eq.~1 in][for the similar discussion]{bond2010milky}. 
Our estimation has a good match to the expectation on $r\gtrsim 22$. On the other hand, the red points are the errors estimated from the centroid
errors of individual galaxies, quoted from the HSC and S82 catalogues (however, the HSC errors are negligible as shown by the purple dots). The 
errors in the catalogues underestimate the measurement errors over all the ranges of magnitudes.

Our measurement errors are taken from those of quasars at $r<22$, and from those of small-size galaxies at $\geq22$. For each star given its 
$r$-band apparent PSF magnitude from S82, we can estimate its measurement error, $\epsilon_i$, according to Figure~\ref{fig:measureerr}. For a
group of $N$ stars, the measurement error of their velocity dispersion is estimated as $\frac{1}{N}\sum_{i=1}^{N}\epsilon_i^2$.

\section{Correction of the DCR effect}
\label{app:dcr}

\begin{figure}
\begin{center}
\includegraphics[width=0.9\columnwidth]{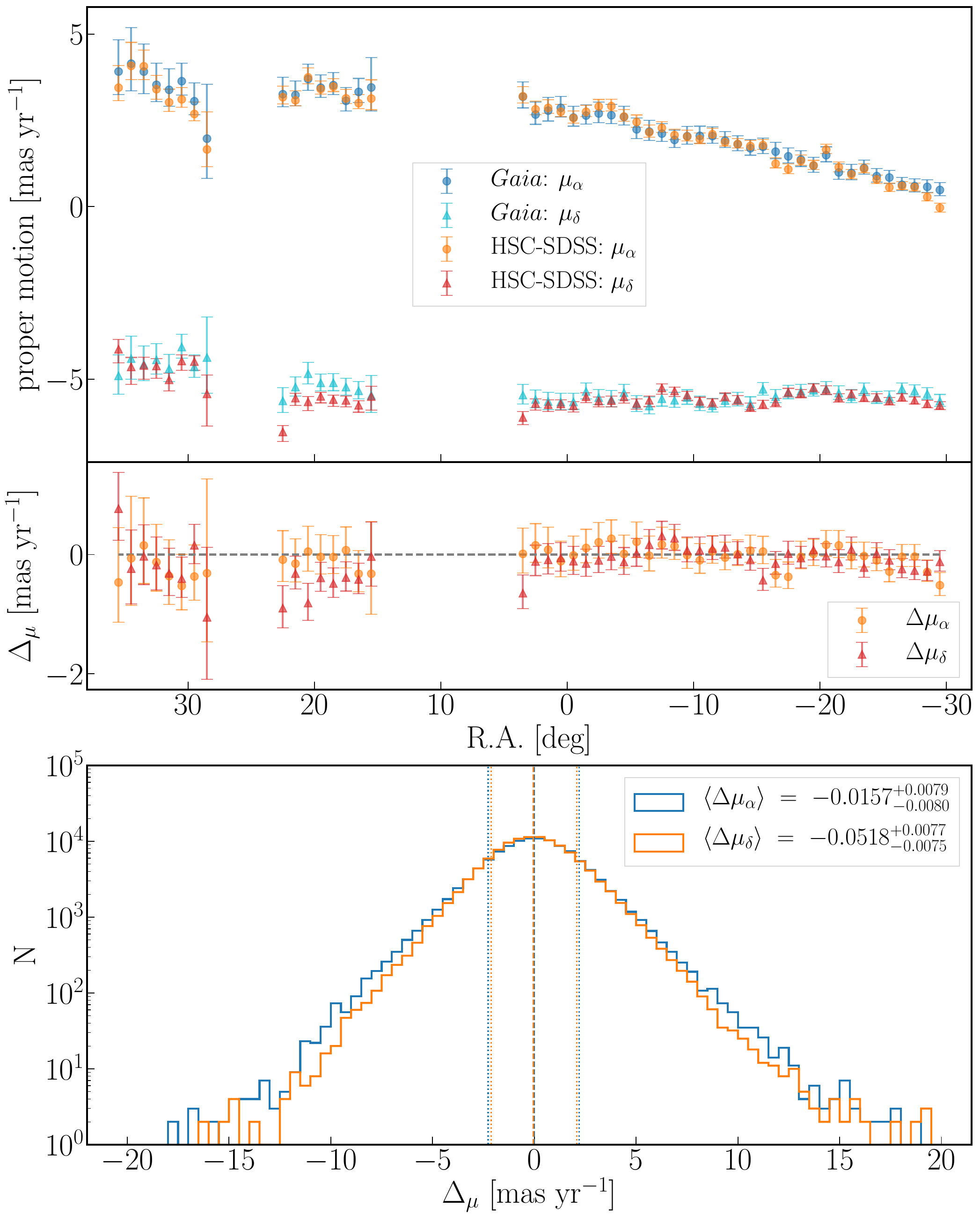}
 \caption{Similar to Figure~\ref{fig:gaia} in the main text, but shows the comparison with {\it Gaia} DR2 proper motions 
 before correcting the DCR effect. The median difference of $\mu_\alpha$ is $-0.0157$~mas~$\mathrm{yr^{-1}}$ 
 while that of $\mu_\delta$ is $-0.0518$~mas~$\mathrm{yr^{-1}}$, marked by the vertical dashed lines. The median difference 
 is about 50 times larger than those in Figure~\ref{fig:gaia}.}
 \label{fig:gaia_nocorr}
 \end{center}
\end{figure}

\begin{figure*}
\begin{center}
\includegraphics[width=0.9\columnwidth]{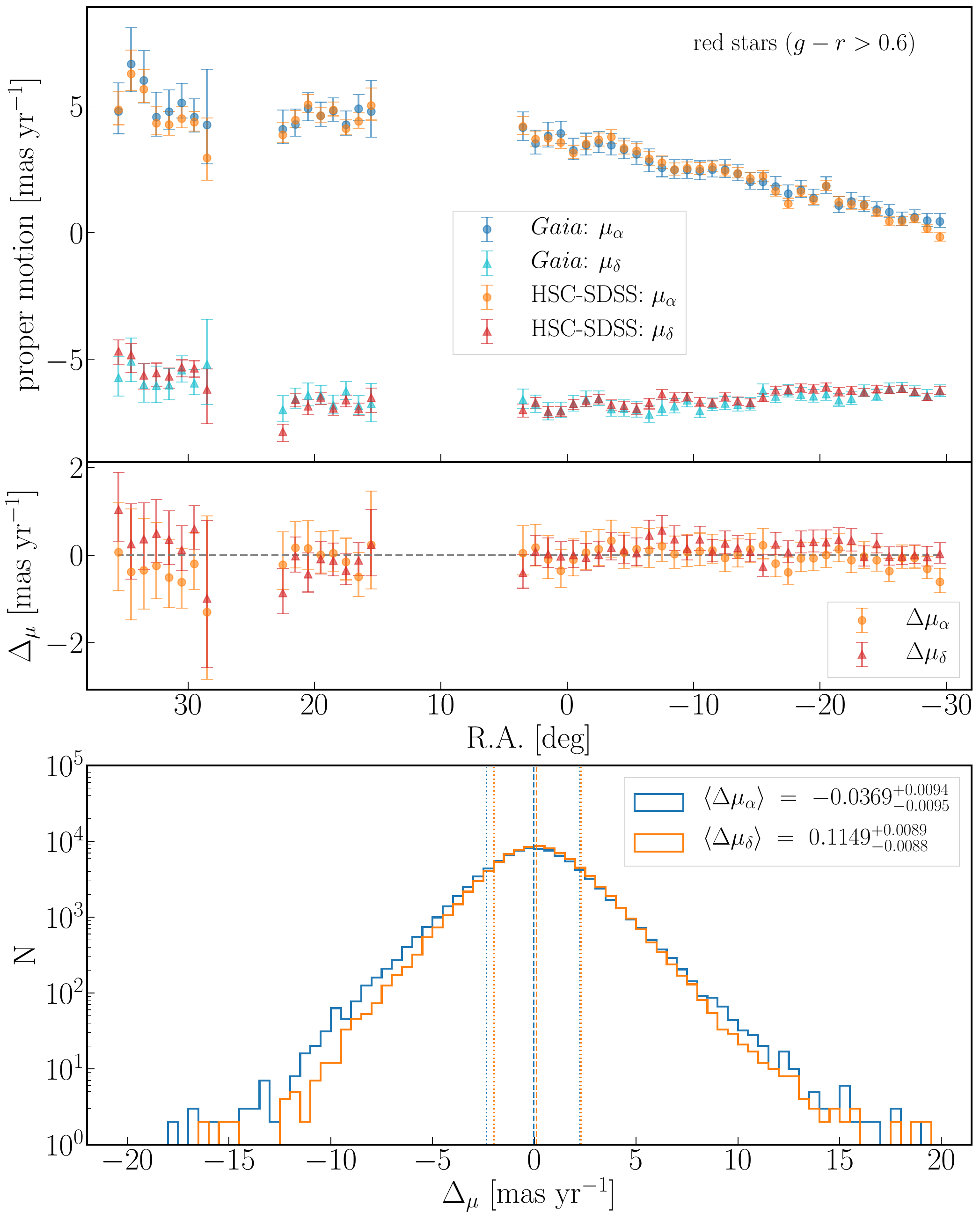}%
\includegraphics[width=0.9\columnwidth]{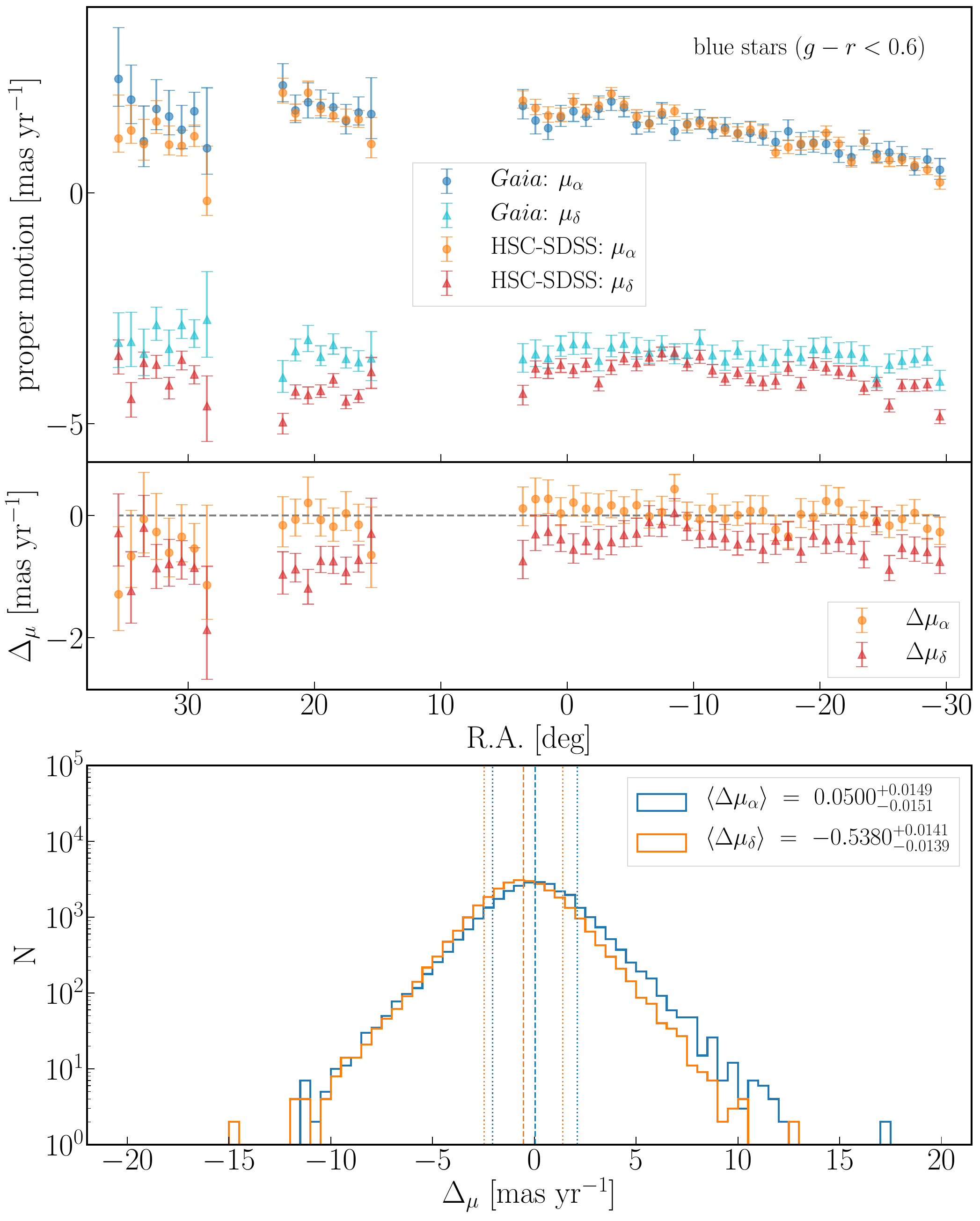}
\caption{Similar to Figure~\ref{fig:gaia}, but the plot shows the comparison of our proper motion measurements against the {\it Gaia} 
DR2 proper motions for red (left) and blue (right) stars, respectively, that are divided by the colour cut of $g-r=0.6$.}
\label{fig:gaia_DCR}
\end{center}
\end{figure*}

\begin{figure}
\begin{center}
\includegraphics[width=0.9\columnwidth]{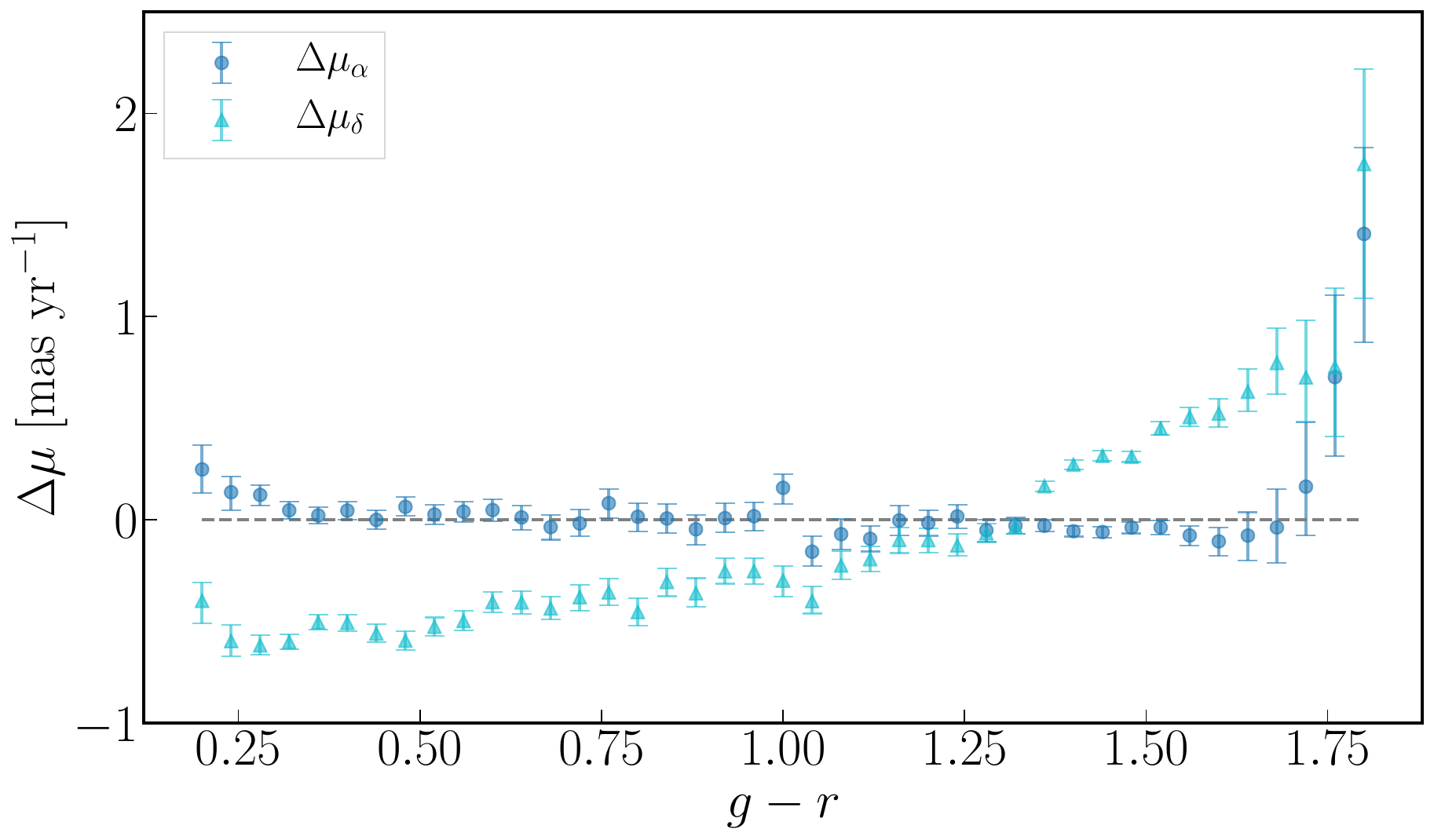}
\includegraphics[width=0.9\columnwidth]{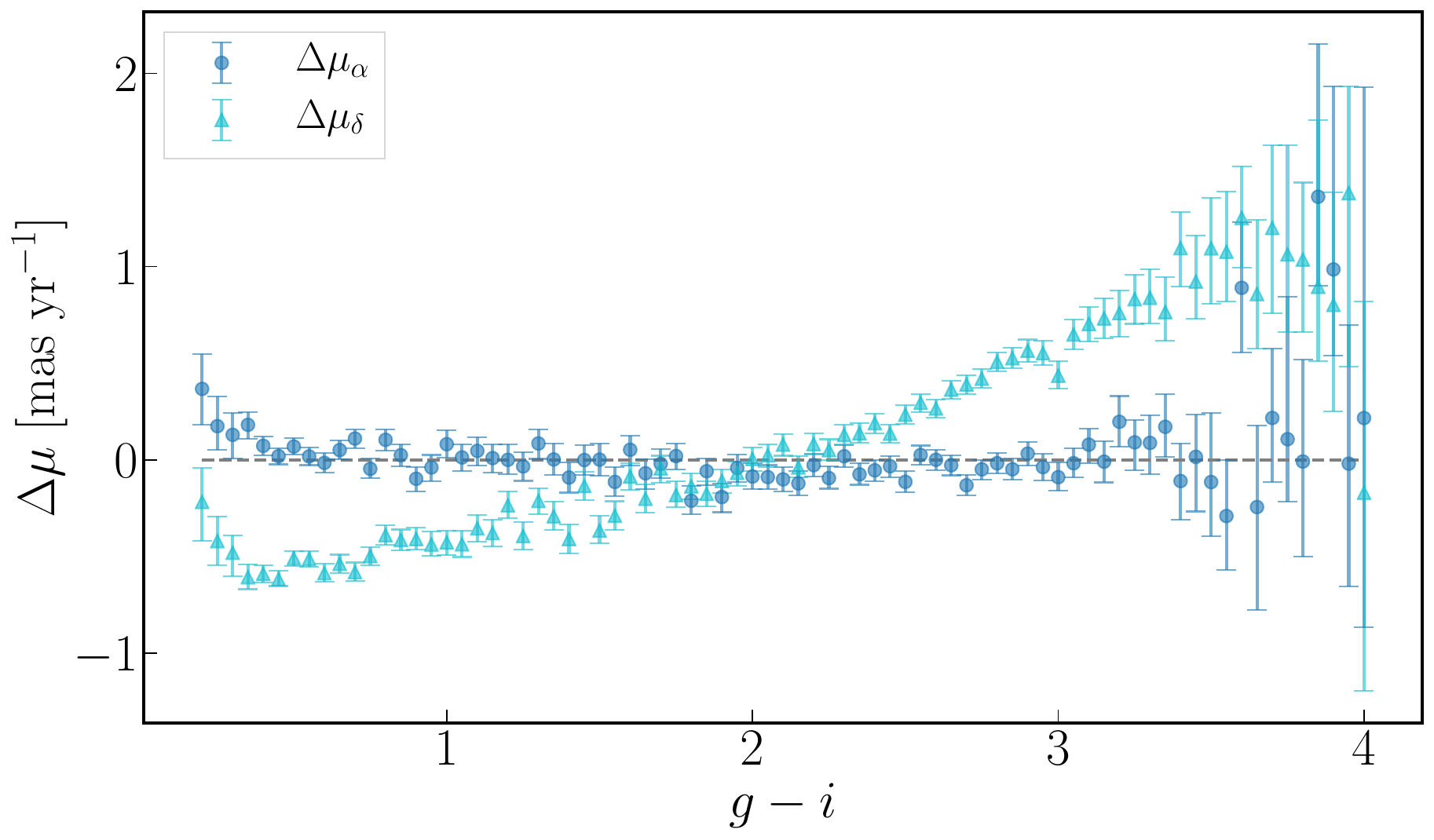}
\caption{The proper motion differences between the HSC-S82 and {\it Gaia} for stars as a function of the colours $(g-r)$ (upper panel)
or $(g-i)$ (lower), where
we estimate the averaged proper motion difference over the survey footprints for subsamples of stars divided in each bin of the colour ($x$-axis). 
Note that samples of blue and red stars we use in our main analysis are defined by the colour cut, $g-r<0.6$ or $>0.6$, respectively. 
\label{fig:DCR_gi}
}
\end{center}
\end{figure}

\begin{figure}
\begin{center}
 \includegraphics[width=0.95\columnwidth]{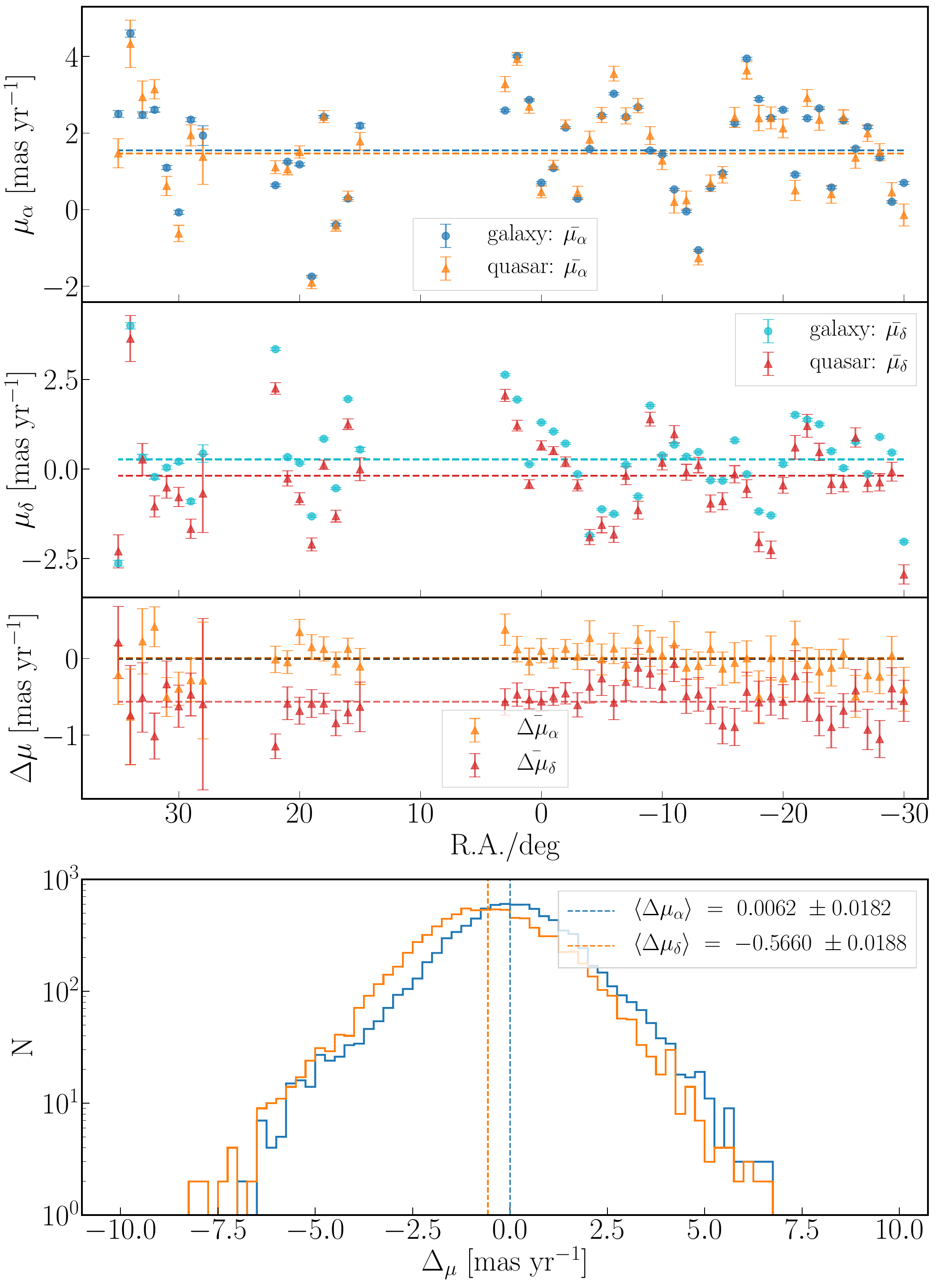}
 \caption{Angular offsets in R.A. and Dec. directions, in units of [${\rm mas}~{\rm yr}^{-1}$] as in Figure~\ref{fig:cali}, 
 measured from the galaxies and quasars in the HSC and S82 catalogues (see Table~\ref{tab:num}) in each R.A. range ($x$-axis).
 Horizontal lines with the corresponding colour denote the level of systematics averaged over all bins of R.A.. The middle 
 panel, with $\Delta\mu$ in the $y$-axis label, denotes the differences between the angular offsets of galaxies and quasars 
 in each R.A. bin. There is a clear offset in the Dec. direction, as expected by the effect of differential chromatic diffraction 
 (DCR) (see text for details). The lower panel shows the distribution of the differences between the angular offsets for 
 individual quasars between HSC and S82, after performing the recalibration of astrometric solutions using the galaxy offsets 
 as in Figure~\ref{fig:cali}. There is a residual systematic error in the Dec. direction, given by $-0.5660$~mas~$\mathrm{yr^{-1}}$
 that is significant compared to the error (more than $2\sigma$). 
 }
 \label{fig:qso}
 \end{center}
\end{figure}

\begin{figure}
\begin{center}
\includegraphics[width=0.98\columnwidth]{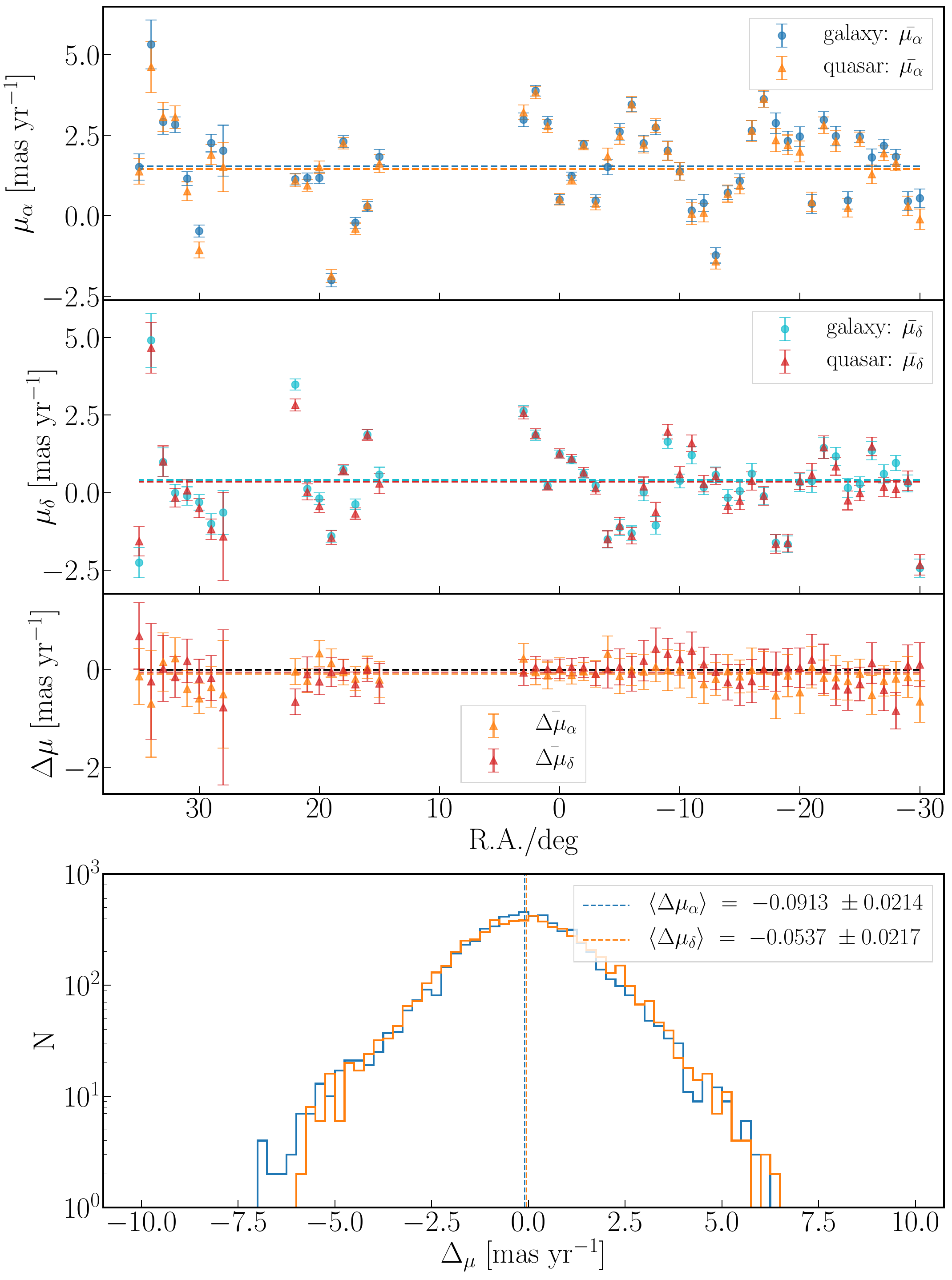}
\caption{Validation of the DCR correction method using the quasars. Here we apply the DCR correction method to individual quasars 
based on their HSC colour $(g-i)$; we infer each quasar's colour in $g-i$ to infer the DCR effects by matching the colour to the DCR-$(g-i)$ 
colour for the {\it Gaia} matched stars in Figure~\ref{fig:DCR_gi}. For the range of $(g-i)$ in Figure~\ref{fig:DCR_gi}, we have 6,288 quasars 
in the HSC-S82 region. The three panels are similar to the previous figure. The upper and lower panels show the angular offsets between the 
HSC and S82 catalogues for the galaxies and the quasars in R.A. and Dec. directions, after implementing the DCR correction. The lower panel 
show the distribution, and the numbers in the legend give the averaged differences between the angular offsets. In doing this, we used 
galaxies in the grids of the calibration map in Figure~\ref{fig:cali} where the quasars in the sample exist.  
\label{fig:qso_dcr}}
\end{center}
\end{figure}

\subsection{DCR correction using the matched stars with {\it Gaia}}

As we stressed in the main text, a correction of the DCR effect is important to achieve the precision of 0.1~mas~yr$^{-1}$ in our proper motion 
measurements, which is important to study the proper motions at large distances $>10~$kpc. As we will show below we find that the centroid positions
of S82 objects are affected by the DCR effect. Here we describe our correction method using the matched stars with {\it Gaia}. 

Figure~\ref{fig:gaia_nocorr} shows the comparison against {\it Gaia} DR2 proper motion measurements, before correcting for the DCR 
effect. Note that here we have applied the recalibration of astrometry using galaxies shown in Figure~\ref{fig:cali}.
The overall agreement looks very good, with the precision of $0.05$~mas~yr$^{-1}$. 
However, we found that the agreement is a coincidence to some extent, after cancellation of the DCR effects for red and blue 
stars as we describe below.

We identified subtle systematic differences as follows. If we repeat the comparison against {\it Gaia} for red 
and blue stars separately, using the colour division of $g-r=0.6$, we find discrepancies for each 
sample from {\it Gaia} as demonstrated in Figure~\ref{fig:gaia_DCR}. The figure shows systematic offsets in the proper motion 
along the Dec. direction for both red and blue stars, which are almost constant across the R.A. regions. The median offsets
averaged over the R.A. regions are $\langle\mu_\delta\rangle\simeq$ 0.115~mas~yr$^{-1}$ and $-0.538$~mas~yr$^{-1}$, for red 
and blue stars, respectively. On the other hand, the measurement of $\mu_\alpha$ is consistent with {\it Gaia}. That is, 
there is no sign of residual systematic error in the R.A. direction.

The systematic effects in Figure~\ref{fig:gaia_DCR} are consistent with the DCR effect. The SDSS camera does not have such a
DCR corrector. DCR stretches an observed image of source along the zenith direction due to the wavelength-dependent refraction 
by atmosphere and, as a result, each object appears to be observed at a shifted position in the camera image by a different 
amount in different filters. For SDSS, the zenith direction is mainly along the Dec. direction for the Stripe~82 region, 
which is consistent with our findings. Since the spectral energy distribution (SED) of galaxies and stars are different, 
our astrometry recalibration method using the galaxy positions (Figure~\ref{fig:cali}) cannot fully correct for the
DCR effects for stars.

Figure~\ref{fig:DCR_gi} further examines how the DCR effects depend on colours of stars. Thanks to the large sample of stars even 
after subdivision based on the colour, the comparison with {\it Gaia} clearly shows that the DCR effects vary depending on colours 
of stars relative to those of galaxies. The figure shows there is almost no DCR effect in the R.A. direction. The $(g-i)$ colour in
Figure~\ref{fig:DCR_gi} displays a wider dynamic range for the DCR effect compared to the $(g-r)$ colour, because very red stars 
would have very similar colour in $g-r$ ($g-r\sim  1.6$), but still have different colours in $g-i$. For blue stars with 
$g-r<0.6$, which roughly corresponds to $g-i<0.6$ as well, the averaged DCR offset is consistent with the amount of difference 
from {\it Gaia} in Figure~\ref{fig:gaia_DCR}. The smaller amount of difference from {\it Gaia} for red and all stars, compared that 
for blue stars, is a result of averaging both positive and negative DCR effects in Figure~\ref{fig:DCR_gi}.

We use the measured offsets in Dec. direction, $\Delta \mu_\delta$, against the $(g-i)$ colour to infer the DCR effect for the matched 
stars with {\it Gaia}. Then, for a given main-sequence star in our main analysis, we infer the DCR effect from the $(g-i)$ colour of the 
star, obtained by linearly interpolating the result of Figure~\ref{fig:DCR_gi}, and then correct for the DCR effect for the star. This is 
our default analysis we use in this paper. Figure~\ref{fig:gaia} shows the results after including the DCR correction. The median differences 
from {\it Gaia} is reduced by construction: $\mu_\alpha\simeq -0.0007$~mas~$\mathrm{yr^{-1}}$ and $\mu_\delta\simeq -0.0010$~mas~$\mathrm{yr^{-1}}$. 
In particular, comparing Figures~\ref{fig:gaia} and \ref{fig:gaia_nocorr} manifests that the discrepancy from {\it Gaia} at R.A.$\sim 20~{\rm deg.}$ 
is significantly improved by our DCR correction method.

\subsection{Quasar validation of the DCR correction method}
\label{app:qso}

As we described in the preceding subsection, the DCR effect is important to take into account. In this subsection we use 
quasars to make an independent confirmation of the DCR effect and also discuss the robustness of astrometry recalibration 
in Section~\ref{sec:recalibration}. 

We adopt spectroscopic quasars from SDSS DR14 to ensure the sample purity. We match their positions in DR14 to those matched HSC-S82 
objects via the source positions from S82. Since the images of most DR14 quasars are taken before DR7, we do not expect their astrometric
positions to be very different from those of S82. We adopt a small angular separation of 0.2$\arcsec$ for the matching. In the end, 
8,757 quasars are matched after a 3-$\sigma$ clipping over the angular offsets between HSC and S82 (see Table~\ref{tab:num}).

We show the median angular offset per year in each R.A. bin for both quasars and galaxies in the upper two panels of Figure~\ref{fig:qso}. 
Note this is before applying any astrometry recalibrations using galaxies (see Section~\ref{sec:recalibration} for more details).
The figure shows a nearly constant offset in the Dec. direction, without strong R.A. dependence, while the offset in R.A. direction 
is consistent with no offset (zero).

The lower panel shows the histogram distributions of angular offsets of quasars, after applying the astrometry recalibration using
galaxies, i.e., the same method as how we recalibrate the astrometry for stars, but we did not include any correction for DCR yet. 
Since distant quasars do not move, any observed proper motions reflect residual systematics in our astrometry recalibration.
There is almost no bias in R.A., while it shows a non-negligible bias in the Dec. direction that amounts to $-0.566$~mas~yr$^{-1}$.

The level of residual systematics is similar to the difference between quasars and galaxies shown in the top and middle panels of
Figure~\ref{fig:qso}, and is in good agreement with the median difference against {\it Gaia} DR2 proper motions for blue stars in
Figure~\ref{fig:gaia_DCR}, which is caused by the effect of DCR. Quasars are quite blue objects suffering more from DCR while 
galaxies can have very different SED compared with quasars. Because we use galaxies to recalibrate the astrometry for quasars, the
difference in quasar and galaxy positions due to DCR remains in the recalibrated results, revealing it as residual systematic errors. 

The reference bands for astrometric measurements are $r$-band and $i$-band for S82 and HSC, respectively. We repeat the comparison 
between quasars and galaxies for the positions in other SDSS filters. We find the difference between quasars and galaxies is larger
in $g$-band, which is about $-1$~mas~yr$^{-1}$ in $\mu_\delta$, but decreases significantly in $r$ and $i$-bands. This is consistent 
with the expected effect caused by DCR, which is more severe in blue bands. With these results we conclude that the discrepancy 
between quasars and galaxies is resulted from the effect of DCR. 

In Figure~\ref{fig:qso_dcr} we further validate the DCR effect for quasars. Here, as we did for the main results, we apply the DCR 
correction to individual quasars based on their colour $(g-i)$ in Figure~\ref{fig:DCR_gi}. If the angular offsets of quasars in Figure~\ref{fig:qso} 
are due to the DCR effect,  we expect that our method can correct for the angular offsets, because the DCR effects depend primarily on 
colours of objects (even if the SEDs of quasars are different from those of the {\it Gaia}-matched stars used to construct the DCR 
correction template). We have 6,288 quasars in the range of the $(g-i)$ colour in Figure~\ref{fig:DCR_gi} in the HSC-S82 region. 
Figure~\ref{fig:qso_dcr} shows that our method remarkably well corrects for the angular offsets of quasars to the accuracy of 
less than $0.1~$mas~yr$^{-1}$ in both R.A. and Dec. directions. With these results, we conclude that the primary systematic errors 
in our astrometric solutions are the DCR effect, and our correction method has a very good control of the DCR effect by less 
than 0.1~mas~yr$^{-1}$.


\bsp	
\label{lastpage}
\end{document}